\documentclass[prd,preprint,tightenlines,floatfix,showpacs,preprintnumbers,nofootinbib,eqsecnum]{revtex4}

 \usepackage[dvips,final]{graphicx}
  \usepackage{amssymb}
   \usepackage{amsmath}
    \usepackage{amsfonts}
     \usepackage{epsfig}
     \usepackage{color}
      \usepackage{bm} % bold math

\usepackage{booktabs}
\usepackage{multirow}
\usepackage{siunitx}

\usepackage{mathrsfs}
\usepackage{slashed}
\usepackage{comment}

 \newcommand\beq{\begin{equation}}
 
 \newcommand\eeq{\end{equation}}
 \newcommand\beqn{\begin{eqnarray}}
 \newcommand\eeqn{\end{eqnarray}}

 \def\gsim{\mathrel{\rlap{\lower4pt\hbox{\hskip1pt$\sim$}}
 \raise1pt\hbox{$>$}}}

 \def\fm{\,\mbox{fm}}
 
 \def\GeV{\,\mbox{GeV}}
 
 \def\lsim{\mathrel{\rlap{\lower4pt\hbox{\hskip1pt$\sim$}}
     \raise1pt\hbox{$<$}}}         %less than or approx. symbol
 \def\gsim{\mathrel{\rlap{\lower4pt\hbox{\hskip1pt$\sim$}}
     \raise1pt\hbox{$>$}}}         %greater than or approx. symbol

 \def\beq{\begin{equation}}
 \def\eeq{\end{equation}}   
 
 \def\beqy{\begin{eqnarray}}
 \def\eeqy{\end{eqnarray}}
 \def\dfr{\mathrm{d}}   
\def\Jpsi{J\!/\!\psi}   
\def\psip{\psi'}
\def\Y{\Upsilon}
\def\Yp{\Upsilon'}
\def\Ypp{\Upsilon''}
\def\r{\tilde{r}}
\begin{document}

\title{Theoretical uncertainties in exclusive electroproduction\\ 
of $S$-wave heavy quarkonia}

\author{Jan Cepila$^{1}$}
\email{jan.cepila@fjfi.cvut.cz}

\author{Jan Nemchik$^{1,2}$}
\email{nemcik@saske.sk}

\author{Michal Krelina$^{1,3}$}
\email{michal.krelina@usm.cl}

\author{Roman Pasechnik$^{4,5,6}$}
\email{roman.pasechnik@thep.lu.se}

\affiliation{
{$^1$\sl 
Czech Technical University in Prague, FNSPE, B\v rehov\'a 7, 11519 
Prague, Czech Republic
}\vspace{0.5cm}\\
{$^2$\sl
Institute of Experimental Physics SAS, Watsonova 47, 04001 Ko\v 
sice, Slovakia
}\vspace{0.5cm}\\
{$^3$\sl
Departamento de F\'{\i}sica,
Universidad T\'ecnica Federico Santa Mar\'{\i}a,
Casilla 110-V, Valpara\'{\i}so, Chile
}\vspace{0.5cm}\\
{$^4$\sl
Department of Astronomy and Theoretical Physics, Lund
University, SE-223 62 Lund, Sweden
}\vspace{0.5cm}\\
{$^5$\sl
Nuclear Physics Institute ASCR, 25068 \v{R}e\v{z}, Czech Republic
}\vspace{0.5cm}\\
{$^6$\sl Departamento de F\'isica, CFM, Universidade Federal 
de Santa Catarina, C.P. 476, CEP 88.040-900, Florian\'opolis, 
SC, Brazil
}\vspace{0.5cm}
}

%%%%%%%%%%%%%%%%%%%%%%%%%%%%%%%%%%%%%%%%%%%%%%%%%%%%%%%%%%%%%%%%%%%%%%%
\begin{abstract}
\vspace{0.5cm}
In this work, we revise the conventional description of $\Jpsi(1S)$, $\Y(1S)$, $\psip(2S)$ and $\Yp(2S)$ elastic photo- and electroproduction off a nucleon target within the color dipole picture and carefully study various sources of theoretical uncertainties in calculations of the corresponding electroproduction cross sections. For this purpose, we test the corresponding predictions using a bulk of available dipole cross section parametrisations obtained from deep inelastic scattering data at HERA. Specifically, we provide the detailed analysis of the energy and hard-scale dependencies of quarkonia yields employing the comprehensive treatment of the quarkonia wave functions in the Schr\"odinger equation based approach for a set of available $c-\bar c$ and $b-\bar b$ interquark interaction potentials. Besides, we quantify the effect of Melosh spin rotation, the $Q^2$-dependence of the diffractive slope and an uncertainty due to charm and bottom quark mass variations.
\end{abstract}
%%%%%%%%%%%%%%%%%%%%%%%%%%%%%%%%%%%%%%%%%%%%%%%%%%%%%%%%%%%%%%%%%%%%%%%


\maketitle

%
%
%
%======================
\section{Introduction}
\label{Sec:Intro}
%======================
%
%
%

One of the major widely used probes for interplay between hard and soft QCD physics is by means of bound states of heavy (charm and bottom) quarks known as quarkonia. Among these, the most well-studied are $S$-wave $\Jpsi$, $\psip$ and $\Y$ states produced in high-energy particle collisions (for a detailed review on quarkonia physics, see e.g.~Refs.~\cite{Brambilla:2010cs,Ivanov:2004ax,Brambilla:2004wf}). 

Despite a notable progress in theoretical description of heavy quarkonia production done over past few decades, the quarkonia production mechanism, as well as their propagation and dissociation in a hot medium, is considered to be an important probe for the medium created in heavy-ion collisions \cite{satz}, and is still an actively developing research area. The problem concerns highly uncertain rates of $\Jpsi$ and $\psip$ mesons production in $pp$ and $pA$ collisions. These processes are also considered to be among the main tools for studying the soft QCD effects in hard processes. 

The wealth of existing experimental data and theoretical studies show that the widely used simplifications in the analysis of exclusive quarkonia electroproduction observables may have significant impact on theoretical predictions and thus should be taken with care. One of the important ingredients of quarkonia production observables are the light-cone (LC) wave functions of heavy quarkonia. A popular simple model for the quarkonia wave functions is based upon an assumption that the potential between the bound heavy quarks is perfectly harmonic and no spin rotation in the quarkonia formation is considered (see e.g. Refs.~\cite{Frankfurt:1995jw,Nemchik:1996cw}). Such a treatment is usually performed in the conventional non-relativistic QCD (NRQCD) framework without an account for a non-trivial dependence on intrinsic transverse momenta and longitudinal momentum fractions of heavy quarks. In the case of charmonia production, a significance of non-perturbative and relativistic effects is often underestimated since the charm quark mass is not sufficiently large. Moreover, a spin rotation of heavy quark spinors from the $Q\bar Q$ rest frame to the infinite momentum frame known as Melosh rotation \cite{Melosh:1974cu,Hufner:2000jb}, which influences mainly the angular part of the wave function, has a notable impact on (in particular, $S$-wave) charmonia differential observables \cite{Hufner:2000jb,jan-18} while it is sometimes neglected in the existing calculations. A properly formulated radial part of the quarkonium wave function should be obtained by a numerical solution of the Schr\"odinger equation for a realistic $Q\bar Q$ potential and with an appropriate boosting and spin rotation. This work is aimed, in particular, at a proper accounting for and studying various sources of theoretical uncertainties in modeling the elastic $\Jpsi$, $\psip$ and $\Y$ electroproduction processes in $\gamma p$ collisions in the color dipole picture \cite{Nemchik:1994fp}.

One of the major problems of the QCD scattering theory is to correctly identify the underlying degrees of freedom which are the eigenstates of interaction. In last decades, it became clear that such eigenstates are color dipoles \cite{Kopeliovich:1991pu,Kopeliovich:1993pw,Nemchik:1994fp,Nemchik:1996cw}, the universal elementary building blocks automatically accumulating both the hard and soft fluctuations \cite{Kopeliovich:1981pz,Nikolaev:1994kk}. In particular, the light-cone (LC) color dipole framework has been developed and applied in treatment of both diffractive and inclusive quarkonia electro- and photoproduction processes in QCD in terms of certain superpositions of these eigenstates \cite{Nemchik:1994fp,Nemchik:1996cw}. The dipole picture has turned out to be rather successful in describing various hard processes beyond conventional QCD factorisation \cite{Kopeliovich:1999am}. In particular, it is known to give as precise predictions e.g. for the Drell-Yan cross section as the Next-to-Leading-Order (NLO) collinear factorisation framework (see e.g. Ref.~\cite{Basso:2015pba} and references therein). In deep inelastic scattering (DIS) or in vector meson production the virtual photon is expected to produce the quark-antiquark dipole with a transverse separation depending on the photon virtuality. The dipole formalism relies on a specific type of factorisation (or dipole factorisation) when a scattering cross section is written in impact parameter space as a convolution of the LC wave functions (for e.g. $\gamma^*\to c\bar c$ and $c\bar c\to J/\psi$ fluctuations in the case of $\Jpsi$ electroproduction) and the universal phenomenological dipole cross section fitted to the DIS data. One of the well-known coloured medium effects predicted by QCD is the so called colour transparency \cite{Bertsch:1981py}, an intrinsic feature of the dipole framework, when the medium becomes more transparent for smaller-size dipole configurations \cite{Kopeliovich:1991pu,Kopeliovich:1993pw,Kopeliovich:1993gk}.

Traditionally, exclusive (or elastic) photo ($Q^2\simeq 0$) and electro ($Q^2\gg0$) production of heavy quarkonia receives a lot of attention due to particularly clean signatures and precision measurements of the corresponding observables differential in hard scale, $Q^2$, energy, $W$, and momentum transfer squared, $t$. Such processes are highly relevant for e.g. a better understanding of the gluon density properties and their impact parameter profile in a target at very small $x$ \cite{Ryskin:1995hz,Marquet:2007qa,Dosch:2009fam,Jones:2013pga,Cisek:2014ala,Armesto:2014sma,Jones:2015nna}, as well as for probing the details of the quarkonia production and propagation processes. Indeed, the exclusive photo production cross sections are given by the gluon density in the second power, thus enabling us to probe it to a much better precision than in inclusive reactions whose differential cross sections are proportional to the first power of the (predominantly, gluon) parton density function (PDF). Our current study aims at a comprehensive analysis of these observables and related theory uncertainties in the dipole picture against the available data. The color dipole formalism \cite{Kopeliovich:1981pz,Kopeliovich:1991pu,Kopeliovich:1993pw,Nemchik:1994fp,Nemchik:1996cw}, well known for particularly successful description of various photo and hadro production reactions in both $pp$ and $pA$ collisions, enables to include systematically the QCD factorisation breaking and nuclear coherence effects as well as the initial-state and saturation phenomena \cite{GolecBiernat:1998js,GolecBiernat:1999qd}. Consequently, there is a notable sensitivity of exclusive photoproduction observables to different gluon saturation models at low-$x$ that is worth a careful study providing an efficient discriminating tool for various existing parametrisations for the low-$x$ gluon density at a periphery of the target nucleon.

As a starting point, we would like to test various models of the LC wave functions with different $c-\bar c$ and $b-\bar b$ interquark interaction potentials and phenomenological dipole parametrisations against the recent data on $\Jpsi$, $\psip$ and $\Y$ photo- and electroproduction as well as to study the impact of Melosh spin rotation in these observables. Of particular interest for us is the study of $\psip$ production observables which are known to be highly sensitive to the shape of the wave function (in particular, to the position of its node) than $\Jpsi$ observables \cite{Hufner:2000jb}. Besides, we would like to estimate an overall theoretical uncertainty in the exclusive cross sections due to poorly known gluon density at low-$x$. This is accounted for by using several representative dipole parameterisations widely used in the literature. Finally, such effects as sensitivity to the heavy quark mass, to the skewness in the gluon density, and to the diffractive slope parameterisation are quantified.

The paper is organised as follows. In Section~\ref{Sec:formalism}, we provide the basic details of the dipole approach to the exclusive quarkonia electroproduction with proper definitions of the kinematical variables, the elastic amplitude and the cross section. A thorough description of the normalised LC quarkonia wave functions, the boosting procedure and an overview of the interquark potential models used in our analysis is given in Section~\ref{Sec:QQ-wf}.
Further on, in Section~\ref{Sec:dipole-CS}, we present a brief overview of the most frequently used dipole parameterisations that will be employed in our numerical analysis for estimation of the underlined uncertainties in QCD modeling of the target gluon density evolution at small-$x$. 
In Section~\ref{Sec:results} we compare our model predictions for the photo- and electroproduction cross sections of different quarkonia with available data and analyze the theoretical uncertainties caused by various sources and ingredients coming into the color dipole formalism. Final remarks and conclusions are summarized in Section~\ref{Sec:conclusions}.

%
%
%
%====================================================================
\section{Exclusive quarkonia electroproduction: dipole picture}
\label{Sec:formalism}
%====================================================================
%
%
%

\begin{figure}[!hbt]
\begin{minipage}{0.47\textwidth}
 \centerline{\includegraphics[width=1.0\textwidth]{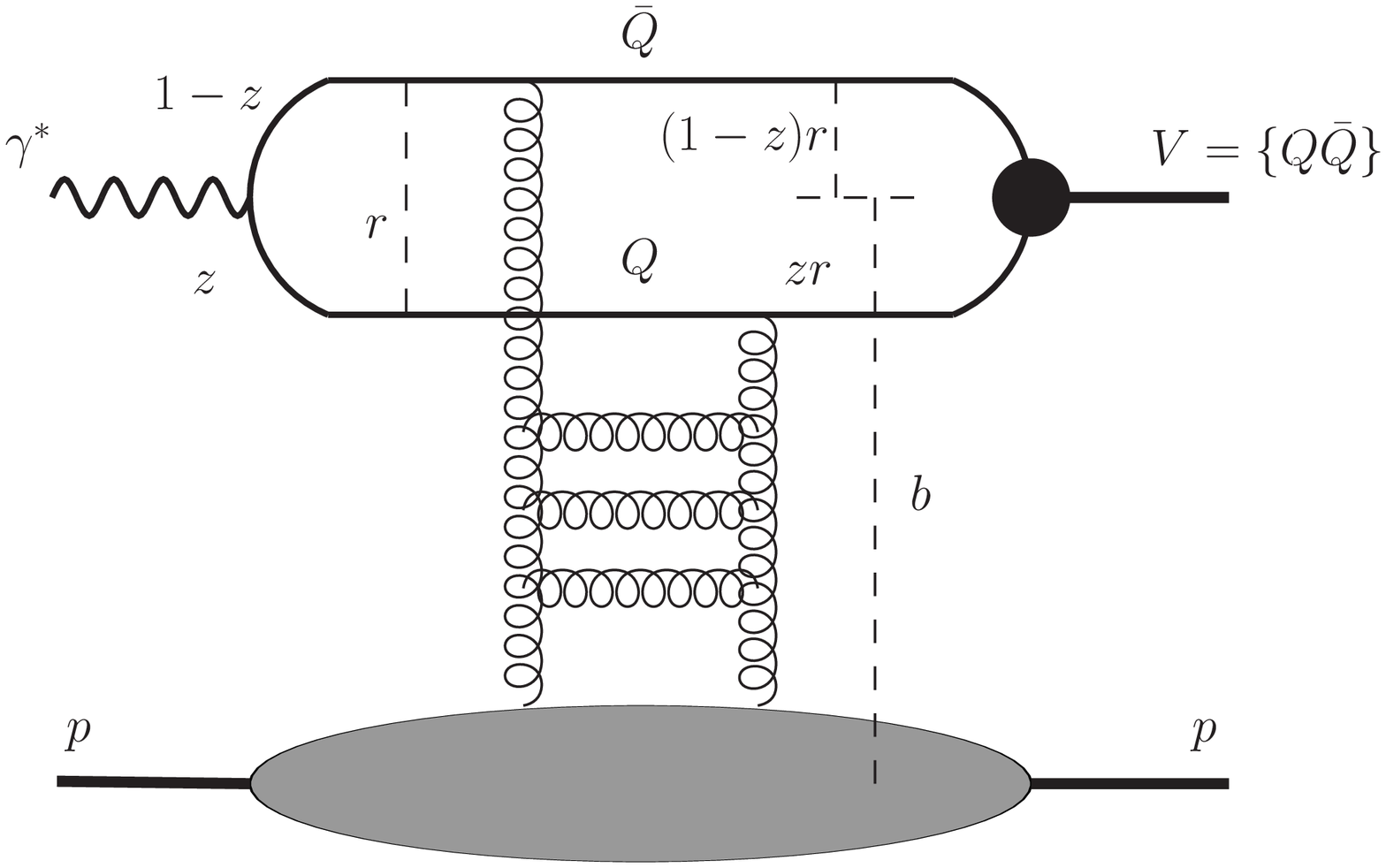}}
\end{minipage}
\begin{minipage}{0.47\textwidth}
 \centerline{\includegraphics[width=1.0\textwidth]{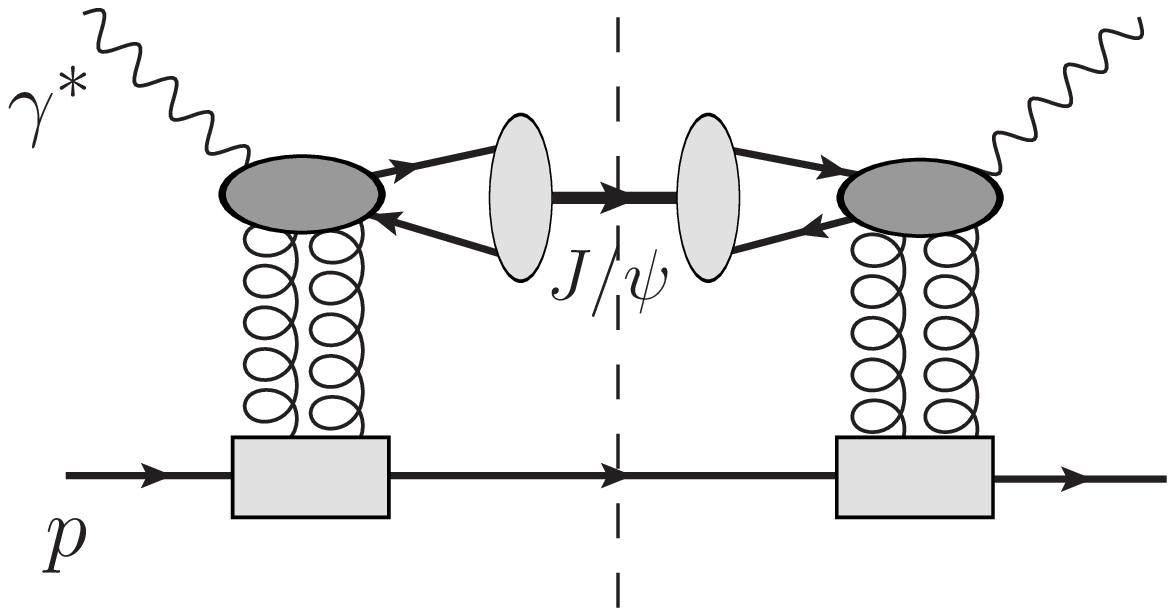}}
\end{minipage}
\caption{A schematic illustration of the exclusive quarkonium 
electroproduction process, $\gamma^*\,p\to V\,p$, in the dipole picture.
On the left panel, the structure of the amplitude and 
kinematic variables in impact parameter space are depicted
while its amplitude squared for the $\Jpsi$ electroproduction 
is shown on the right panel.}
%%%%%%%%%%%%%%%%%%%
\label{fig:diagram}
%%%%%%%%%%%%%%%%%%%
\end{figure}

In the framework of color dipole approach \cite{Kopeliovich:1981pz,Kopeliovich:1991pu,Kopeliovich:1993pw,Nemchik:1994fp,Nemchik:1996cw}, the projectile (real or virtual, with $q^2=-Q^2$) photon undergoes strong interactions via its Fock components containing quarks and gluons with the target proton in the frame where the target proton is at rest. In the dipole picture, such interactions are described by the universal dipole cross section, which is not derivable from the first principles but, instead, is fitted to e.g.~HERA data (for more details, see below). In the case of exclusive vector meson electroproduction illustrated in Fig.~\ref{fig:diagram} (left panel), such a lowest Fock state corresponds to the $Q\bar Q$ dipole whose transverse size $r$ is nearly frozen in the high energy limit. Once the dipole scattering occurs, a coherent $Q\bar Q$ state forms a vector meson by means of a projection of the $Q\bar Q$ production amplitude on to a given LC quarkonium wave function. Let us now briefly describe the main ingredients of the dipole formulation of this process.
%
%                      Fig.1
%================================================
%================================================
%
%

The forward amplitude for exclusive electroproduction of a vector meson $V$ with mass $M_V$ in the target rest frame is given by (see e.g.~Ref.~\cite{Hufner:2000jb} and references therein)
%
%**************************************************
\begin{eqnarray}
\mathrm{Im}\mathcal{A}^{\gamma^{\ast}p\rightarrow Vp}_{T,L}(x,Q^2)=\int\dfr^{2}r\int\limits_{0}^{1}\dfr z \,\Psi^{\dagger}_{V}(r,z)\,
\Psi_{\gamma^{\ast}_{T,L}}(r,z;Q^2)\sigma_{q\bar q}(x,r) \,, \;\; x=\frac{M_V^2+Q^2}{s} \,,
%%%%%%%%%%%%%%%%%
\label{prod-amp}
%%%%%%%%%%%%%%%%
\end{eqnarray}
%**************************************************
%
where $x$ is the standard Bjorken variable \cite{Ryskin:1995hz}, $s=Q^2+W^2$ is the square of the $ep$ center-of-mass energy (with $W$ being the $\gamma^*p$ center-of-mass energy), $\Psi_{V}(r,z)$ is the vector meson $V$ wave function, $\Psi_{\gamma^{\ast}_{T,L}}(r,z;Q^2)$ is the LC distribution (or wave) function of a transversely ($T$) or longitudinally ($L$) polarized virtual photon for a $Q\bar Q$ fluctuation, $\vec r$ is the transverse size of the $Q\bar Q$ dipole, and $z=p_Q^+/p_{\gamma}^+$ is the boost-invariant fraction of the photon momentum $p_{\gamma}^+=E_{\gamma}+p_{\gamma}$ carried by a heavy quark (or anti-quark). The universal dipole cross section $\sigma_{q\bar q}(x,r)$ describes the dipole scattering off the target. It is typically fitted to the precision inclusive DIS data at HERA and then is used to describe a variety of other processes in $ep$ and $pp$ collisions (see below). In the NRQCD limit, one neglects relative motion of $Q$ and $\bar Q$ such that $z=1/2$, and the LC wave function reduces to $\Psi_{V}(r,z)\propto \delta(z-1/2)$ \cite{Kopeliovich:1991pu}. In what follows, we go beyond this approximation.

The perturbative LC $\gamma^{\ast} \to Q\bar Q$ wave function is given by \cite{Kogut:1969xa,Bjorken:1970ah}
%
%**************************************************
\begin{eqnarray}
\Psi^{(\mu,\bar\mu)}_{\gamma^{\ast}_{T,L}}(r,z;Q^2)=\frac{\sqrt{N_c\alpha_{\rm em}}}{2\pi}Z_Q\,\chi_Q^{\mu\dag}
\hat{\mathcal{O}}_{T,L}\tilde\chi_{\bar Q}^{\mu}\,K_0(\varepsilon r) \,,
\qquad
\varepsilon^2=z(1-z)Q^2+m_Q^2 \,,
%%%%%%%%%%%%%%%%%
\label{gamma-wf}
%%%%%%%%%%%%%%%%%
\end{eqnarray}
%**************************************************
%
where $\varepsilon$ and $Z_Q$ are the energy and the electric charge of the heavy quark ($Z_c=2/3$, $Z_b=1/3$), $\chi^\mu_Q$ and $\tilde\chi^{\bar\mu}_{\bar Q}\equiv 
i\sigma_y\chi^{\bar\mu\ast}_{\bar Q}$ are the two-component spinors in the infinite momentum frame normalized as follows \cite{Kopeliovich:2001hf},
%
%**************************************************
\begin{eqnarray}
\sum\limits_{\mu,\bar \mu}\left(\chi^{\mu\dag}_Q\hat A\tilde\chi^{\bar\mu}_{\bar Q}\right)^\ast\left(\chi^{\mu\dag}_Q\hat B\tilde\chi^{\bar\mu}_{\bar Q}\right)=
\mathrm{Tr}(\hat A^\dag\hat B)\,,
\end{eqnarray}
%**************************************************
%
and $K_0(\varepsilon r)$ is the modified Bessel function of the second kind. The operators $\hat{\mathcal{O}}_{T,L}$ are defined as follows,
%
%**************************************************
\begin{eqnarray}
\mathcal{O}_T &=& m_Q\vec\sigma\cdot\vec e_\gamma + i(1-2z)(\vec\sigma\cdot\vec n)(\vec e_\gamma\cdot\vec\nabla_r) +
(\vec n\times\vec e_\gamma)\vec\nabla_r \,, \nonumber\\
\mathcal{O}_L &=& 2Qz(1-z)\vec\sigma\cdot \vec n \,, \qquad \vec\sigma=(\sigma_x,\sigma_y,\sigma_z) \,, \qquad \vec\nabla_r \equiv \partial/\partial \vec{r} \,,
\end{eqnarray}
%**************************************************
%
where $\vec e_\gamma$ is the transverse photon polarisation vector, $\vec n=\vec p_{\gamma}/|\vec p_{\gamma}\,|$ is a unit vector along the photon momentum, and $\sigma_{x,y,z}$ are the Pauli matrices. In what follows, following Ref.~\cite{Hufner:2000jb} we neglect the effects associated with non-perturbative interactions within the heavy quark pair (including charm quarks) and that are not included into the corresponding interaction potential since $m_Q$ provides a sufficiently perturbative scale even in the photoproduction limit $Q^2 \to 0$.

The quarkonium wave function is properly defined only in the $Q\bar Q$ rest frame where it can be directly found by solving the Schr\"odinger equation. Below, we discuss such solutions for several distinct interquark potentials. In order to obtain the production amplitude given by Eq.~(\ref{prod-amp}), the quarkonium wave function should be found in the infinite momentum frame. For a classical $Q\bar Q$ configuration, such a wave function could be computed from that in the $Q\bar Q$ rest frame by a applying a Lorentz boost. The quantum effects, however, are relevant such that a tower of Fock states emerges as a result of such a transformation \cite{Hufner:2000jb}, and the lowest Fock $|Q\bar Q\rangle$ components in these frames do not represent the same configuration. This issue is further discussed in the next Section.

In what follows, we assume a simple factorization of spatial and spin-dependent parts of the vector meson $V$ wave function such as
%
%***********************************************
\begin{eqnarray}
\Psi_{V}^{(\mu,\bar\mu)}(z,\vec p_T)=U^{(\mu,\bar\mu)}(z,\vec p_T)\Psi_{V}(z,p_T) \,,
%%%%%%%%%%%%%%
\label{PsiQQ}
%%%%%%%%%%%%%
\end{eqnarray}
%***********************************************
%
where
%
%***********************************************
\begin{eqnarray}
U^{(\mu,\bar\mu)}(z,\vec p_T)=\frac{1}{\sqrt{2}}\xi_Q^{\mu\dag}\vec\sigma\vec e_{V}\tilde\xi_{\bar Q}^{\bar \mu}\,,\qquad 
\tilde\xi_{\bar Q}^{\bar\mu}=i\sigma_y\xi_{\bar Q}^{\bar\mu\ast} \,, 
%%%%%%%%%%%%
\label{Uini}
%%%%%%%%%%%%
\end{eqnarray}
%***********************************************
%
in terms of the vector meson polarisation vector $\vec e_{V}$ and quark spinors $\xi$ in the meson rest frame. The latter are related to spinors $\chi$ in the infinite momentum frame as follows
%
%***********************************************
\begin{eqnarray}
\xi^\mu_Q = R(z,\vec p_T)\chi_Q^\mu \,, \qquad 
\xi_{\bar Q}^{\bar\mu}=R(1-z,-\vec p_T)\chi_{\bar Q}^{\bar\mu} \,,
\end{eqnarray}
%***********************************************
%
known as the Melosh spin rotation \cite{Melosh:1974cu,Hufner:2000jb}. The $R$-matrix of such a rotation is given by
%
%***********************************************
\begin{eqnarray}
R(z,\vec p_T)=\frac{m_Q + zM_V - i(\vec\sigma\times\vec n)\vec p_T}{\sqrt{(m_Q+zM_V)^2+p_T^2}} \,.
\end{eqnarray}
%***********************************************
%

Using the quarkonium wave function given by Eq.~(\ref{PsiQQ}) we assume that the vertex $\psi \to c\bar{c}$ differs from the photon-like $\gamma^{\ast} \to c\bar{c}$ vertex with the structure $\psi_\mu \bar{u} \gamma^\mu u $ used in Refs.~\cite{Ryskin:1992ui,Brodsky:1994kf,Frankfurt:1995jw,Nemchik:1996cw}. As was noticed in Ref.~\cite{Hufner:2000jb}, the latter in the $c\bar c$ rest frame contains both $S$- and $D$-wave states whereas the $D$-wave weight is correlated strongly with the structure of the vertex and cannot be proved by any reasonable nonrelativistic $c-\bar c$ interaction potential.

Substituting
%
%***********************************************
\begin{eqnarray}
\tilde\xi_{\bar Q}^{\bar\mu}
=
i\sigma_yR^\ast(1-z,-\vec p_T)(-i)\sigma_y^{-1}\tilde\chi_{\bar Q}^{\bar\mu} \,, \qquad
\xi_Q^{\mu\dag}=\chi_Q^{\mu\dag}R^{\dag}(z,\vec p_T) \,,
\end{eqnarray}
%***********************************************
%
into Eq.~(\ref{Uini}) one gets finally
%
%***********************************************
\begin{eqnarray}
U^{(\mu,\bar\mu)}(z,\vec p_T)
=
\frac{1}{\sqrt{2}}\,\chi_Q^{\mu\dag}R^\dag(z,\vec p_T)\,
\vec\sigma\cdot\vec e_{V}\,\sigma_y R^\ast (1-z,-\vec p_T)\,
\sigma_y^{-1}\tilde\chi_{\bar Q}^{\bar \mu} \,.
\end{eqnarray}
%************************************************
%

The resulting dipole formula for the amplitude of photo and electroproduction of heavy quarkonia reads
%
%************************************************
\begin{eqnarray}
\mathrm{Im}\mathcal{A}^{\gamma^{\ast}p\rightarrow Vp}_{T,L}(x,Q^2)
=
\int\limits_0^1\dfr z \int\dfr^2r\, \Sigma_{T,L}(z,\vec r;Q^2)\, \sigma_{q\bar q}(x,r)
\end{eqnarray}
%*************************************************
%
where
%
%*************************************************
\begin{eqnarray}
\Sigma_{T,L}(z,\vec r;Q^2)=\int\frac{\dfr^2 p_T}{2\pi}e^{-i\vec p_T\vec r}\,\Psi_{V}(z,p_T)\sum\limits_{\mu,\bar\mu}U^{\dagger(\mu,\bar\mu)}(z,\vec p_T)\,
\Psi^{(\mu,\bar\mu)}_{\gamma^{\ast}_{T,L}}(r,z;Q^2)\,.
\end{eqnarray}
%*************************************************
%
The $T$ and $L$ amplitudes in a more explicit form are shown in Appendix~\ref{App:Amps}. 

The total $\gamma^{\ast}p\rightarrow Vp$ cross section is conventionally represented as sum of $T$ and $L$ contributions \cite{Hufner:2000jb}
%
%*************************************************
\begin{eqnarray}
\sigma^{\gamma^{\ast}p\rightarrow Vp}(x,Q^2)
&=&
\sigma_T^{\gamma^{\ast}p\rightarrow Vp}(x,Q^2) +
\tilde{\varepsilon}\,
\sigma_L^{\gamma^{\ast}p\rightarrow Vp}(x,Q^2)
\nonumber\\
&=&
\frac{1}{16\pi B}\left(\Big\vert \mathrm{Im}\mathcal{A}^{\gamma^{\ast}p\rightarrow Vp}_{T}(x,Q^2)\Big\vert^{2}+
\tilde{\varepsilon}\Big\vert \mathrm{Im}\mathcal{A}^{\gamma^{\ast}p\rightarrow Vp}_{L}(x,Q^2)\Big\vert^{2}\right) \,,
%%%%%%%%%%%%%%%%
\label{total-cs}
%%%%%%%%%%%%%%%%
\end{eqnarray}
%************************************************
%
 with $\tilde{\varepsilon} = 0.99$. Here, $B$ is the slope parameter fitted to the exclusive quarkonia electroproduction data. In the energy-independent approximation \cite{Adloff:2000vm} it is taken to be $B=4.73$ GeV$^{-2}$, while its possible energy and $Q^2$ dependence is discussed in more detail in Sections~\ref{Sec:psi1S-slope} and \ref{Sec:psi2S-slope}.

Derivation of above formulas relies on the assumption that the $S$-matrix is purely real and so the amplitude $\mathcal{A}$ is purely imaginary. Following Refs.~\cite{bronzan-74,Nemchik:1996cw,Forshaw:2003ki}, the real part of the amplitude can be accounted for by multiplying the cross section $\sigma^{\gamma^{\ast}p\rightarrow Vp}_{T,L}(x,Q^2)$ by a factor $1+\tan^2(\pi\lambda_{T,L}/2)$, where
%
%************************************************
\begin{eqnarray}
\lambda_{T,L}=\Bigg|\frac{\partial\ln
\mathrm{Im}\mathcal{A}^{\gamma^{\ast}p\rightarrow Vp}_{T,L}(x,Q^2)}
{\partial\ln x}\Bigg|=\Bigg|\frac{1}
{\mathrm{Im}\mathcal{A}^{\gamma^{\ast}p\rightarrow Vp}_{T,L}}
\int\limits_0^1\dfr z \int\dfr^2r \Sigma_{T,L}(z,\vec r;Q^2) 
\frac{\partial\sigma_{q\bar q}(r,x)}{\partial\ln x}\Bigg|,\quad 
%%%%%%%%%%%%%%
\label{lambda}
%%%%%%%%%%%%%%
\end{eqnarray}
%************************************************
%
provided that only the dipole cross section depends on the Bjorken $x$.

%
%
%
%===========================================
\section{Light-cone quarkonia wave function}
\label{Sec:QQ-wf}
%===========================================
%
%
%

One yet missing ingredient of the formula (\ref{prod-amp}) is the LC quarkonium wave function $\Psi_{V}(r,z)$. Alike the LC photon-quark wave function $\Psi^{T,L}_{\gamma^{\ast}}(r,z;Q^2)$, it is defined in the infinite momentum frame. Let us discuss if and how this object can be obtained from the first principles.

In the $Q\bar Q$ rest frame and in impact parameter representation, the quarkonia wave function is typically found by solving the Schr\"odinger equation for a given choice of the heavy quark interaction potentials as discussed in Appendix~\ref{App:Schrodinger}. In this work, we have employed five well-known parametrisations for the heavy quark interaction potentials illustrated in Fig.~\ref{fig:cb-V} for $c-\bar c$ (left panel) and $b-\bar b$ (right panel) cases and described in detail in Appendix~\ref{App:potentials}. 
%
%                   Fig.2
%==============================================
\begin{figure}[!htbp]
\begin{center}
\includegraphics[width=0.47\textwidth]{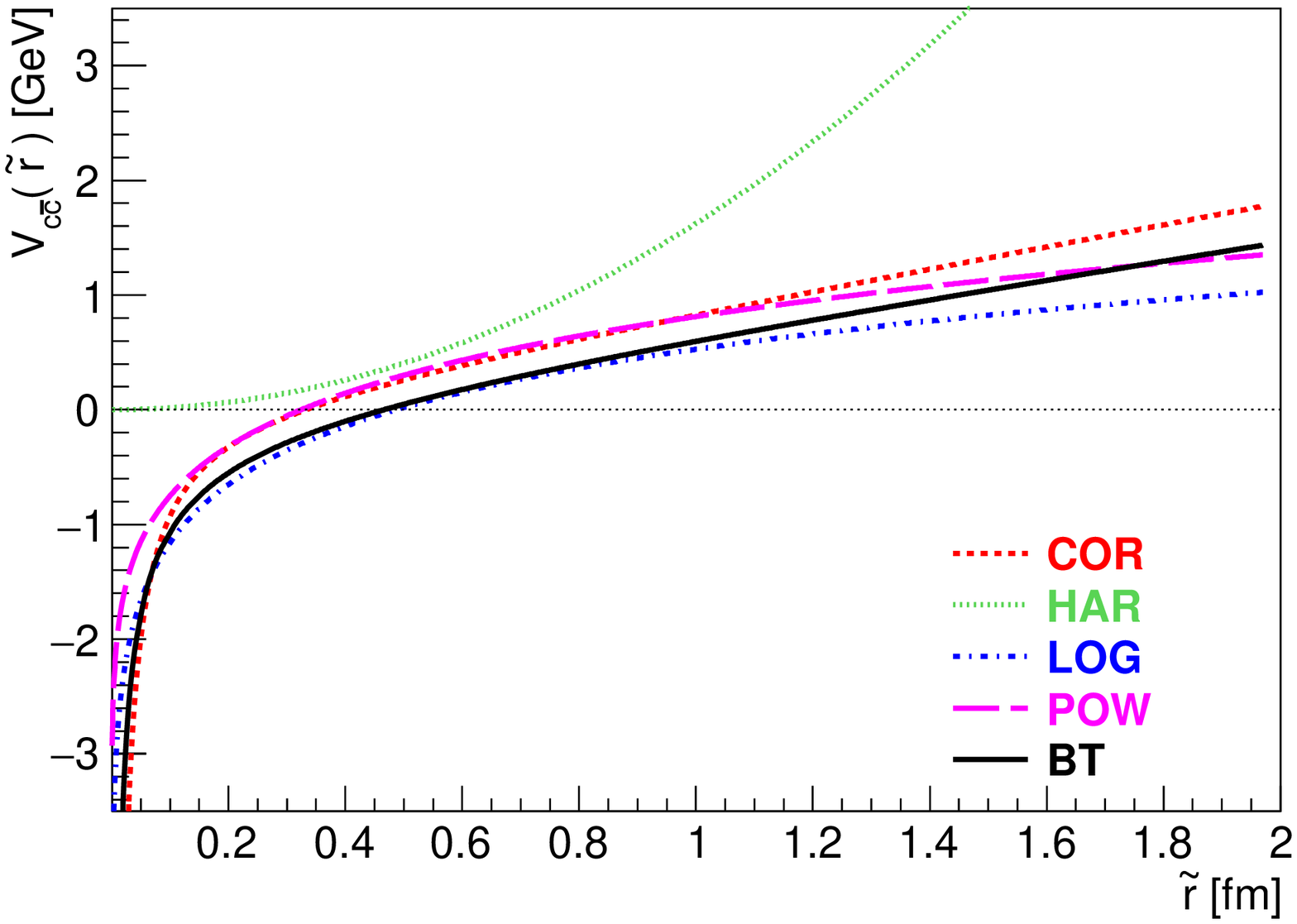}
\includegraphics[width=0.47\textwidth]{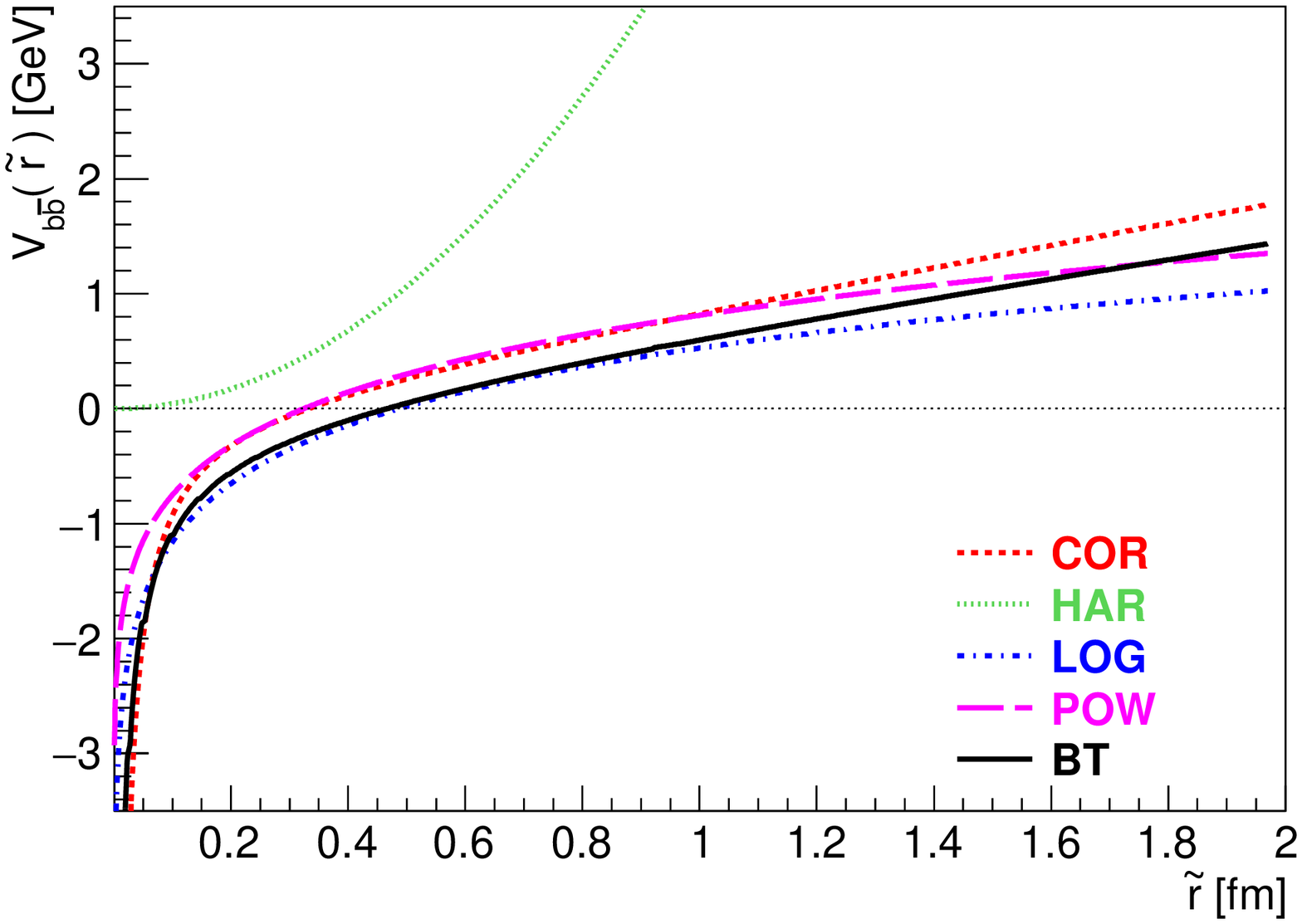}
  \caption{An illustration of five distinct $c-\bar c$ (left panel) and $b-\bar b$ (right panel) interaction potentials, used in this work, as functions of 3D interquark distance $\r$. For a detailed description and characteristic parameters of these potentials, see Appendix~\ref{App:potentials}.}
%%%%%%%%%%%%%%%%%%
  \label{fig:cb-V}
%%%%%%%%%%%%%%%%%%
 \end{center}
 \end{figure}
%============================================== 
%
%

Since, in general, there is no direct relation between the rest-frame wave function of the lowest Fock $|Q\bar Q\rangle$ component and that in the infinite-momentum frame, the problem of building the latter is rather difficult and there is no generally acceptable solution yet. In the literature, there are recipes towards finding such a wave function, and in what follows we employ one particular widely used recipe of Ref.~\cite{Terentev:1976jk} known to give rather accurate predictions in the relevant kinematical regions (cf.~Ref.~\cite{Kopeliovich:2015qna}).

For practical purposes, it is convenient to turn to the momentum-space wave function,
%
%***********************************************
\begin{eqnarray}
\psi(p)=\frac{2}{\sqrt{2\pi}p}\int\limits_0^{\infty} \dfr \r\,\r\,\psi(\r)\sin(p\,\r) \,, \qquad \int|\psi(p)|^2\,\dfr^3 p=1 \,,
\end{eqnarray}
%***********************************************
%
in terms of the quark 3-momentum $p\equiv |\vec p\,|$ and the 3D interquark distance, $\r\equiv |\vec{\r}\,|$. Since the quarkonium production amplitude (\ref{prod-amp}) is written in the infinite momentum frame, 
the corresponding wave function $\psi(p)$ should first be appropriately boosted to that frame. In terms of the LC variables, $z$ and $p_T$, 
the invariant mass squared of the $Q\bar Q$ pair reads
%
%***********************************************
\begin{eqnarray}
M_{Q\bar Q}^2 = \frac{p_T^2+m_Q^2}{z(1-z)} \,,
\end{eqnarray}
%***********************************************
%
while the same quantity in the $Q\bar Q$ rest frame is given by
%
%***********************************************
\begin{eqnarray}
M_{Q\bar Q}^2 = 4(p^2+m_Q^2)\,, \qquad p^2 = p_L^2 + p_T^2 \,,
\end{eqnarray}
%***********************************************
%
where $p_L$ is the longitudinal component of the quark 3-momentum $\vec p$. These two relations, therefore, yield
%
%***********************************************
\begin{eqnarray}
p^2=\frac{p_T^2+(1-2z)^2m_Q^2}{4z(1-z)}\,, \qquad p_L^2=\frac{(p_T^2+m_Q^2)(1-2z)^2}{4z(1-z)} \,,
\label{boost}
\end{eqnarray}
%***********************************************
%
providing an appropriate conversion of kinematical variables between the infinite momentum and $Q\bar Q$ rest frames. Besides, following the recipe of Ref.~\cite{Terentev:1976jk} the conservation of probability density upon such a boosting
%
%**********************************************
\begin{eqnarray}
\dfr^3 p |\psi(p)|^2
=
\dfr^2 p_T\dfr z|\psi(p_T,z)|^2 \,, \qquad \dfr^3 p = \dfr p_L\dfr^2p_T
\end{eqnarray}
%**********************************************
%
results in the following Terent'ev relation \cite{Terentev:1976jk} between the LC wave function $\psi(p_T,z)$ and its counterpart in the target rest frame $\psi(p)$
%
%**********************************************
\begin{eqnarray}
\psi(p_T,z)
=
\left(\frac{p_T^2+m_Q^2}{16(z(1-z))^3}\right)^{\frac{1}{4}}\,\psi(p) \,, \qquad \int|\psi(p_T,z)|^2\dfr^2 p_T\dfr z = 1 \,,
%%%%%%%%%%%%%%%%%%
\label{VMwave_pz}
%%%%%%%%%%%%%%%%%%
\end{eqnarray}
%**********************************************
%
where $p=p(p_T,z)$ is given by Eq.~(\ref{boost}). 
Note, the Terent'ev prescription for the Lorentz boosting presented above has been discussed and compared with exact calculations using the sophisticated Green function approach in Ref.~\cite{Kopeliovich:2015qna}. It has been shown that for the $\Jpsi$ wave function the Terent'ev prescription gives very accurate results for the averaged $\langle z \rangle \sim 0.5$. The LC wave function in the impact parameter representation is then given by
\vspace*{-0.2cm}
%**********************************************
\begin{eqnarray}
\Psi_{V}(r,z)
=
\int\limits_0^{\infty}\dfr p_T\, p_T J_0(p_Tr)\,\psi(p_T,z)\,.
%%%%%%%%%%%%%%%%%
\label{VMwave_rz}
%%%%%%%%%%%%%%%%%
\end{eqnarray}
%**********************************************
%
%
%                Fig.3
%===============================================
\begin{figure}[!bhpt]
\begin{center}
\vspace*{-0.60cm}
\includegraphics[width=0.47\textwidth]{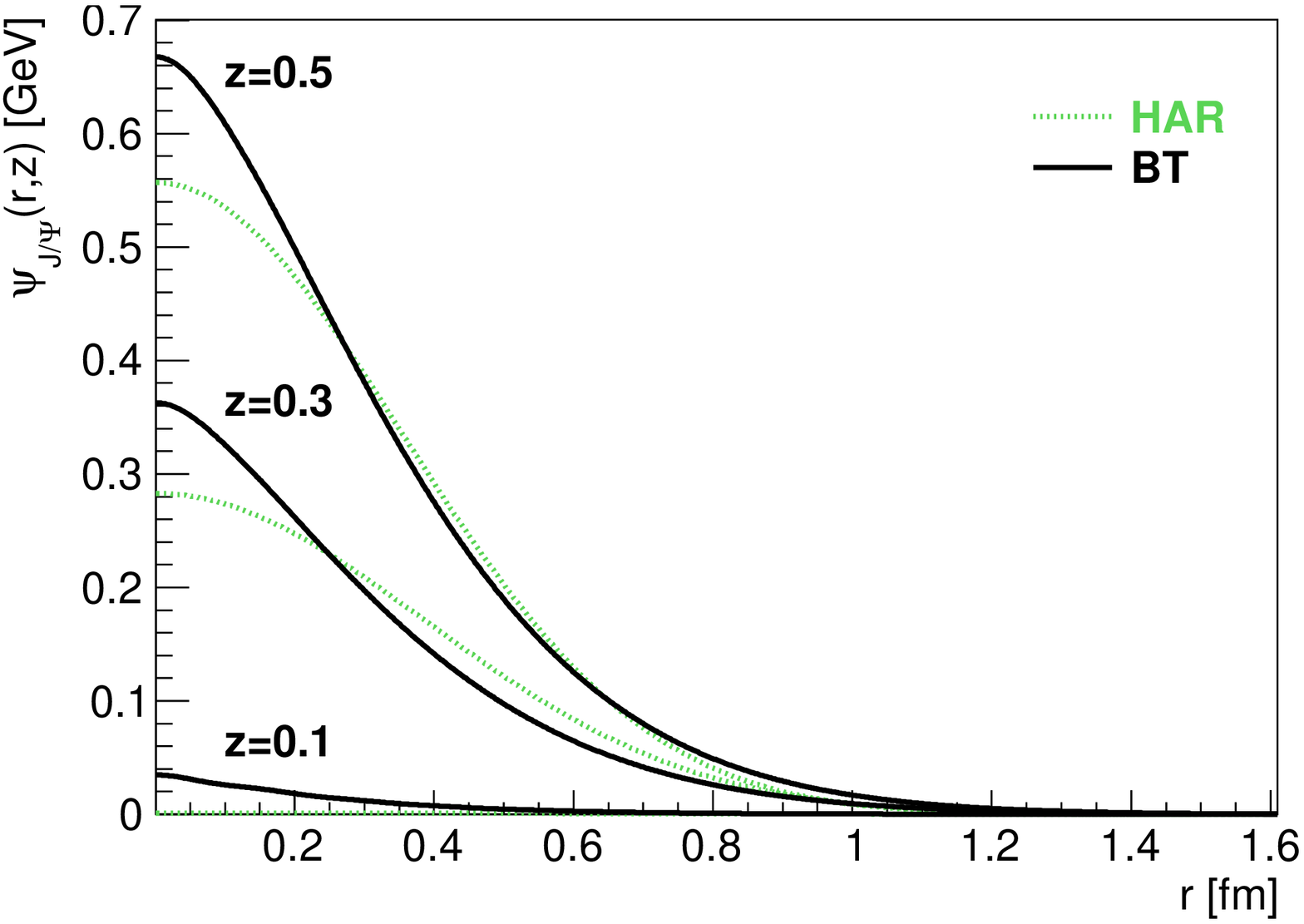}
\includegraphics[width=0.47\textwidth]{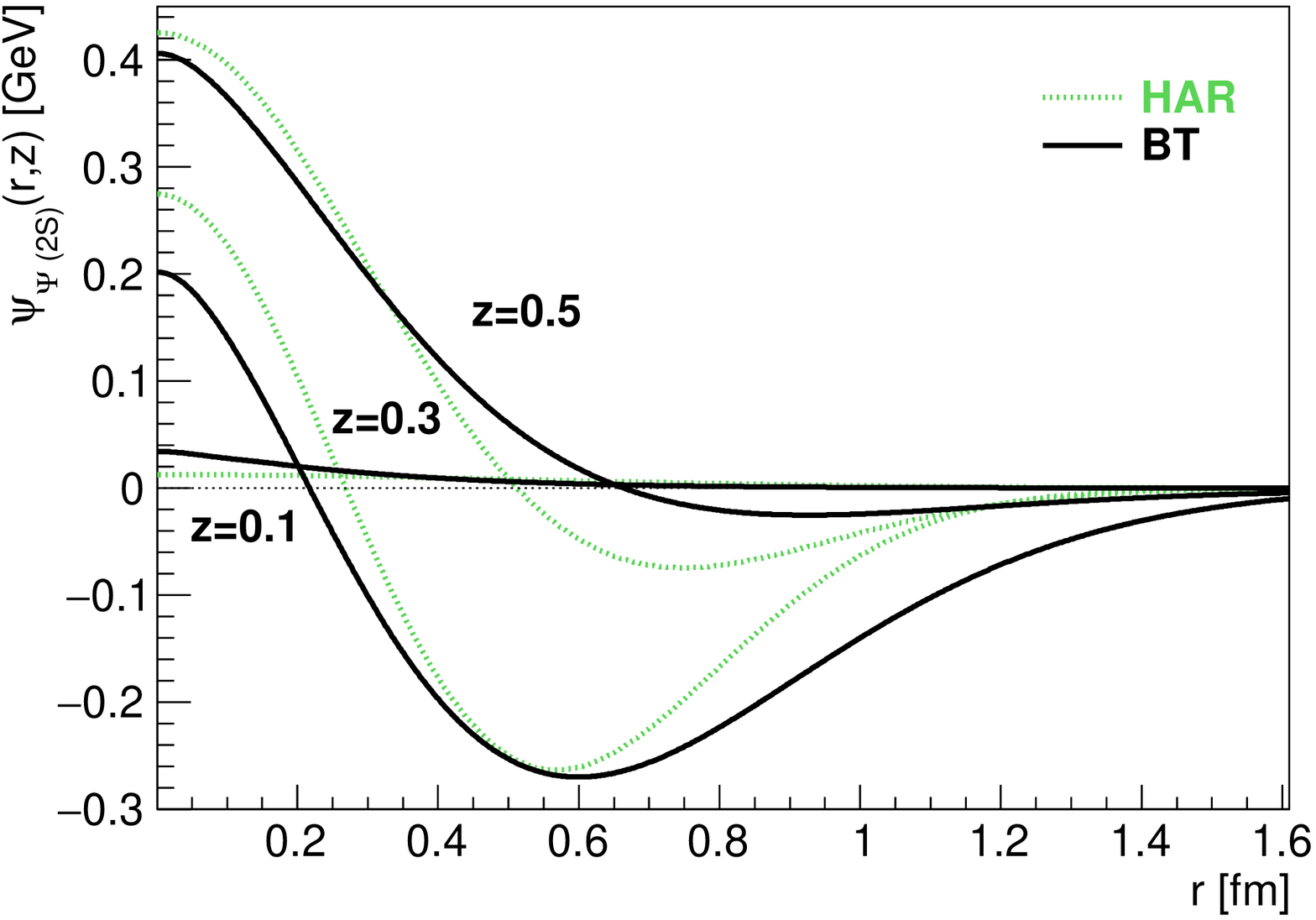}
\vspace*{-0.2cm}
  \caption{The LC wave function $\Psi_{V}(r,z)$ for $\Jpsi(1S)$ (left panel) and $\psip(2S)$ (right panel) mesons for different quark momentum fractions $z$. The distribution function $\Psi_{V}(r,z)$ is generated by two models for the $c-\bar c$ interaction potential: harmonic oscillator model denoted as HAR (green dotted lines) 
   and Buchm\"uller-Tye parameterisation, or BT (black solid lines).}
%%%%%%%%%%%%%%%%%%% 
  \label{fig:psiLC}
%%%%%%%%%%%%%%%%%%%
 \end{center}
 \end{figure}
%==============================================
%
%
%
%                  Fig.4
%==============================================
\begin{figure}[!htbp]
\begin{center}
\includegraphics[width=0.31\textwidth]{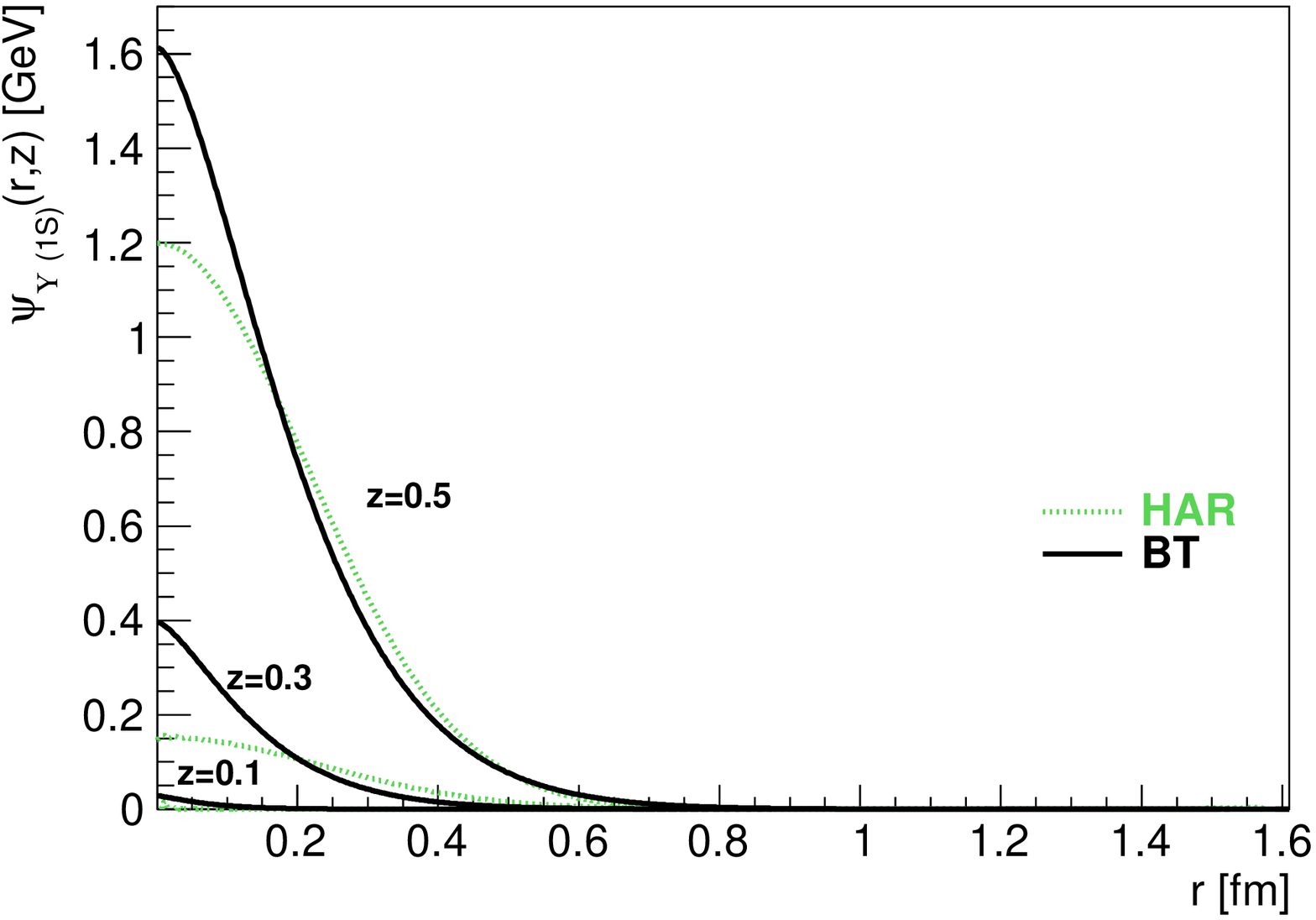}
\includegraphics[width=0.31\textwidth]{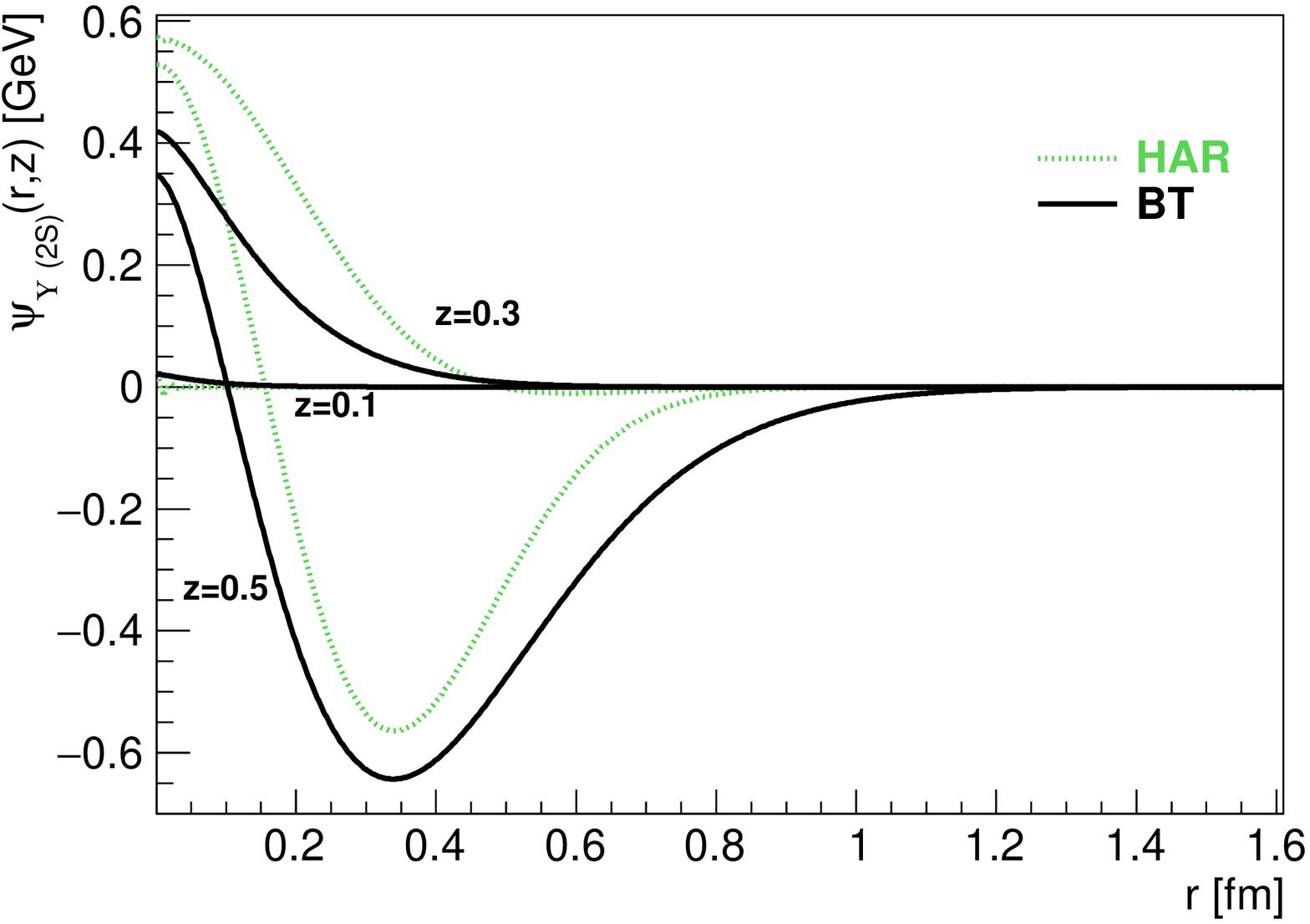}
\includegraphics[width=0.31\textwidth]{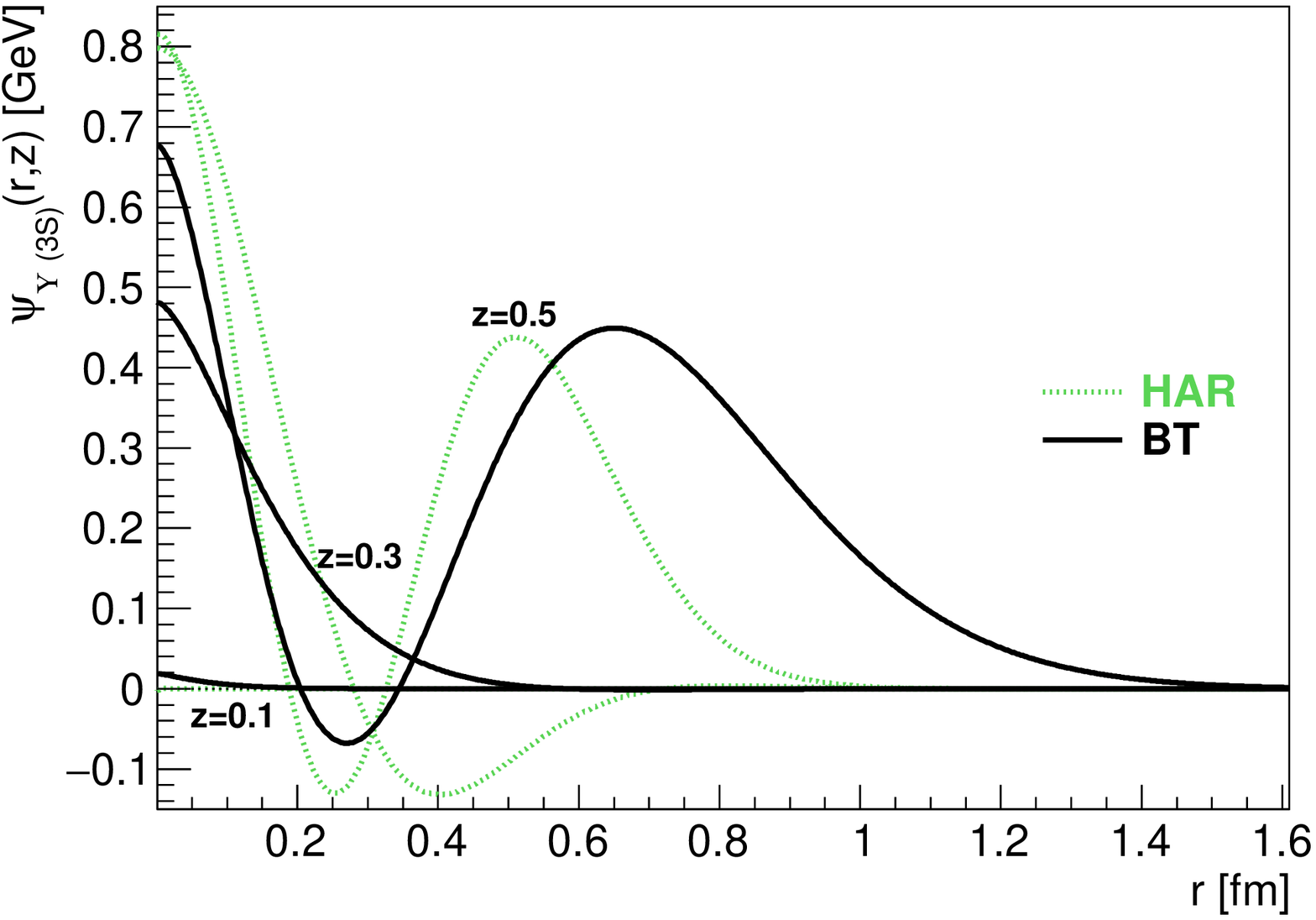}
\caption{The LC wave function $\Psi_{V}(r,z)$ for $\Y(1S)$ (left panel), $\Yp(2S)$ (middle panel) and $\Ypp(3S)$ (right panel) mesons for different quark momentum fractions $z$. The distribution function $\Psi_{V}(r,z)$ is generated by two models for the $b-\bar b$ interaction potential: harmonic oscillator model denoted as HAR (green dotted lines) and Buchm\"uller-Tye parameterisation, or BT (black solid lines).}
%%%%%%%%%%%%%%%%
  \label{fig:LC}
%%%%%%%%%%%%%%%%
 \end{center}
 \end{figure}
%==============================================
%
%

In Fig.~\ref{fig:psiLC} we show the numerical results for the boosted LC wave functions $\Psi_{V}(r,z)$ for two charmonia states, $\Jpsi(1S)$ and $\psip(2S)$, and which are obtained starting from numerical solutions of the Schr\"odinger equation for two models of the $c-\bar c$ interaction potential -- the harmonic oscillator and Buchm\"uller-Tye parametrisation, discussed in Appendix~\ref{App:potentials}. While the overall shape of the wave functions appears to be consistent between the two models, they yield notable quantitative differences, especially for $\psip(2S)$, where the positions of the node and the minimum are quite sensitive to the choice of the potential.

We have also performed calculations of the wave functions and total elastic electroproduction cross sections for a number of different $c\bar c$ ($1S$ and $2S$) and $b\bar b$ ($1S$, $2S$ and $3S$) vector meson states for five distinct parameterisations of the interquark potentials and five different parameterisations for the dipole cross sections, $\sigma_{q\bar q}(r,x)$. Since the number of possible combinations of the parameterisations and states can be rather extensive, as a part of this project, we created a webpage on \href{https://hep.fjfi.cvut.cz/vm.php}{https://hep.fjfi.cvut.cz/vm.php}, where one can find numerical datasets (grids) for each vector meson state, interquark potential and the dipole parameterisation. 

The datasets are available for vector meson wave functions in the forms of a 3D radial solution of the Schr\"odinger equation in the $Q\bar Q$ rest frame, $\psi(\tilde{r})$ (shown in Eq.~(\ref{VMwave_r})), the boosted LC wave function in momentum space, $\psi(p_T,z)$, given by Eq.~(\ref{VMwave_pz}), and the boosted LC wave function in impact parameter space, $\psi(r,z)$, given by Eq.~(\ref{VMwave_rz}). An interpolating routine written in C++ (including also an example for calculations) is also available on the same webpage. The web interface enables to generate plots for the electroproduction cross sections for a selected combination of the quarkonium wave function generated by the explicit $Q-\bar Q$ potential with the explicit dipole model for $\sigma_{q\bar q}(r,x)$. Calculations can be performed including or neglecting the Melosh spin rotation effects. Numerical results can be presented also in the form of a table, which can be used for practical purposes.

%
%
%
%===================================================
\section{Dipole cross section}
\label{Sec:dipole-CS}
%===================================================
%
%
%

The essential ingredient of the color dipole approach, the universal dipole cross section $\sigma_{q\bar q}(r,x)$, has been 
first introduced a long ago in Ref.~\cite{Kopeliovich:1981pz}. During past three decades, its kinematic (and energy) 
dependence has undergone remarkable development largely promoted by precision experimental information from the HERA collider. 
While an exact theoretical modelling of the dipole cross section (and the corresponding partial dipole amplitude) is not 
nearly close to its complete understanding, a number of phenomenological parametrisations accounting for the saturation
phenomenon and the QCD-inspired Bjorken $x$- and hard-scale evolution have been proposed in the literature \cite{Goncalves:2006yt,GolecBiernat:1998js,Iancu:2003ge,Kharzeev:2004yx,Dumitru:2005gt,Boer:2007ug,Kowalski:2006hc,deSantanaAmaral:2006fe,Soyez:2007kg,Bartels:2002cj,Kowalski:2003hm,Rezaeian:2012ji,Rezaeian:2013tka}) and rely on the fits to the HERA DIS data.

Introducing the partial dipole amplitude ${\cal N} (\vec{r}, x, \vec{b})$, one conventionally determines the universal dipole 
cross section $\sigma_{q\bar{q}}(r,x)$ as an integral over the impact parameter $\vec{b}$:
%==========================================
\begin{eqnarray}
\sigma_{q\bar{q}}(r,x) = 2\,\int d^2b\,  
{\cal N} (\vec{r}, x, \vec{b}\,)\,, \qquad r=|\vec{r}\,| \,.
\label{sigda}
\end{eqnarray}
%==========================================
The evolution in $x$- or in rapidity $Y = \ln (1/x)$ in the high-energy ($x\ll 1$) regime is treated e.g. 
by an infinite hierarchy of the so-called Balitsky-JIMWLK equations \cite{JalilianMarian:1997gr,JalilianMarian:1997dw,Kovner:2000pt,Weigert:2000gi,Balitsky:1995ub,Balitsky:1998kc} 
in the framework of the Color Glass Condensate (CGC) formalism \cite{McLerran:1993ka,McLerran:1994vd}. These equations 
reduce to the Balitsky-Kovchegov (BK) equation \cite{Balitsky:1995ub, Kovchegov:1999yj} in the mean-field approximation. 
As it is rather difficult to obtain the $\vec{b}$-dependent solutions of the BK equation \cite{GolecBiernat:2003ym} 
while the impact-parameter profile is determined by essentially non-perturbative QCD phenomena, one usually imposes 
such approximations as the translational invariance of the amplitude disregarding the impact parameter dependence 
in numerical solutions. An alternative way usually admitted in the literature is to consider more phenomenological 
models for the $\vec{b}$ dependence, as well as accounting for the saturation phenomenon and the hard-scale evolution 
via DGLAP, that are fit to precision data e.g. from HERA. A naive comparison of the predictions of the dipole model 
calculations using several distinct parametrisations for the universal dipole cross section is a commonsense tool 
for a rough estimation of the associated theoretical uncertainties.

Since a long ago, it was understood that at small dipole separations $r$ the dipole cross section is essentially 
proportional to the gluon PDF in the target \cite{Blaettel:1993rd,Frankfurt:1993it,Frankfurt:1996ri}, i.e.
\begin{eqnarray}
\sigma_{q\bar q}(r,x)\simeq \frac{\pi^2}{3}\alpha_s\Big(\frac{\Lambda}{r^2}\Big)\,r^2\,xg\Big(x,\frac{\Lambda}{r^2}\Big)\,,
\label{dip-pdf}
\end{eqnarray}
with $\Lambda\approx 10$ \cite{Nikolaev:1994cn}. Later on, in Ref.~\cite{Bartels:2002cj} it was suggested to merge 
this asymptotics with a naive saturated ansatz for the dipole cross section. Later on, in Ref.~\cite{Kowalski:2003hm}
it was proposed to introduce explicitly the $\vec{b}$-dependence into the corresponding ansatz for the partial 
dipole amplitude. The latter yields the following widely used parametrisation enabling for a description of exclusive observables at HERA and is known as the {\it IP-Sat} model,
%=======================================================
\begin{eqnarray}
\sigma_{q\bar{q}}(r, x) = 2\,\int d^2b\,\left[1-\exp\left(-\frac{\pi^2}{2N_c}\,r^2\,\alpha_s(\mu^2)\, xg(x,\mu^2)T_G(b)\right)\right]
\label{ipsat}
\end{eqnarray} 
%=======================================================
given in terms of the number of colors in QCD, $N_c=3$, the strong coupling constant $\alpha_s(\mu^2)$ determined at 
the hard scale $\mu$ connected to the size of the dipole $r$ in a simple way as $\mu^2=C/r^2 + \mu_0^2$. Here, the model 
parameters $C$, $\mu_0$ and $\sigma_0$ are extracted by fitting to the HERA data. Besides, the gluon PDF in the target 
nucleon $xg(x, \mu^2)$ at small $x$ is found as a solution of the conventional DGLAP equation \cite{Gribov:1972ri,Altarelli:1977zs,Dokshitzer:1977sg} 
which takes into account the gluon splitting function $P_{gg}(z)$ only,
%=======================================================
\begin{equation}
\frac{\partial xg(x,\mu^2)}{\partial \ln \mu^2 } = \frac{\alpha_s(\mu^2)}{2\pi} \int_x^1 dz\,  
P_{gg}(z) \frac{x}{z} g\Big(\frac{x}{z}, \mu^2\Big)\,.
\label{dglap}
\end{equation}
%=======================================================
Here, the starting value of the target gluon density at $\mu^2=\mu_0^2$ is given by
%=======================================================
\begin{equation}
xg(x,\mu_0^2) = A_g x^{-\lambda_g} (1-x)^{5.6}\,.
\end{equation}
%=======================================================
The $\vec{b}$-dependence is accounted for by means of a simple Gaussian profile
\begin{eqnarray}
T_G(b)=\frac{1}{2\pi B_G}\,\exp\left(-\frac{b^2}{2 B_G}\right) \,,
\end{eqnarray}
where $B_G$ is an additional free parameter in the model that can be extracted, in particular, 
from the measured $t$-dependent elastic electron-proton scattering. In general, $B_G$ in the IP-Sat model
is taken to be different from the slope of the exclusive quarkonia electroproduction cross section defined in Eq.~(\ref{total-cs}) (see e.g.~a discussion in Ref.~\cite{Kowalski:2006hc}).
A comprehensive IP-Sat model fit of the complete (inelastic and elastic) set of 
HERA data has been performed in Ref.~\cite{Rezaeian:2012ji} leading to
\begin{eqnarray}
A_g = 2.373\,, \quad \lambda_g = 0.052\,, \quad \mu_0^2 = 1.428\,{\rm GeV}^2\,, \quad B_G = 4.0\,{\rm GeV}^2\,, 
\quad C = 4.0 \,,
\end{eqnarray}

Needless to mention, a practically simple saturated ansatz well-known as the Golec-Biernat-Wusthoff (GBW) 
model \citep{GolecBiernat:1998js} 
%=======================================================
\begin{equation}
\label{gbw}
\sigma_{q\bar{q}}(r,x) = \sigma_0\,\left(1 - e^{-\frac{r^2Q_s^2(x)}{4}}\right) \,, \qquad Q_s^2(x) \equiv 
R_0^{-2}(x) = Q_0^2\left( \frac{x_0}{x} \right)^\lambda \,,
\end{equation}
%=======================================================
with $Q_s(x)$ being the $x$-dependent (and $\mu$-independent) saturation scale, gives rise to a fairly good description of a large variety of observables 
in high-energy $ep$ and $pp$ collisions, as well as those on nuclear targets and for both inclusive and exclusive final states at not-so-large transverse 
momenta (or $Q^2$) and small $x\lesssim 0.01$. This model resembles the Glauber model of multiple interactions and can also be straightforwardly used to 
incorporate the saturation effects. The global of the DIS HERA data accounting for the charm quark contribution provides different sets of parameters. We use two sets - the one taken from \cite{GolecBiernat:1999qd} we label as {\it GBWold} and the one from \cite{Kowalski:2006hc} we label as {\it GBWnew}
\begin{eqnarray}
{\mathrm{GBWold:}}& Q_0^2 = 1\, {\rm GeV}^2\,, \quad x_0 = 3.04 \times 10^{-4}\,, \quad 
\lambda = 0.288\,, \quad \sigma_0 = 23.03\, {\rm mb} \nonumber\\
{\mathrm{GBWnew:}}& Q_0^2 = 1\, {\rm GeV}^2\,, \quad x_0 = 1.11 \times 10^{-4}\,, \quad 
\lambda = 0.287\,, \quad \sigma_0 = 23.9\, {\rm mb} \,.
\end{eqnarray}
Following the tradition and for the sake of completeness, we use this simple model as a reference in comparison with other more complicated parametrisations.
Besides, we will also consider the solution to the running coupling BK equation calculated according to the procedure in Ref.~\cite{Cepila:2015qea}. The BK equation describes the evolution in rapidity $Y$ of the scattering amplitude ${\cal N} (\vec{r}, x)$. This formulation is based on the work of \cite{Albacete:2007yr,Albacete:2009fh,Albacete:2010sy}. 
\begin{equation}
\frac{\partial N(\vec r,x)}{\partial Y}=\int d \vec{r}_1 K(\vec r,\vec r_1,\vec r_2)\Bigg(N(\vec r_1,x)+N(\vec r_2,x)-N(\vec r,x)-N(\vec r_1,x)N(\vec r_2,x)\Bigg)
\end{equation}
where $\vec{r}_2 = \vec{r}-\vec{r}_1$. The kernel incorporating the running of the QCD coupling \cite{Albacete:2010sy} is given by
\begin{equation}
K(\vec r,\vec r_1,\vec r_2)=\frac{\alpha_s(r^2) N_c}{2\pi^{2}}\Bigg(\frac{r^2}{r_1^2 r_2^2}+\frac{1}{r_1^2}\left(\frac{\alpha_s(r_1^2)}{\alpha_s(r_2^2)}-1\right)+\frac{1}{r_2^2}\left(\frac{\alpha_s(r_2^2)}{\alpha_s(r_1^2)}-1\right)\Bigg),
\end{equation}
with
\begin{equation}
\alpha_{s}(r^2)=\frac{4\pi}{(11-\frac{2}{3}N_f)\ln\left(\frac{4C^2}{r^2\Lambda^2_{\rm QCD}}\right)}
\end{equation}
where $N_f$ is the number of active flavours and $C$ is a parameter to be fixed from data. We use the fixed number of flavours scheme with $ \Lambda_{\rm QCD}=0.241$ MeV. For the initial form of the dipole scattering amplitude the McLerran-Venugopalan (MV) model \cite{McLerran:1997fk} is used: 
 \begin{equation}
N(\vec r,x=0.01)=1-\exp\Bigg(-\frac{\left(r^2Q^2_{s0}\right)^\gamma}{4} \ln \left(\frac{1}{r\Lambda_{\rm QCD}}+e\right)\Bigg)
\end{equation}
with the values of the parameters $Q^2_{s0}$, $C$ and $\gamma$ taken from \cite{Albacete:2010sy} yielding $ \sigma_0=32.895$ mb, $Q^2_{s0} = 0.165$ GeV$^2$, $\gamma=1.135$ and $C = 2.52$. The fit was performed under the assumption that $\alpha_s(r^2)$ freezes for values of $r$ larger than $r_0$ defined by $\alpha_s(r^2_0)=0.7$. This model is denoted below as {\it rcBK}.

We have also included the collinearly improved kernel \cite{Iancu:2015joa} given by
\begin{equation}
K(\vec r,\vec r_1,\vec r_2)=\frac{\bar \alpha_s N_c}{2\pi^{2}}\Bigg(\frac{r^2}{r_1^2 r_2^2}\left( \frac{r^2}{\mathrm{min}(r_1^2,r_2^2)}\right)^{\pm\bar\alpha_s A_1}\frac{J_1(2\sqrt{\bar\alpha_s |\ln(r_1^{2}/r^{2})\ln(r_2^{2}/r^{2})|})}{\sqrt{\bar\alpha_s |\ln(r_1^{2}/r^{2})\ln(r_2^{2}/r^{2})|}}\Bigg),
\end{equation}
with $ A_1=11/12$, the positive sign refers to the situation where $ r < \mathrm{min}(r_1,r_2) $ and $ \bar\alpha_s=\alpha_s(\mathrm{min}(r^{2},r_1^{2},r_2^{2})) \frac{N_c}{\pi} $. This kernel was used with variable number of flavours scheme \cite{Albacete:2010sy}, each flavour has its $ \Lambda_{\rm QCD} $ calculated from the recurrent relation 
\begin{equation}
\Lambda_{N_f-1}=m_f^{1-\frac{\beta_{N_f}}{\beta_{N_f-1}}}\Lambda_{N_f}^{\frac{\beta_{N_f}}{\beta_{N_f-1}}},
\end{equation}
where $ \beta_{N_f}=(11N_c-2N_f)/3 $ and $ m_f $ is the mass of the quark of flavour $ f $. As a starting point one can take measured $ \alpha_s(r^{2}=4C^{2}/M_Z^{2})=0.1189 $ for $ n_f=5 $ at a scale of Z boson mass $ M_Z=91.187 $ GeV. This leads to the formula
\begin{equation}
\Lambda_{5}=M_Ze^{-\frac{2\pi}{\alpha_s(r^{2}=4C^{2}/M_Z^{2})\beta_{5}}}.
\end{equation}
A collinear version of MV initial conditions was published in \cite{Iancu:2015joa}
 \begin{equation}
N(\vec r,x=0.01)=\left[1-\exp\left(-\left[\frac{r^2Q^2_{s0}}{4}\bar\alpha_s(r^{2}) \left( 1+\ln \left(\frac{\alpha_{sat}}{\bar\alpha_s(r^{2})}\right)\right)\right]^{p}\right)\right]^{1/p},
\end{equation}
where $ \bar\alpha_{sat}=\frac{N_c}{\pi}\alpha_{sat}, \bar\alpha_{s}(r^{2})=\frac{N_c}{\pi}\alpha_{s}(r^{2}) $ and $ \alpha_{sat} $ is fixed to 1. Parameters were fitted in \cite{Iancu:2015joa} with $ \sigma_0=31.4055 $ mb, $ Q^2_{s0}=0.4 $ GeV$^{2} $, $ C=2.586 $ and $ p=0.807 $. This model is denoted as {\it colBK}.
The dipole cross section is obtained from the scattering amplitude as $\sigma_{q\bar{q}}(r,x) = \sigma_0\,{\cal{N}}(r,x)$, where the normalisation $\sigma_0$ is fitted to the HERA data.

When decreasing the hard scale $Q^2\to\Lambda_{\rm QCD}^2$ relevant for e.g. photoproduction, one may reach small $x$ values even for moderate and low energies. As was argued e.g. in Ref.~\cite{Kopeliovich:1999am}, the Bjorken variable $x$ becomes inappropriate in the soft limit. For such processes as e.g. the pion-proton scattering as well as the diffractive Drell-Yan and gluon radiation processes the saturation scale $Q_s^2 \gtrsim Q^2$ becomes a function of the dipole-target collision c.m. energy squared $\hat{s}$, and not Bjorken $x$. The corresponding parametrisation of the dipole cross section based upon the same saturated ansatz as in Eq.~(\ref{gbw}) is found by a replacement $\sigma_0 \to \overline{\sigma}_0(\hat{s})$ and $R_0 \to \overline{R}_0(\hat{s})$
where \cite{Kopeliovich:1999am}
\begin{eqnarray}\nonumber
 \overline{R}_0(\hat s)=0.88\,\mathrm{fm}\,(s_0/\hat s)^{0.14}\,,\quad
 \overline{\sigma}_0(\hat s)=\sigma_{\rm tot}^{\pi p}(\hat s)
 \Big(1+\frac{3\overline{R}_0^2(\hat s)}{8\langle r_{\rm ch}^2 \rangle_{\pi}}\Big)\,.
 \label{KST-params}
\end{eqnarray}
Here, $\sigma_{\rm tot}^{\pi p}(\hat s)=23.6(\hat s/s_0)^{0.08}$ mb is the pion-proton total cross section \cite{Barnett:1996yz}, $\langle r_{\rm ch}^2 \rangle_{\pi}= 0.44$ fm$^2$ is the mean pion radius squared \cite{Amendolia:1986wj}, and $s_0=1000\,{\rm GeV}^2$. Interestingly enough, this parametrisation known as the {\it KST} model has been shown to give the correct description of the pion-proton cross section at scales up to $Q^2\sim 20$ GeV$^2$. This parametrisation, together with the ones above, will be used in our analysis of exclusive real and virtual photoproduction of quarkonia.

Another parametrization was proposed by Iancu, Itakura and Munier \cite{Iancu:2003ge} 
\begin{eqnarray}
\sigma_{q\bar q}(r,x)&=&\sigma_0 N_0 \left(\frac{rQ_s(x)}{2}\right)^{2\gamma_{eff}(r,x)}\qquad rQ_s(x)\leq 2\nonumber\\
&=&\sigma_0 \left( 1-e^{-A\ln^{2}(BrQ_s(x))}\right)\qquad\, rQ_s(x)> 2 \nonumber\\
\gamma_{eff}(r,x)&=&\gamma+\frac{1}{\kappa\lambda Y}\ln\left(\frac{2}{rQ_s(x)}\right) \qquad Y=\ln\left(\frac{1}{x}\right),
\end{eqnarray}
where $\gamma_{eff}(r,x)$ is an effective anomalous dimension, $ \kappa=9.9 $, $ N_0=0.7 $, $ Q_s^{2}(x)=\left(\frac{x_0}{x}\right)^\lambda $ GeV$^{2}$ and $ \sigma_0=2\pi R_p^{2}$. Parameters $ A $ and $ B $ are chosen to ensure continuity between both parts of the parametrization at $ rQ_s(x)= 2 $ as
\begin{eqnarray}
A&=&-\frac{N_0^{2}\gamma^{2}}{(1-N_0)^{2}\ln(1-N_0)}\nonumber\\
B&=&\frac{1}{2}(1-N_0)^{-\frac{1-N_0}{N_0\gamma}}.
\end{eqnarray}
Parameters $ \sigma_0, R_p, \gamma, x_0 $ and $ \lambda $ have to be fitted to data. We use the fit performed in \cite{Soyez:2007kg} with $ \gamma=0.7376 $, $ \lambda=0.2197 $, $ x_0=0.1632\times 10^{-4} $ and $ \sigma_0=27.33 $ mb. This model will be denoted as {\it IIM}.

The last parametrization used in our analysis was proposed in Refs.~\cite{Kowalski:2006hc,Watt:2007nr} as a modification of the IIM parametrization to include the explicit impact parameter dependence by introducing the modified saturation scale 
\begin{equation}
Q_s^{2}(x)\rightarrow Q_s^{2}(x,b)=\left(\frac{x_0}{x}\right)^\lambda \left( e^{-\frac{b^{2}}{2B_{G}}}\right)^{\frac{1}{\gamma}}.
\end{equation}  
Parameters $ B_{G},N_0, \gamma, x_0 $ and $ \lambda $ has to be fitted to data. The most recent set of parameters comes from \cite{Rezaeian:2013tka} and sets $ \gamma=0.6492 $, $ N_0=0.3658 $, $ \lambda=0.2023 $, $ x_0=0.00069 $ and $ B_{G}=5.5 $ GeV$^{-2}. $ This model is denoted as {\it b-CGC}.
%
%              Fig.5
%=============-===================
\begin{figure}[!hbt]
\begin{center}
\includegraphics[width=1.0\textwidth]{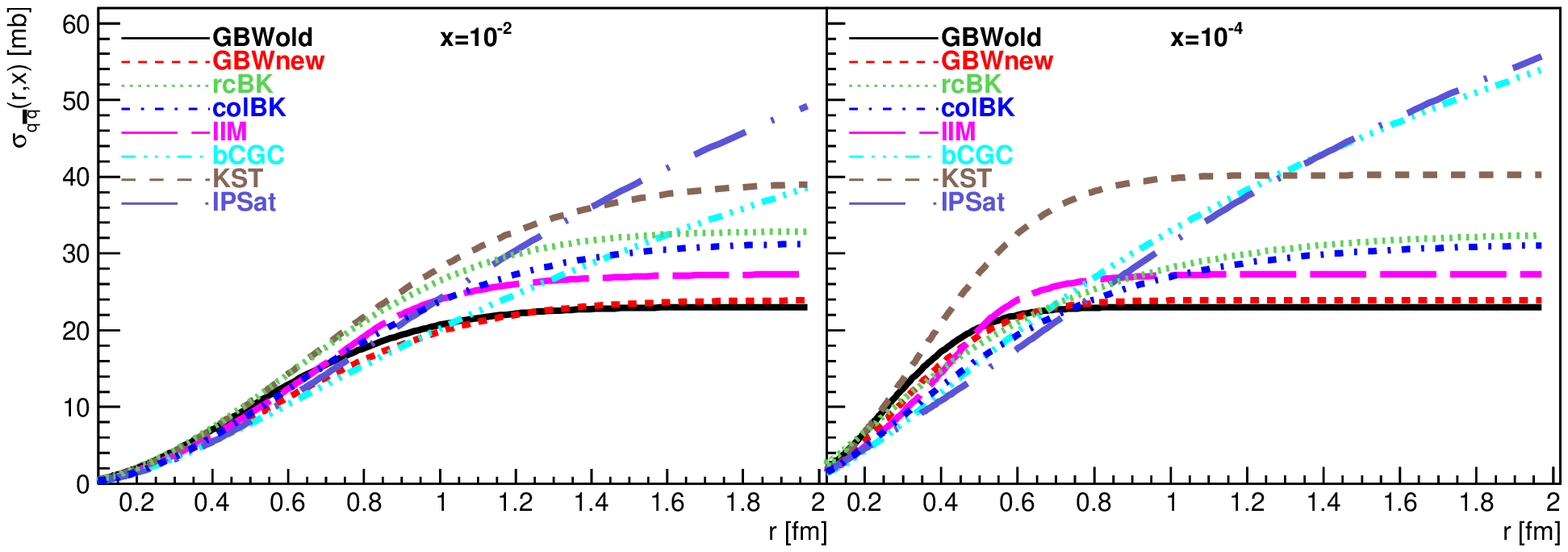}
\includegraphics[width=1.0\textwidth]{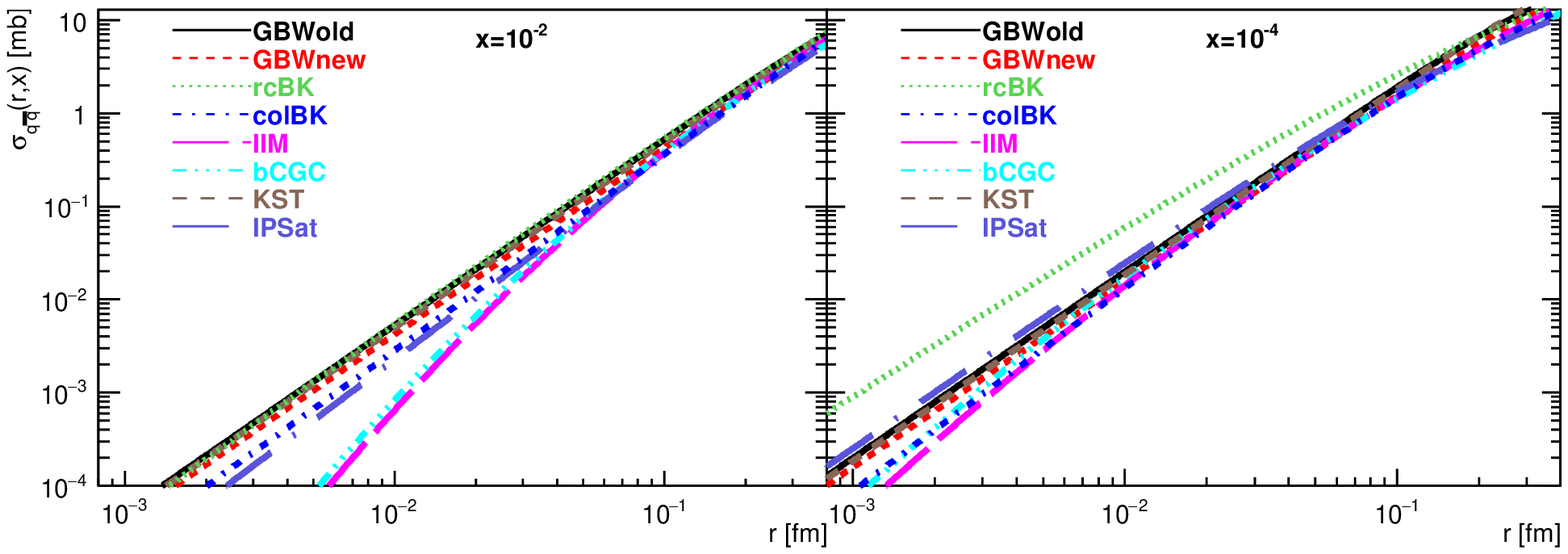}
  \caption{Comparison of different parametrisations for the dipole cross section 
  used in our calculations as described in the text.}
  \label{fig:sigqq}
 \end{center}
 \end{figure}
%==============----===============

In order to get the impact parameter independent dipole cross section from IPSat and b-CGC parametrizations an integral over the impact parameter was performed. 
As an illustration, in Fig.~\ref{fig:sigqq} we show the shape of different parametrisations for $\sigma_{q\bar q}(r,x)$ as a function of the dipole transverse
separations $r$ as two fixed values of the Bjorken variable $x=10^{-2}$ (left panels) and $x=10^{-4}$ (right panels). 

At large dipole separations, we observe a substantial variation between both the shapes and magnitudes of $\sigma_{q\bar q}(r,x)$, and the differences in dipole models tend to rise with decreasing $x$. Quite interestingly, such differences become also large at very small dipole sizes $r\lesssim 0.05\div 0.06\,\fm$, i.e. in the perturbative region. Thus, the measurements of exclusive electroproduction of quarkonia at very large scales $Q^2\gtrsim 300\div 400\,\GeV^2$ may provide additional constraints on the dipole parametrizations and means to further reduce theoretical uncertainties in the small-$x$ treatment of the gluon density. Using the precision data in the hard and soft limits, one could ultimately start ruling out the models.

%
%
%
%=======================
\section{Numerical results vs data}
\label{Sec:results}
%=======================
%
%
%

%
%
%====================================
\subsection{Theoretical uncertainties caused by determination of the diffraction slope}
\label{Sec:slope}
%====================================
%
%

Let us turn to discussion of numerical results on the
$\gamma^*\,p \to V\,p$ process in comparison with the data available from HERA.
In order to calculate the total photo- and electroproduction cross section
Eq.~(\ref{total-cs})
with amplitudes given by Eqs.~(\ref{AT}) and (\ref{AL}) one should know
the magnitudes of the slope parameter as a function of the photon
energy $W$ and virtuality $Q^2$.

%====================================
\subsubsection{Diffraction slope for the process $\gamma^*\,p\to\Jpsi (\Y)\,p$}
\label{Sec:psi1S-slope}
%====================================

For the c.m. energy behavior of the diffraction slope $B(W)$ we use the
standard form based on the Regge phenomenology,
%
%*************************************
\beqn
B(W)=B_0 + 4\,\alpha'(0) \ln\Big(\frac{W}{W_0}\Big) \,, \qquad 
W_0 = 90 \, \GeV \,,
%%%%%%%%%%%
\label{BW}
%%%%%%%%%%%
\eeqn
%*************************************
%
where $\alpha'(0)$ represents the slope of the Pomeron trajectory.

\begin{figure}[!htbp]
\begin{center}
\includegraphics[width=0.47\textwidth]{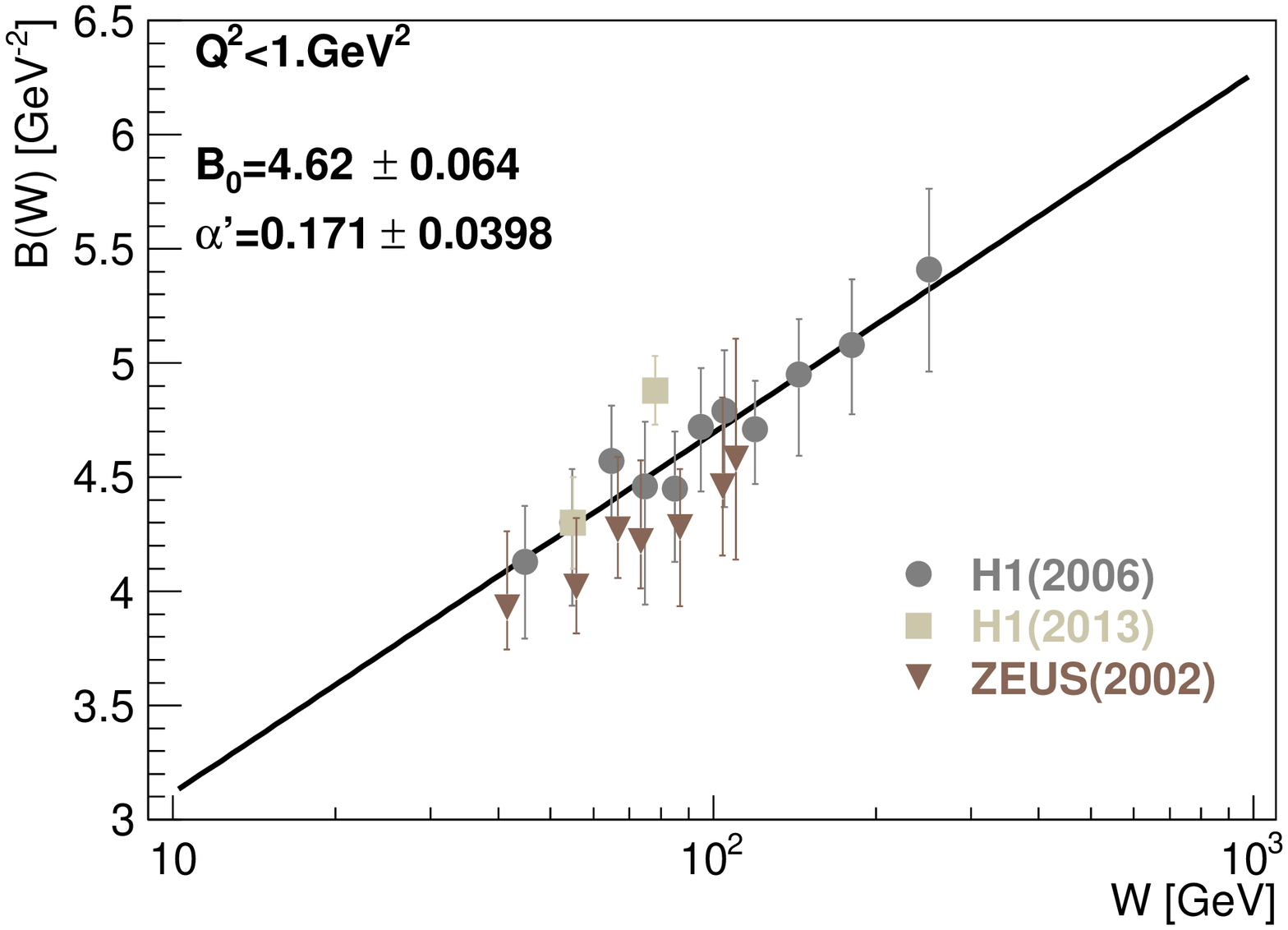}
\includegraphics[width=0.47\textwidth]{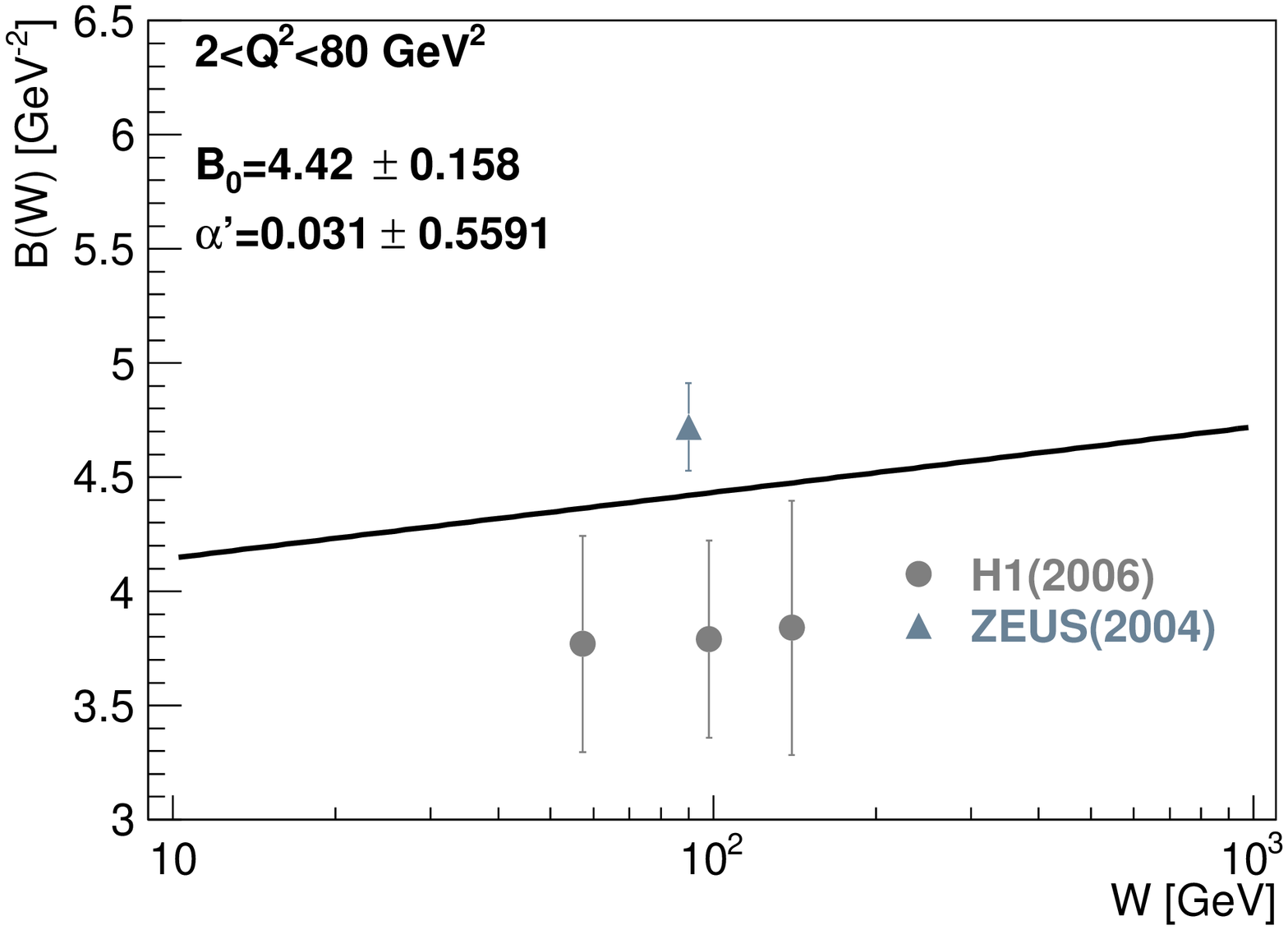}
  \caption{The dependence of the diffractive slope on c.m. energy $B(W)$ given by Eq.~(\ref{BW}) for the process $\gamma^*\,p\to \Jpsi\,p$ with two characteristic parameters $B_0$ and $\alpha'(0)$ determined by the fit to combined H1 \cite{Aktas:2005xu,Alexa:2013xxa} and ZEUS \cite{Chekanov:2002xi,Chekanov:2004mw} data for two distinct $Q^2$ regions  -- low-$Q^2$ (photoproduction domain, left) and high-$Q^2$ (electroproduction domain, right).}
%%%%%%%%%%%%%%%%%%%%%%%%%%
  \label{fig:psi1S-Bslope}
%%%%%%%%%%%%%%%%%%%%%%%%%%
 \end{center}
 \end{figure}

Both parameters $B_0$ and $\alpha'(0)$ for the process $\gamma^*\,p\rightarrow\Jpsi\,p$
have been obtained by a fit to data from H1 \citep{Aktas:2005xu,Alexa:2013xxa} and ZEUS \citep{Chekanov:2002xi,Chekanov:2004mw} collaborations at HERA as well as by our overall fit to the combined data from both collaborations as is shown in Fig.~\ref{fig:psi1S-Bslope}. Our fit resulted with $\chi^2/ndf = 0.6$ for photoproduction and $\chi^2/ndf = 3.75$ for electroproduction. The corresponding values are presented in Table~\ref{Tab:slope-pars}.
%
%                Table 1
%===========================================
\begin{table}[!htbp]
\begin{center}
  \begin{tabular}{|l|SS|SS|}
    \hline
    \multirow{2}{*}{Parameters} &
      \multicolumn{2}{c|}{$Q^2 < 1$ GeV$^2$} &
      \multicolumn{2}{c|}{$Q^2 > 1$ GeV$^2$}  \\
      & {$B_0$} & {$\alpha'(0)$} & {$B_0$} & {$\alpha'(0)$}  \\ \hline
      fixed B \cite{Hufner:2000jb}   &   4.73    &  0  &  3.86   &  0          \\
      H1    \citep{Aktas:2005xu}     &   4.63    &  0.164  &  3.86   &  0.019        \\ 
      ZEUS \citep{Chekanov:2002xi}   &   4.15    &  0.116  &  4.72   &  0.07          \\  
      this work                      &   4.62    &  0.171  &  4.42   &  0.031          \\        \hline
  \end{tabular}
\end{center}
\caption{Parameters $B_0$ and $\alpha'(0)$ of the diffraction slope $B$ obtained by a fit 
         to different data sets at HERA in photo- ($Q^2 < 1\,\GeV^2$) and electroproduction 
         ($Q^2 > 1\,\GeV^2$) of ground-state $1S$-charmonium.}
%%%%%%%%%%%%%%%%%%%%%%
\label{Tab:slope-pars}
%%%%%%%%%%%%%%%%%%%%%%
\end{table}
%===========================================
%
%
The values of $\alpha'(0)$ extracted from the available HERA data are in accordance with theoretical predictions in Ref.~\cite{nnn-94} based on the color dipole formalism and presented already in 1994. It was shown that the slope of the Pomeron trajectory is strongly correlated with the magnitude of the gluon correlation radius.

Since the data for the diffraction slope at $Q^2\gg 0$ are scarce,
for the $Q^2$ dependence of the slope parameter $B(Q^2)$ we use the empirical parametrization from Ref.~\cite{jan-98} based on the color dipole model calculations and valid for production of $\Jpsi$ and $\Y$ within the range of $Q^2\lesssim 500\,\GeV^2$,
%
%********************************************
\beqn
B(W,Q^2)\approx B(W,Q^2=0) - 
B_1\,\ln\,\Bigl (\frac{Q^2 + M_V^2}{M_{\Jpsi}^2}\Bigr )\, ,
%%%%%%%%%%%
\label{BQ2}
%%%%%%%%%%%
\eeqn
%********************************************
%
where the energy dependence of $B(W,Q^2=0)$ is determined using Eq.~(\ref{BW}) with parameters found in Tab.~\ref{Tab:slope-pars}, and $B_1 = 0.45\,\GeV^{-2}$. 
We tested that such a parametrization gives values of the slope parameter in a reasonable agreement with the existing data \cite{Aktas:2005xu,Chekanov:2002xi} on electroproduction of $\Jpsi$ at HERA.

Here we would like to emphasize that for the photo- and electroproduction of $1S$ bottomonium the corresponding diffraction slope can be estimated also from Eq.~(\ref{BQ2}) as $B_{\Y}(W,Q^2)\approx B_{\Jpsi}(W,Q^2+M_{\Y}^2)$.

%
%
%====================================
\subsubsection{Diffraction slope for the process $\gamma^*\,p\to\psip (\Yp)\,p$}
\label{Sec:psi2S-slope}
%====================================
%
%

Detailed analysis of the diffraction slope in photo- and electroproduction of $2S$-radially excited heavy quarkonia $\psip(2S)$ and $\Yp(2S)$
is presented in Ref.~\cite{jan-98}. It was shown within the color dipole formalism
that the inequality $B(2S) < B(1S)$ comes from the nodal structure of corresponding quarkonium wave functions. This is a direct consequence of the destructive interference of the contributions to the production amplitude coming from regions of small and large dipole separations. For production of bottomonia states, the node effect is negligibly small and one can safely use the same magnitudes of the slope parameter for both $1S$ and $2S$ states, i.e. $B_{\Yp}(2S)\sim B_{\Y}(1S)$.

However, for production of $2S$-radially excited charmonium, the difference of diffraction slopes $\Delta_B=B(1S)-B(2S)$ can not be neglected. Model calculations within the color dipole formalism \cite{jan-98} at $W=90\,\GeV$ give the values $\Delta_B^T\sim 0.25\,\GeV^{-2}$ and $\Delta_B^L\sim 0.45\,\GeV^{-2}$
for photoproduction of $T$ and $L$ polarized $\psip(2S)$, respectively, as a clear
manifestation of the node effect. The quantity $\Delta_B$ gradually vanishes with $Q^2$ and can be neglected at $Q^2\gtrsim 20\,\GeV^2$ as a result of a weak node effect at small dipole sizes. However, $\Delta_B$ rises towards small energies   
and at $W=15\,\GeV$ reaches much higher values, i.e.
$\Delta_B^T\sim 0.38$ and $\Delta_B^L\sim 0.9\,\GeV^{-2}$ \cite{jan-98}  
for photoproduction of $T$ and $L$ polarized $\psip(2S)$, respectively.

In our calculations, we employ the following parametrization of the color dipole model predictions of the positive-valued part of $\Delta_B$ \cite{jan-98},
%
%********************************************
\beqn
\Delta_B^{T,L} (W,Q^2)= 
c^{T,L}(W) \Biggl [1 - d(W)\,
\ln \biggl (\frac{Q^2+M_{\psip}^2}{M_{\psip}^2}\biggr)\Biggr ]\,\geq 0\, .
%%%%%%%%%%%%%%%%
\label{delta-b}
%%%%%%%%%%%%%%%%
\eeqn
%********************************************
%
Otherwise, $B(1S) = B(2S)$ for $\Delta_B^{T,L} (W,Q^2) \lesssim 0$ is adopted.
Here, the energy-dependent coefficients are $c^T(W)=0.24-0.08\,\ln(W/W_0)\,\GeV^{-2}$ and $c^L(W)=0.45-0.24\,\ln(W/W_0)\,\GeV^{-2}$ for production of $T$ and $L$ polarized $\psip(2S)$ state, respectively, and the factor $d(W) = 1.65 + 0.3\,\ln(W/W_0)$.

In what follows, in all figures we denote by ``out fit'' the model calculations that use the parametrization of the slope parameter given by Eq.~(\ref{BQ2}), where the first term $B(W,Q^2=0)$ is determined from Table~\ref{Tab:slope-pars}.
%
%                 Fig.6
%===========================================
%===========================================
%
%

First, we test how uncertainties in determination of the diffraction slope for the process $\gamma^*\,p\to \Jpsi\,p$ lead to uncertainties in model predictions for the real and virtual photoproduction cross sections. For this purpose, we use the realistic BT potential \cite{Buchmuller:1980su}
(see also Appendix~\ref{App:potentials})
in determination of the charmonium wave functions as well as the phenomenological KST dipole cross section \cite{Kopeliovich:1999am}, which provides a good description of hadronic processes, also at small scales corresponding to the nonperturbative region of large dipole sizes.  

\begin{figure}[!htbp]
\begin{center}
\includegraphics[width=0.47\textwidth]{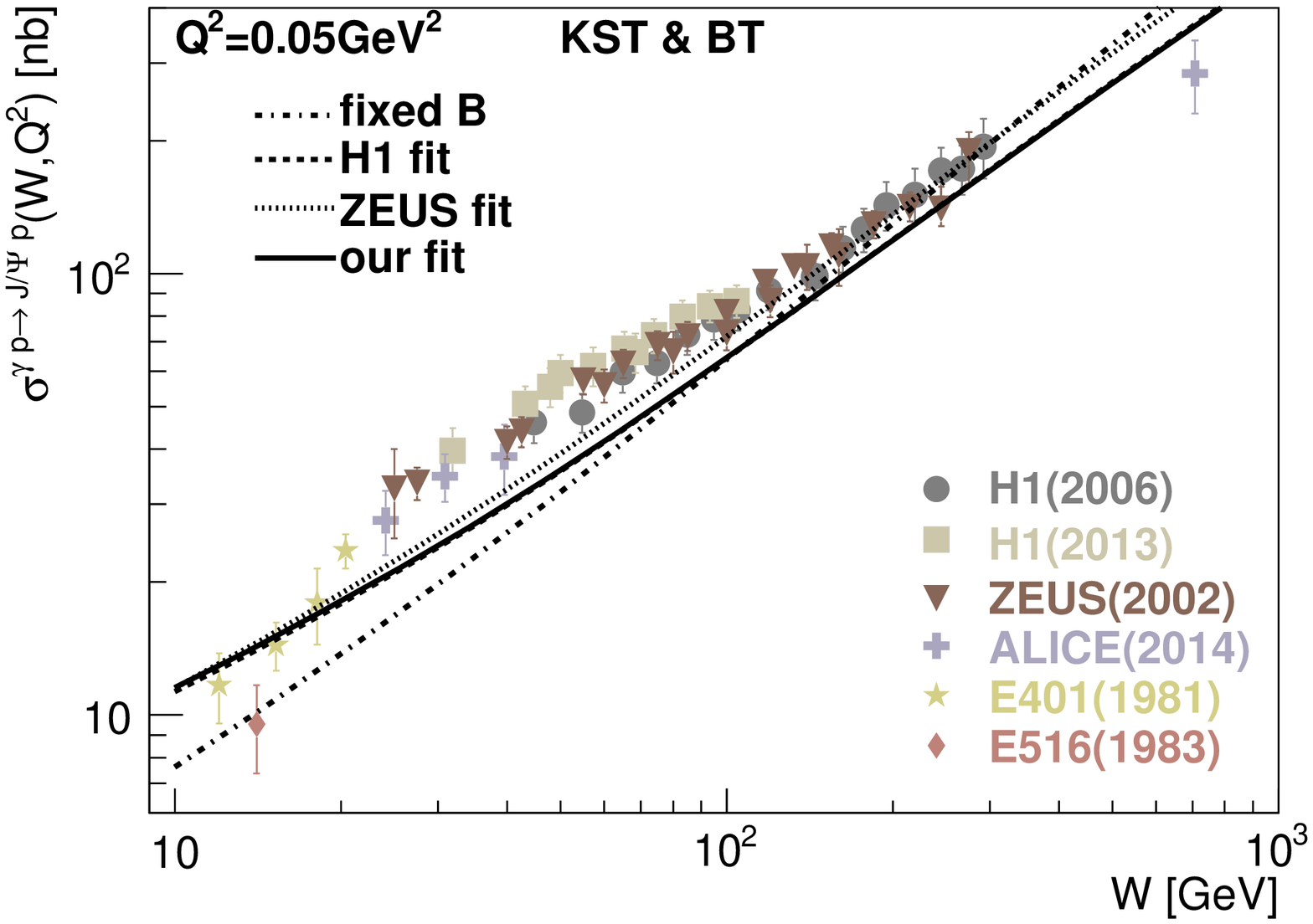}
\includegraphics[width=0.47\textwidth]{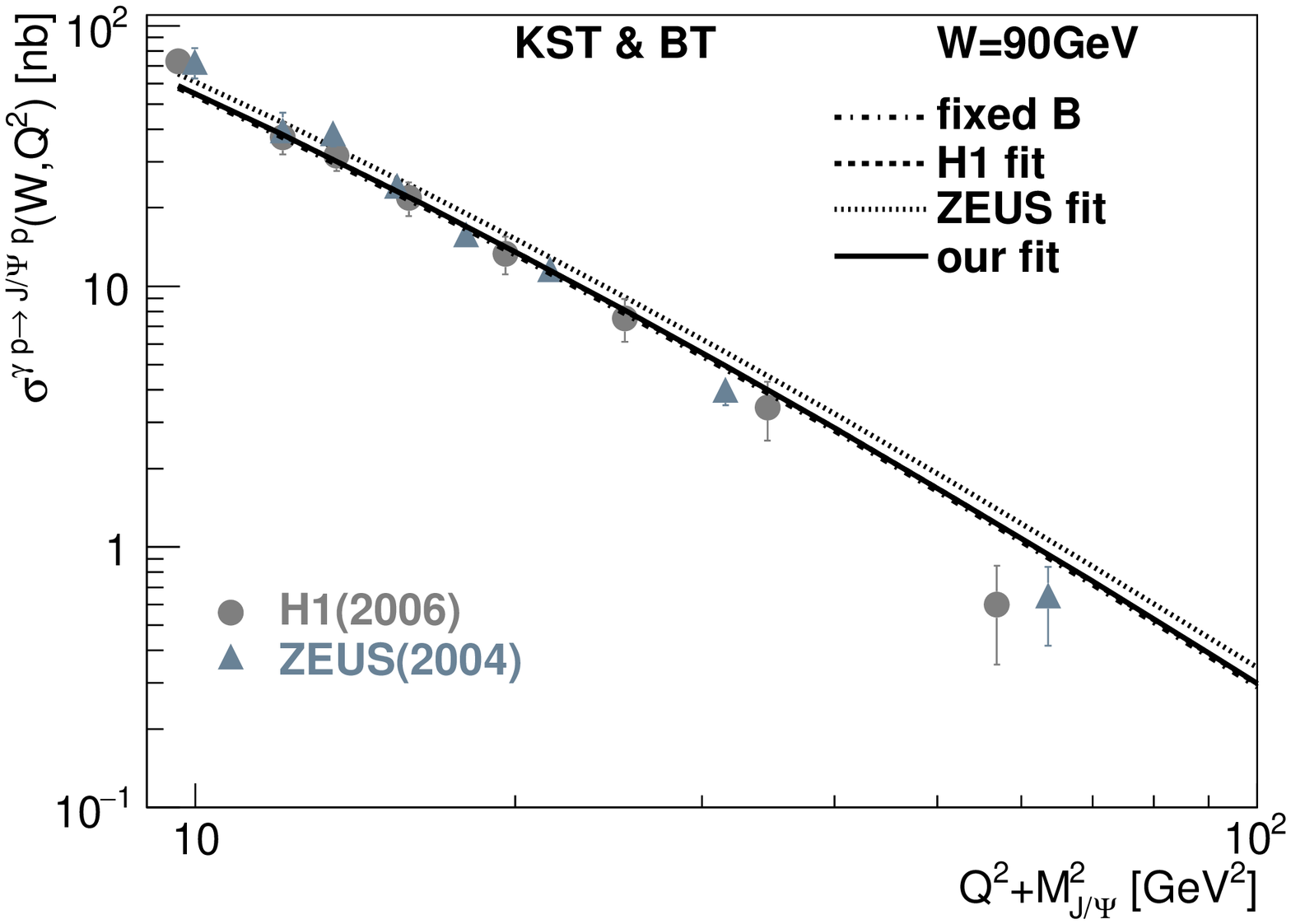}
  \caption{The exclusive $\Jpsi$ electroproduction cross section as a function of c.m. energy $W$ at fixed $Q^2=0.05$ GeV$^2$ (left panel) and the scaling variable  $Q^2+M_{\Jpsi}^2$ at fixed $W=90$ GeV (right panel). The model calculations were performed with the $\Jpsi$ wave function, obtained by using the BT potential, \cite{Buchmuller:1980su}
  and with the phenomenological KST dipole cross section \cite{Kopeliovich:1999am}.
  Here and below, the $Q^2$-dependent slope parameter labeled as ``our fit'' is determined from Eq.~(\ref{BQ2}) with the energy behavior at $Q^2\to 0$ found in Table~\ref{Tab:slope-pars} and indicated there as ``this work''.
  The results also incorporate the Melosh spin rotation effects. The data are 
  provided by H1 \cite{Aktas:2005xu,Alexa:2013xxa}, ZEUS  \cite{Chekanov:2002xi,Chekanov:2004mw}, ALICE \cite{TheALICE:2014dwa}, E401 \cite{Binkley:1981kv} and E516 \cite{Denby:1983az} Collaborations.}
%%%%%%%%%%%%%%%%%%%%%%%%%%
  \label{fig:psi1S-Bcomp}
%%%%%%%%%%%%%%%%%%%%%%%%%%
 \end{center}
 \end{figure}

Fig.~\ref{fig:psi1S-Bcomp} shows the color dipole model calculations
versus the HERA data on the photo- and electroproduction cross sections as a function of the c.m. energy $W$ at fixed $Q^2 = 0.05\,\GeV^2$ (left panel) and the scaling variable $Q^2+M_{\Jpsi}^2$ at fixed $W=90\,\GeV$ (right panel) using the different parametrizations for the diffraction slope as depicted in Table~\ref{Tab:slope-pars}. Corresponding amplitudes (\ref{AT}) and (\ref{AL}) for production of $T$ and $L$ polarized charmonia, entering the expression (\ref{total-cs}) for the electroproduction cross section, contain corrections for the Melosh spin rotation effects. Since the data on the $Q^2$ behavior of the diffraction slope are very scarce we took the results of model calculations \cite{jan-98}, which can be simply parametrized by Eq.~(\ref{BQ2}) and provide a reasonable description of the HERA data.

One can see from the left panel of Fig.~\ref{fig:psi1S-Bcomp} that model
predictions using the constant value for the slope parameter $B=4.73\,\GeV^{-2}$ underestimate the data at lower c.m. energies $W\lesssim 100\,\GeV$. However, they lead to an overestimation of the ALICE experimental value \cite{TheALICE:2014dwa} at higher $W\sim 700\,\GeV$. An agreement with the data can be improved by taking the energy-dependent diffraction slope with parameters from Table~\ref{Tab:slope-pars}. All these parametrizations lead to very similar values for the diffraction slope at small energies but start to differentiate from each other at higher energies corresponding to the LHC energy range. Here, the best description of the data is achieved by the fit to only H1 data, as well as by our fit to the combined H1 and ZEUS data sets.

The right panel of Fig.~\ref{fig:psi1S-Bcomp} shows the model predictions for electroproduction cross section $\sigma^{\gamma^*\,p\to\Jpsi\,p}(W,Q^2)$ as a function of the scaling variable $Q^2+M_{\Jpsi}^2$ at fixed value of c.m. energy $W=90\,\GeV$. The $Q^2$ dependence of the slope parameter is given by the empirical formula Eq.~(\ref{BQ2}), whereas for $B(W=90,Q^2=0)$ we take different parametrizations from Table~\ref{Tab:slope-pars}. As a result, the shape of the corresponding theoretical curves is almost identical describing the available data from H1 and ZEUS collaborations reasonably well.

Note that differences in model predictions using various parametrizations for the diffraction slopes can be treated as a measure of the underlined theoretical uncertainty.
%
%                 Fig.7
%===========================================
%===========================================
%
%
%
%                Fig.8
%===========================================
\begin{figure}[!htbp]
\begin{center}
\includegraphics[width=0.47\textwidth]{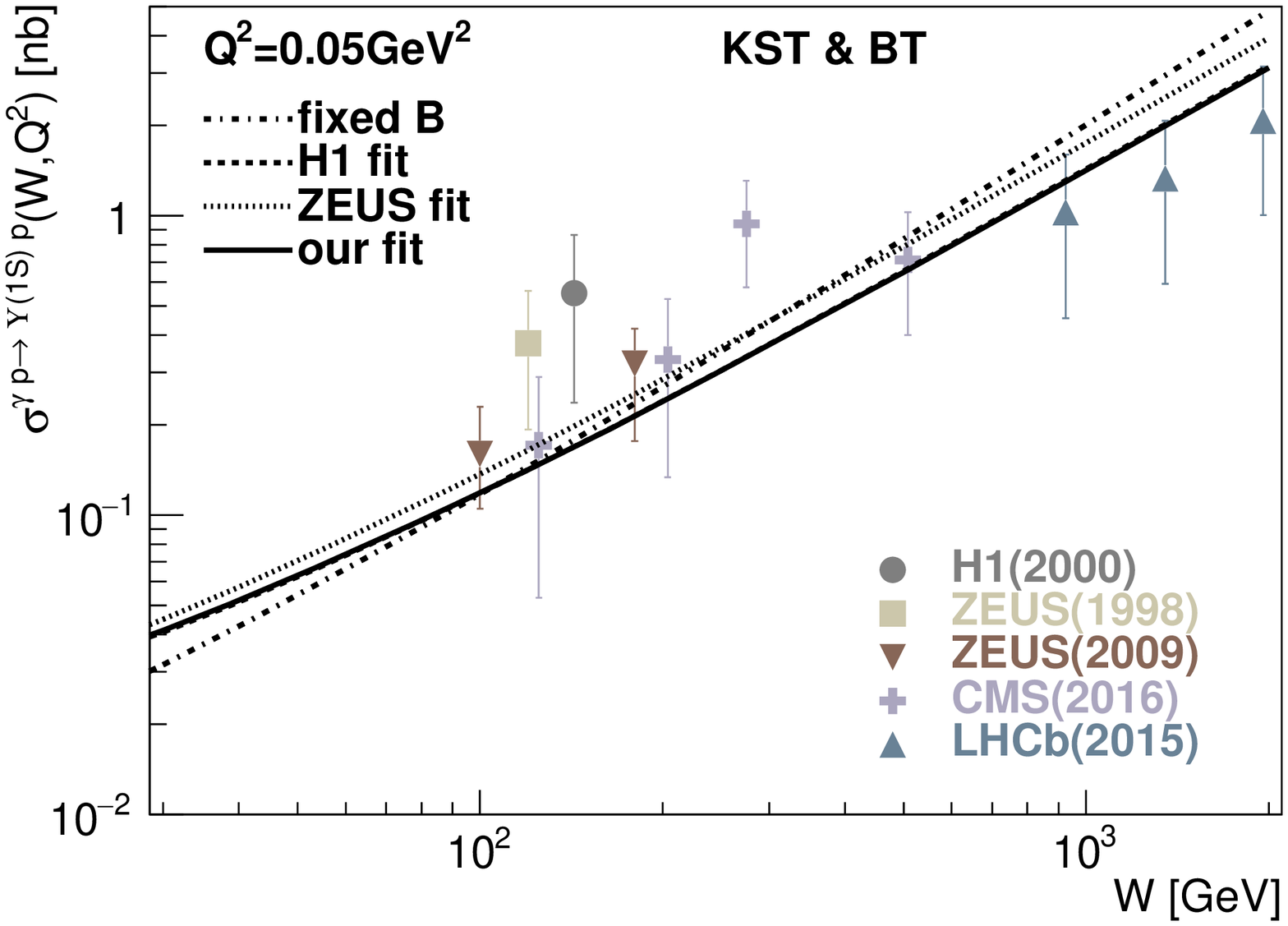}
\includegraphics[width=0.47\textwidth]{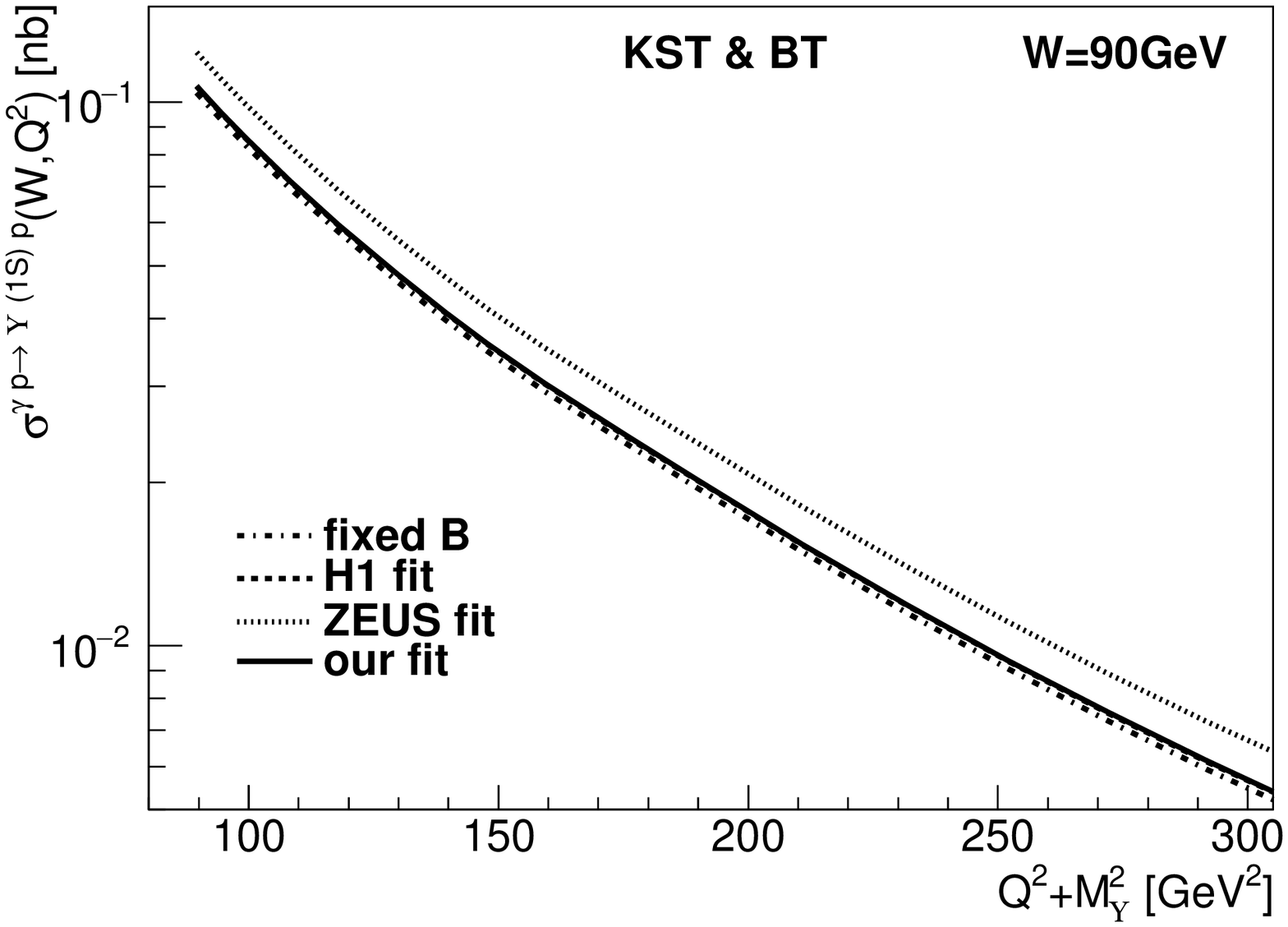}
  \caption{The same as Fig.~\ref{fig:psi1S-Bcomp} but for the photo- and electroproduction of $\Y(1S)$ state. The model predictions for $\sigma^{\gamma^*\,p\to\Y\,p}(W,Q^2)$ are compared with the existing data from H1 \cite{Adloff:2000vm}, ZEUS \cite{Breitweg:1998ki,Chekanov:2009zz}, CMS \cite{CMS:2016nct} and LHCb \cite{Aaij:2015kea}
  experiments.}
%%%%%%%%%%%%%%%%%%%%%%%%%%
  \label{fig:upsi1S-Bcomp}
%%%%%%%%%%%%%%%%%%%%%%%%%%
 \end{center}
 \end{figure}
%===========================================
%
%

Theoretical uncertainties in predictions of the real and virtual photoproduction cross sections of the elastic process $\gamma^*\,p\to\Y(1S)\,p$ inherent for determination of the slope parameter are depicted in Fig.~\ref{fig:upsi1S-Bcomp}. Here, similarly to the case of $1S$-charmonium production, we compare the model predictions for different parametrizations of the slope parameter from Table~\ref{Tab:slope-pars}. In the case of electroproduction of $1S$ bottomonium, the corresponding diffractive slope can be approximately estimated from Eq.~(\ref{BQ2}) as follows: $B_{\Y}(W,Q^2)\approx B_{\Jpsi}(W,Q^2+M_{\Y}^2)$. This is a consequence of the scaling properties in production of different vector mesons \cite{jan-98}. Here, we assume a similar value of the Pomeron trajectory slope $\alpha'(0)$ describing the energy dependence of the diffractive slope, see Eq.~(\ref{BW}), for charmonium as well as for bottomonium production. This is supported by calculations of $\alpha'(0)$ performed in Ref.~\cite{nnn-94} within the color dipole formalism.

The left panel of Fig.~\ref{fig:upsi1S-Bcomp} clearly demonstrates that
inclusion of the energy dependent slope parameters brings our model predictions to a better agreement with the available data. As was already emphasized above, the differences in model predictions corresponding to different parametrizations of the diffractive slope can be considered as a good measure of the underlined theoretical uncertainty.
%
%                  Fig.9
%===========================================
\begin{figure}[!htbp]
\begin{center}
\includegraphics[width=0.47\textwidth]{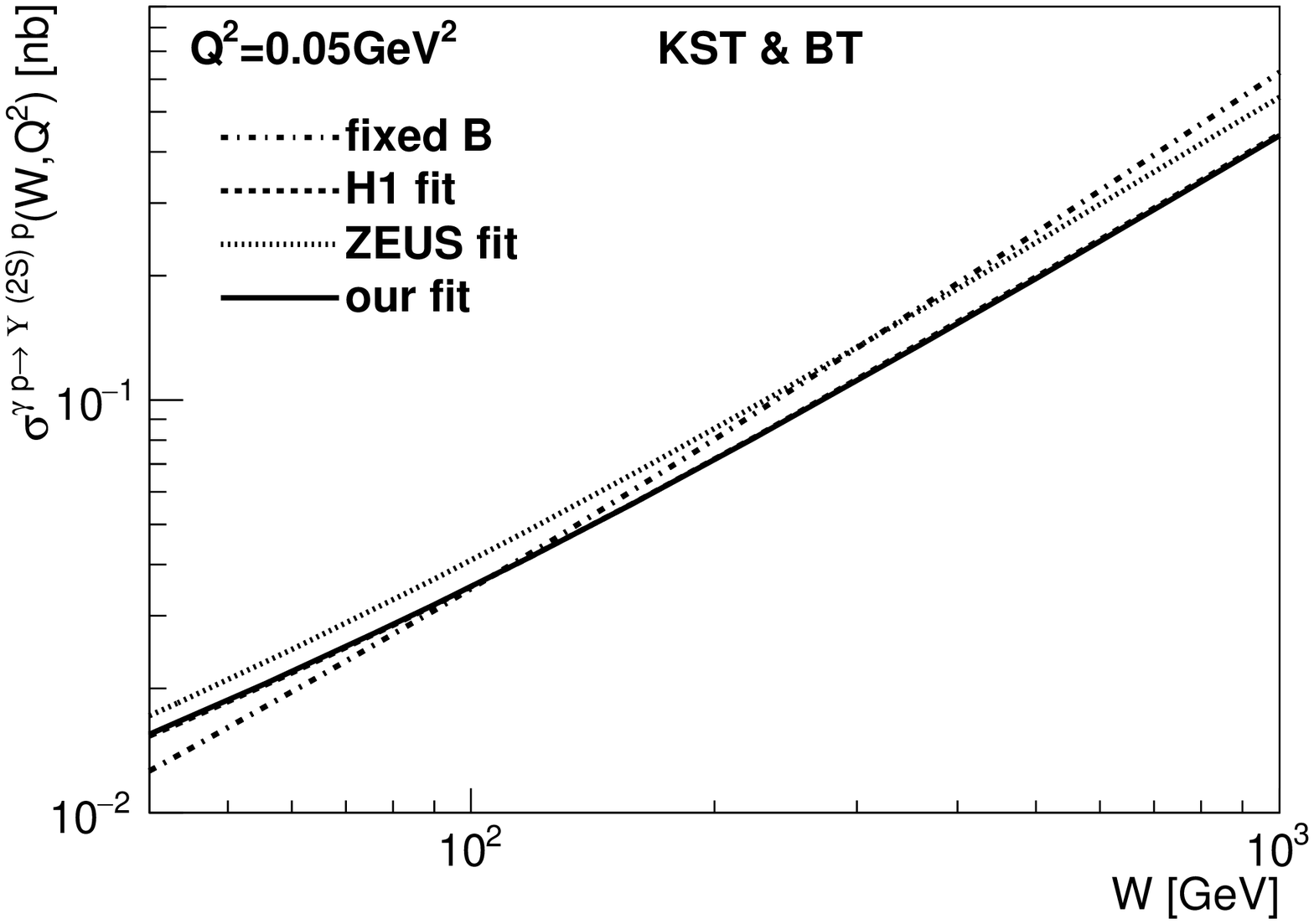}
\includegraphics[width=0.47\textwidth]{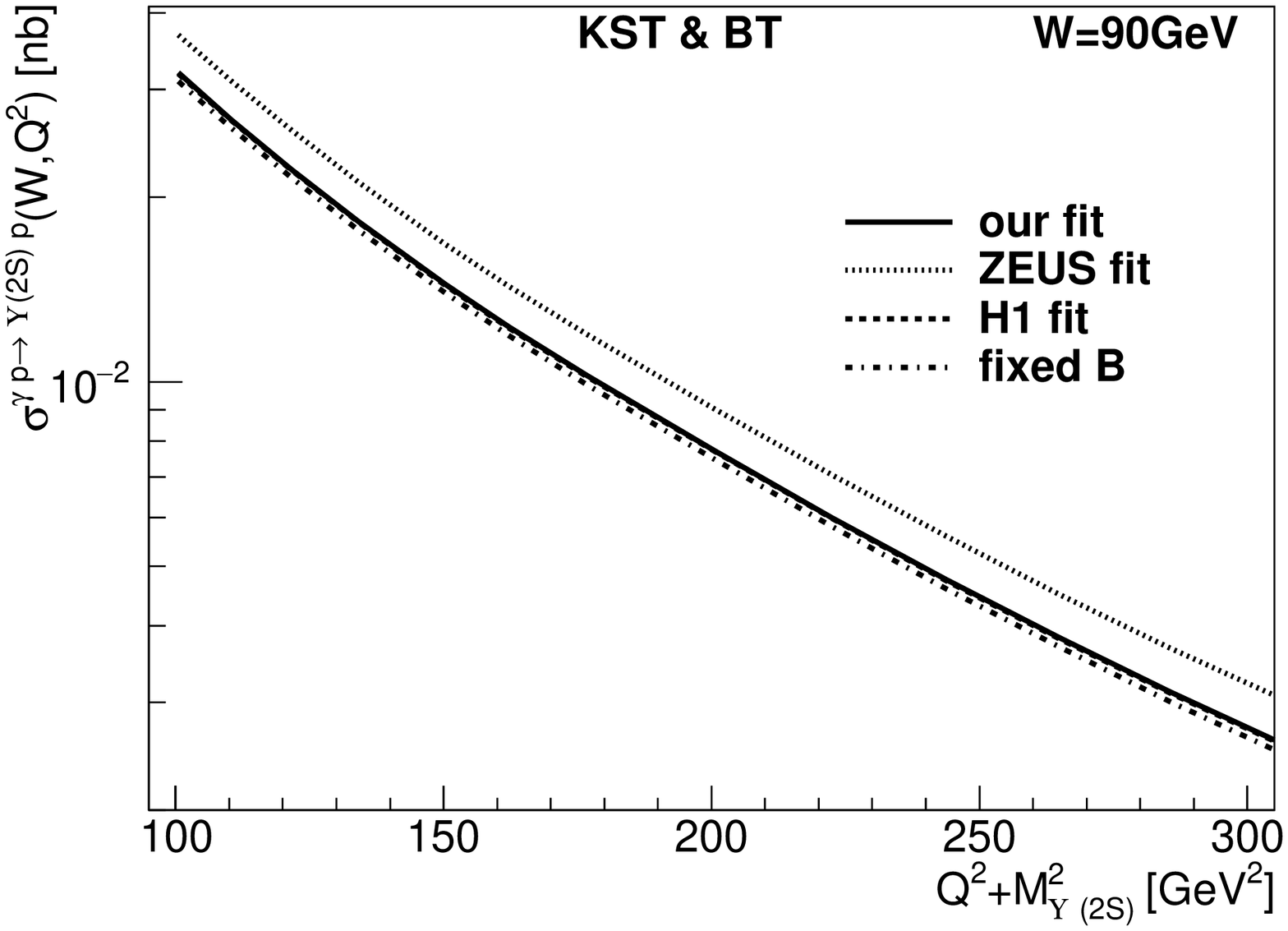}
  \caption{The same as Fig.~\ref{fig:psi1S-Bcomp} but for the real and virtual photoproduction  of $\Yp(2S)$ bottomonia.}
%%%%%%%%%%%%%%%%%%%%%%%%%%%
  \label{fig:upsi2S-Bcomp}
%%%%%%%%%%%%%%%%%%%%%%%%%%%
 \end{center}
 \end{figure}
%===========================================
%
%
%
%                Fig.10
%===========================================
\begin{figure}[!htbp]
\begin{center}
\includegraphics[width=0.47\textwidth]{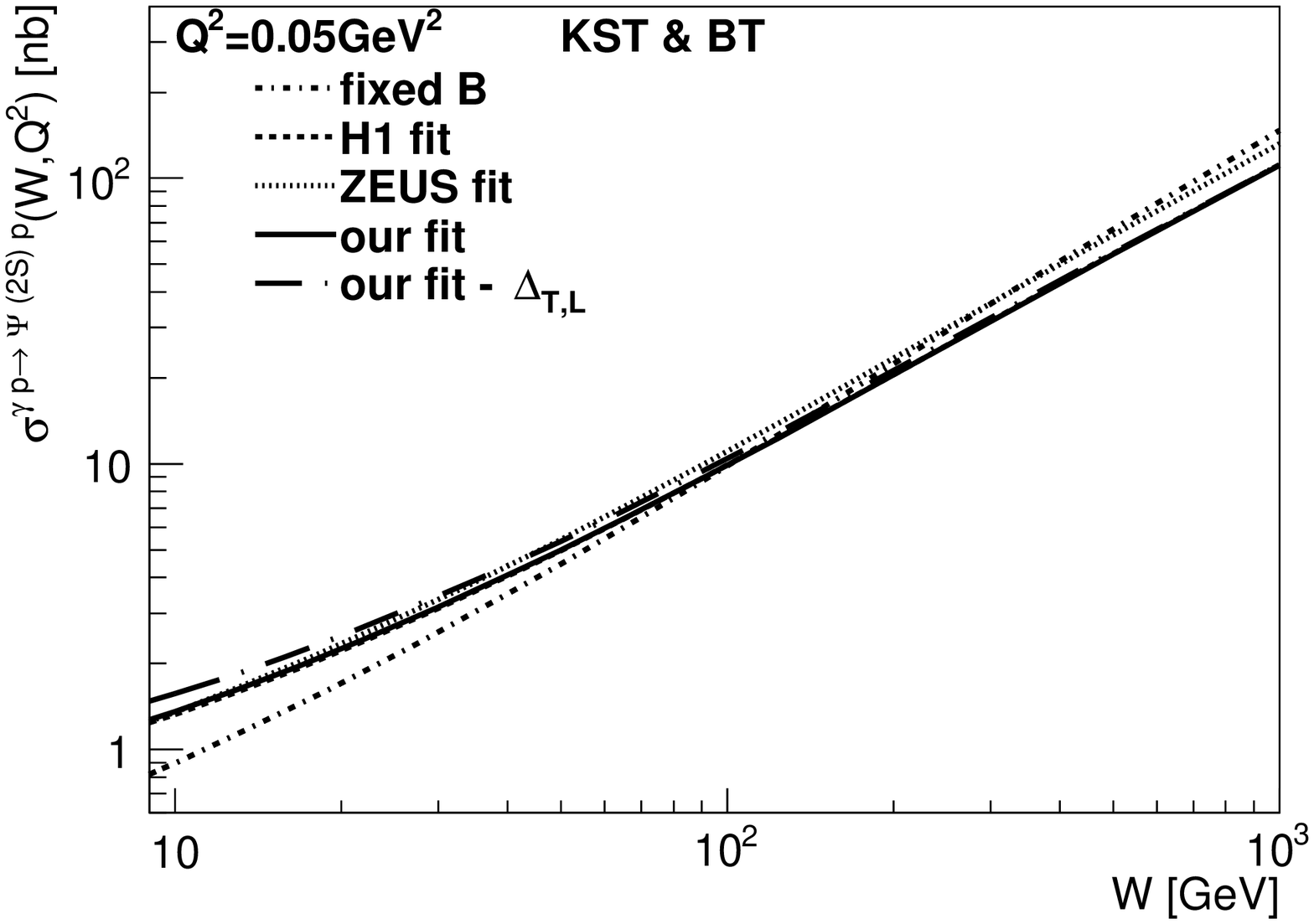}
\includegraphics[width=0.47\textwidth]{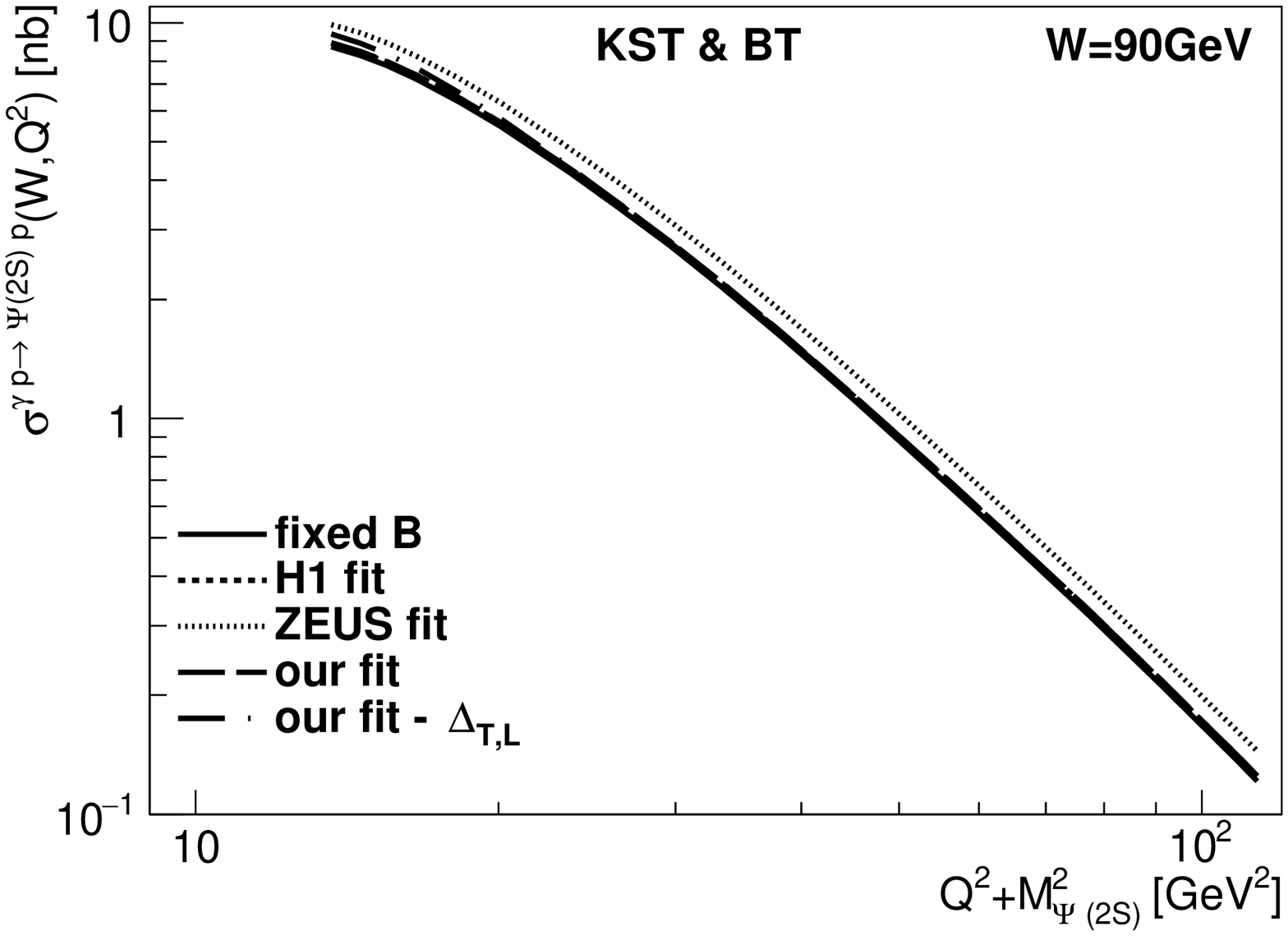}
  \caption{The same as Fig.~\ref{fig:psi1S-Bcomp} but for the real and virtual photoproduction of $\psip(2S)$ charmonium.}
%%%%%%%%%%%%%%%%%%%%%%%%%
  \label{fig:psi2S-Bcomp}
%%%%%%%%%%%%%%%%%%%%%%%%%
 \end{center}
 \end{figure}
%===========================================
%
%

For the photo- and electroproduction of $2S$ radially-excited $\psip(2S)$ and $\Yp(2S)$ states the nodal structure of the corresponding wave functions (see Figs.~\ref{fig:psiLC} and \ref{fig:LC}) causes an inequality $B(2S)\lesssim B(1S)$. The corresponding difference $B(1S)-B(2S)$ was calculated in Ref.~\cite{jan-98} within the color dipole formalism and can be parametrized as is given by Eq.~(\ref{delta-b}). For the photo- and electroproduction of $\Yp(2S)$ the node effect can be neglected and we can safely take the same slope parameter as for the $\Y(1S)$ state, namely, $B_{\Yp}(2S)\sim B_{\Y}(1S)$. The corresponding model predictions, taking four different parametrizations for the diffraction slope from Table~\ref{Tab:slope-pars}, are presented in Fig.~\ref{fig:upsi2S-Bcomp}. 

\begin{figure}[!htbp]
\begin{center}
\includegraphics[width=0.48\textwidth]{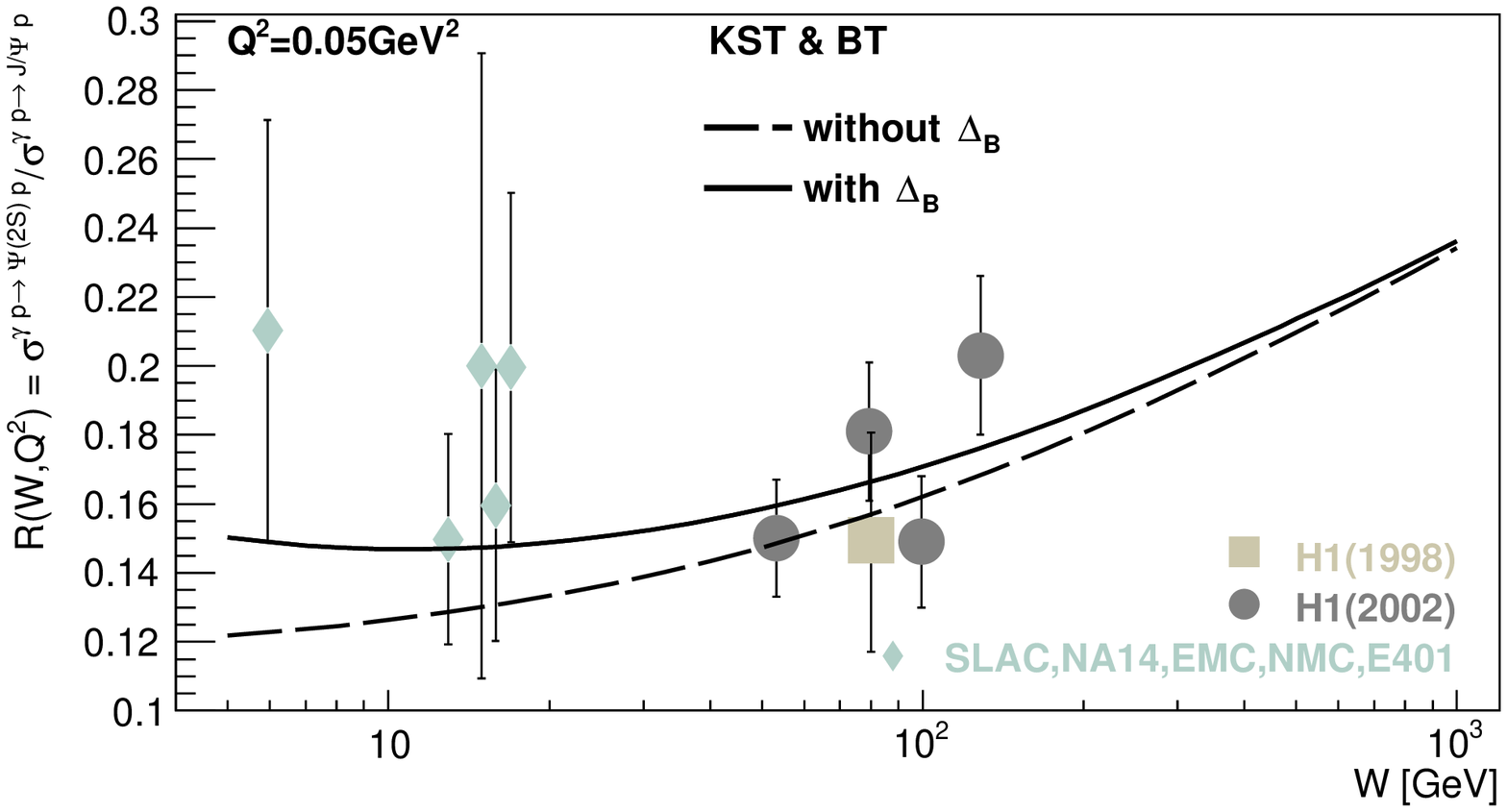}
\includegraphics[width=0.48\textwidth]{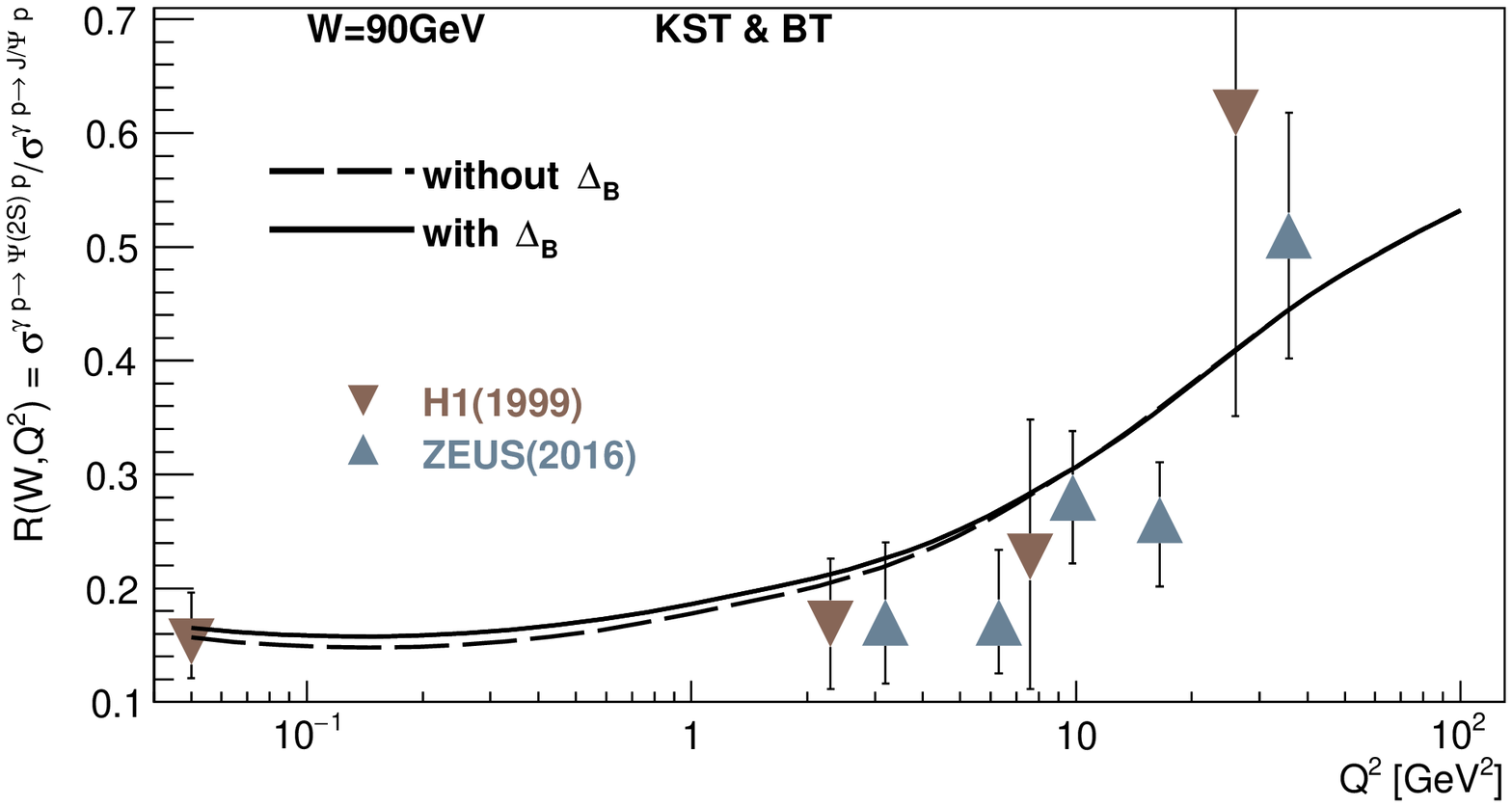}
  \caption{
  The color dipole model predictions for the $\psip(2S)$-to-$\Jpsi(1S)$ ratio of electroproduction cross sections as functions of c.m. energy $W$ at fixed $Q^2=0.05\,\GeV^2$ (left panel), as well as functions of $Q^2$ at fixed $W=90\,\GeV$ (right panel) versus the existing data from H1 \cite{Adloff:1997yv,Adloff:2002re}, ZEUS \cite{Abramowicz:2016xls} and fixed target experiments \cite{Camerini:1975cy,Barate:1986fq,Binkley:1982yn,Aubert:1982tt,Amaudruz:1991sr}. The solid and dashed lines correspond to calculations with and without the correction $\Delta_B$ given by Eq.~(\ref{delta-b}) in determination of the slope parameter for the process $\gamma^*\,p\to\psip(2S)\,p$, respectively.
  The model calculations were performed with the charmonium wave functions obtained by using the BT potential \cite{Buchmuller:1980su}
  and with the phenomenological KST dipole cross section \cite{Kopeliovich:1999am}. The Melosh spin rotation effects are included in this calculation.}
%%%%%%%%%%%%%%%%%%%%%%%%%%%%%%%%
  \label{fig:psi2StoJpsi-Bcomp}
%%%%%%%%%%%%%%%%%%%%%%%%%%%%%%%%
 \end{center}
 \end{figure}

In comparison with $\Yp(2S)$ eletroproduction, a stronger node effect in production of $2S$ radially-excited charmonium causes a larger difference of diffractive slopes given by Eq.~(\ref{delta-b}) such that it can not be neglected. Consequently, one expects that $B_{\Jpsi(1S)}\gtrsim B_{\psip}(2S)$ \cite{jan-98}. The corresponding model predictions for $\sigma^{\gamma^*\,p\to\psip(2S)\,p}(W,Q^2)$ including the different parametrizations for the slope parameter from Table~\ref{Tab:slope-pars}, as well as the corrected diffraction slope in eletroproduction of $2S$ radially-excited charmonium, $B_{\psip}(2S) = B_{\Jpsi}(1S) - \Delta_B$, are shown in Fig.~\ref{fig:psi2S-Bcomp}. One can see that the node effect leads to an enhancement of the $\Jpsi(2S)$ photoproduction cross section, especially at small c.m. energies $W$, as well as at small values of $Q^2$ enabling a better agreement with the data (see also Fig.~\ref{fig:psi2StoJpsi-Bcomp}).
%
%                   Fig.11
%===========================================
%===========================================
%
%

We would like to emphasize that one should distinguish between manifestations of the node effect in amplitude for production of $2S$ radially-excited quarkonia and in the magnitude of the corresponding diffraction slope. The nodal structure of the wave function for radially-excited states causes cancellations in the production amplitude from regions of large and small transverse sizes above and below the node position. Here, investigation of the ratio $R\equiv R(W,Q^2)$ of the $\psip(2S)$-to-$\Jpsi(1S)$ photo- and electroproduction cross sections allows to minimize the theoretical uncertainties connected to a determination of the corresponding slope parameters for $\gamma^*\,p\to\Jpsi(1S)\,p$ and  $\gamma^*\,p\to\psip(2S)\,p$ processes. 

Neglecting the impact of the node effect on the magnitude of the slope parameter $B_{\psip}(2S)$, one can safely use the approximate equality $B_{\Jpsi}(1S)\sim B_{\psip}(2S)$ with a rather good accuracy. Consequently, $B_{\Jpsi}(1S)$ and $B_{\psip}(2S)$ cancel in the ratio $R(W,Q^2)$. Then the rise of $R$ with c.m. energy $W$ and $Q^2$ depicted by dashed lines in Fig.~\ref{fig:psi2StoJpsi-Bcomp} is a characteristic manifestation of the node effect. Since the size of $\psip(2S)$ is larger than $\Jpsi(1S)$, one should naturally expect a stronger energy dependence for the $\Jpsi(1S)$ electroproduction cross section because dipoles with a smaller transverse size have a steeper rise with energy. As a result, the ratio $R(W)$ should decrease with energy. However, despite of this expectation, the nodal structure of the wave function for $2S$ radially-excited states causes an opposite effect, i.e. the rise of $R(W)$ with energy. The steeper energy dependence at smaller dipole sizes below the node position diminishes the node effect at higher energies. This is a result of reduction of a cancellation in the $2S$ production amplitude from regions below and above the node position. This then leads to a steeper energy dependence of $\psip(2S)$ compared to  
$\Jpsi(1S)$ production cross section (compare Fig.~\ref{fig:psi1S-Bcomp} with Fig.~\ref{fig:psi2S-Bcomp}). The rise of the ratio $R(W)$ with c.m. energy $W$ is depicted in the left panel of Fig.~\ref{fig:psi2StoJpsi-Bcomp} where the model predictions are in accordance with the data, especially at higher energies $W\gtrsim 50\,\GeV$. Similarly, the node effect becomes weaker at larger $Q^2$ causing a rise of the ratio $R(Q^2)$ in a reasonable agreement with the existing data as is demonstrated in the right panel of Fig.~\ref{fig:psi2StoJpsi-Bcomp}.

The node effect has some impact also on the magnitude of the diffractive slope for electroproduction of $2S$ radially-excited charmonium as was presented in Ref.~\cite{jan-98} and discussed above. This leads to the following inequality $B(2S)\lesssim B(1S)$. The corresponding difference $\Delta_B$ was computed within the color dipole model in Ref.~\cite{jan-98} and can be parametrized by Eq.~(\ref{delta-b}). This correction $\Delta_B$ rises towards small $W$ and $Q^2$ since the onset of the node effect becomes stronger and leads to an enhancement of the ratio $R(W,Q^2)$ as shown in Fig.~\ref{fig:psi2StoJpsi-Bcomp} by solid lines. Notably, such an effect brings our predictions to a better agreement with the data at smaller energies $W\lesssim 20\,\GeV$.
%
%                      Fig.12
%==========================================================
\begin{figure}[!tbhp]
\begin{center}
\includegraphics[width=0.47\textwidth]{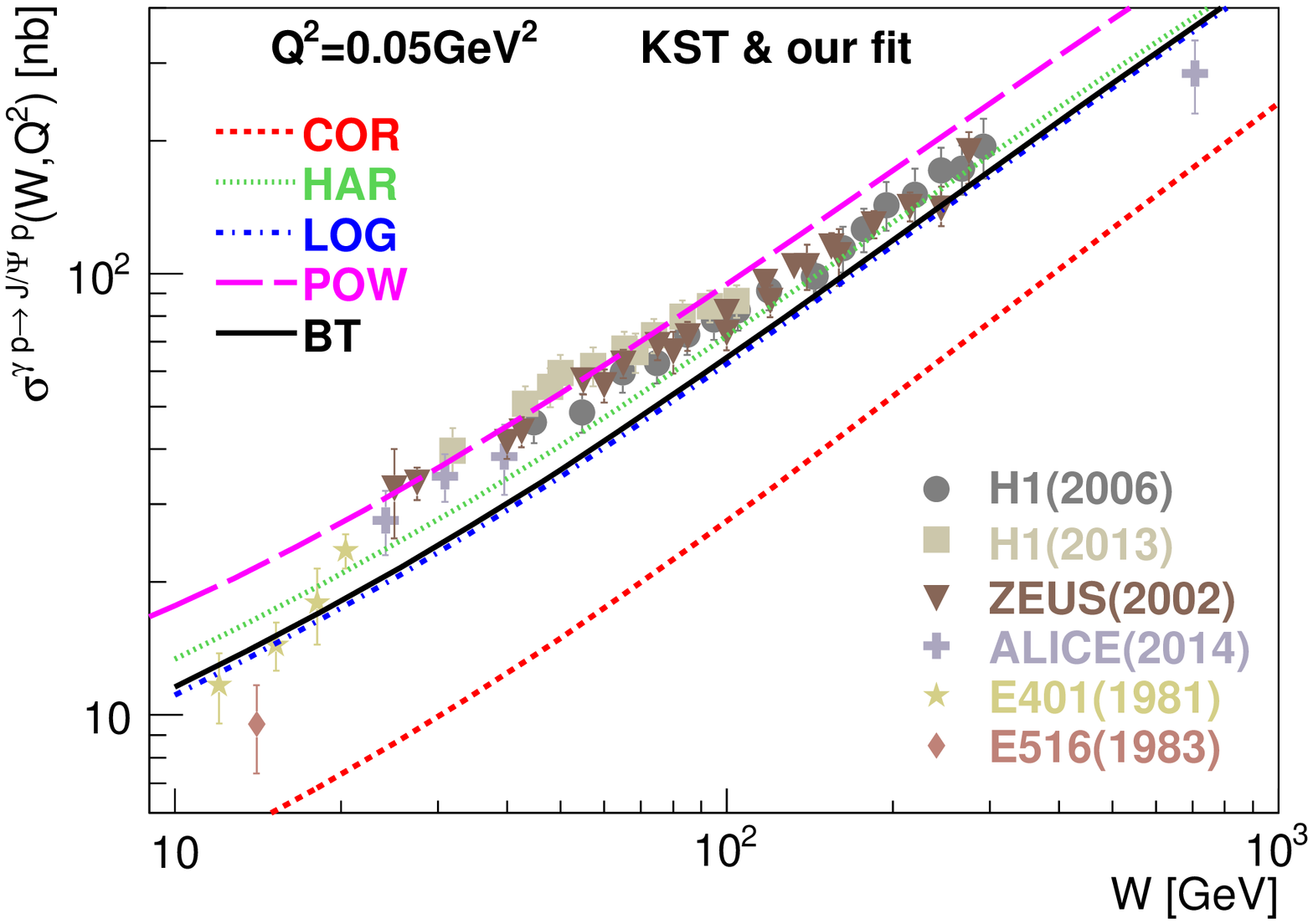}
\includegraphics[width=0.47\textwidth]{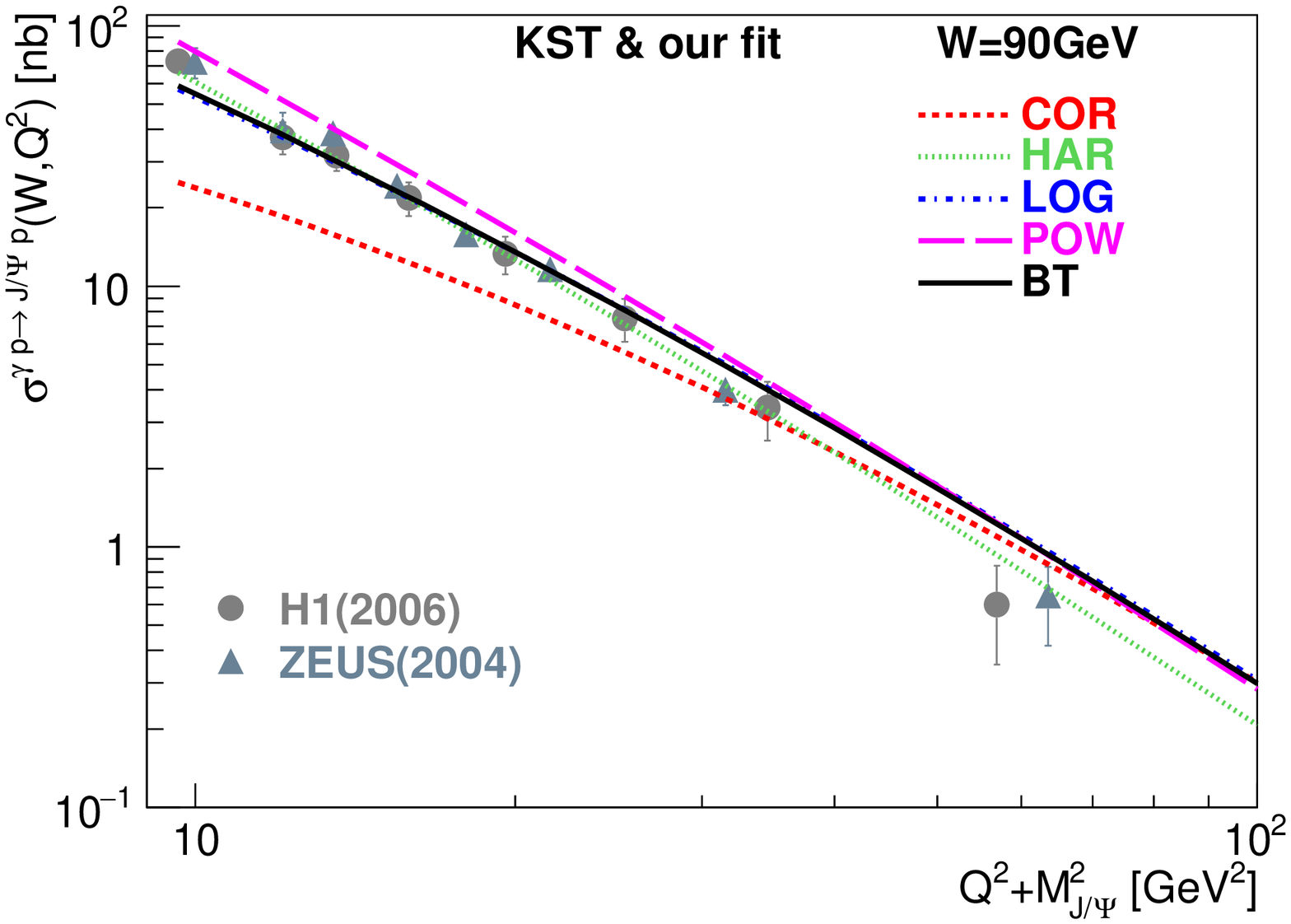}
  \caption{The same as Fig.~\ref{fig:psi1S-Bcomp} but for different realistic
  $c-\bar c$ interaction potentials as described in Appendix~\ref{App:potentials}.
  }
%%%%%%%%%%%%%%%%%%%%%%%
  \label{fig:psi1S-pot}
%%%%%%%%%%%%%%%%%%%%%%%
 \end{center}
 \end{figure}
%==========================================================
%
%
%
%                      Fig.13
%==========================================================
\begin{figure}[!htbp]
\begin{center}
\includegraphics[width=0.47\textwidth]{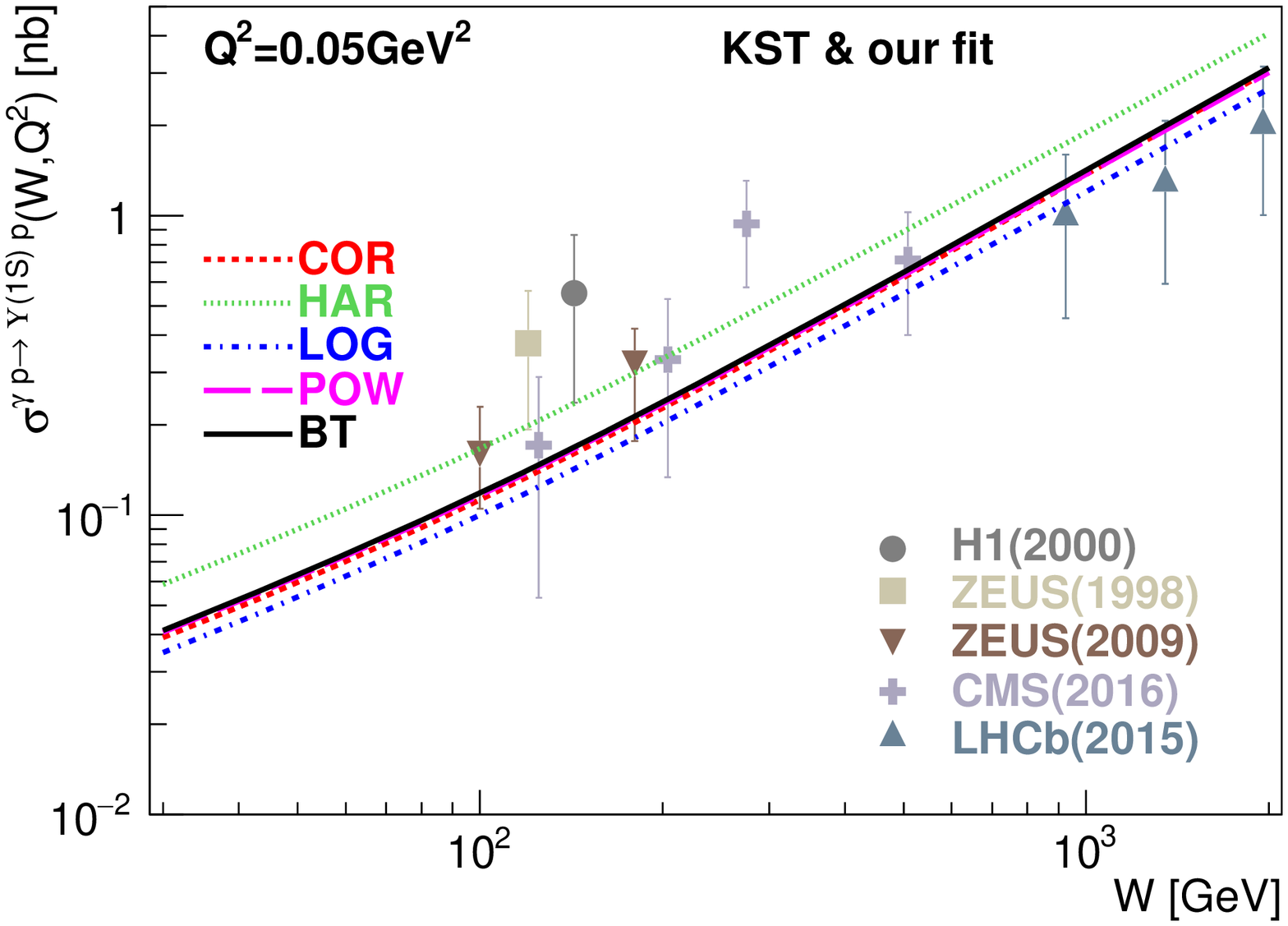}
\includegraphics[width=0.47\textwidth]{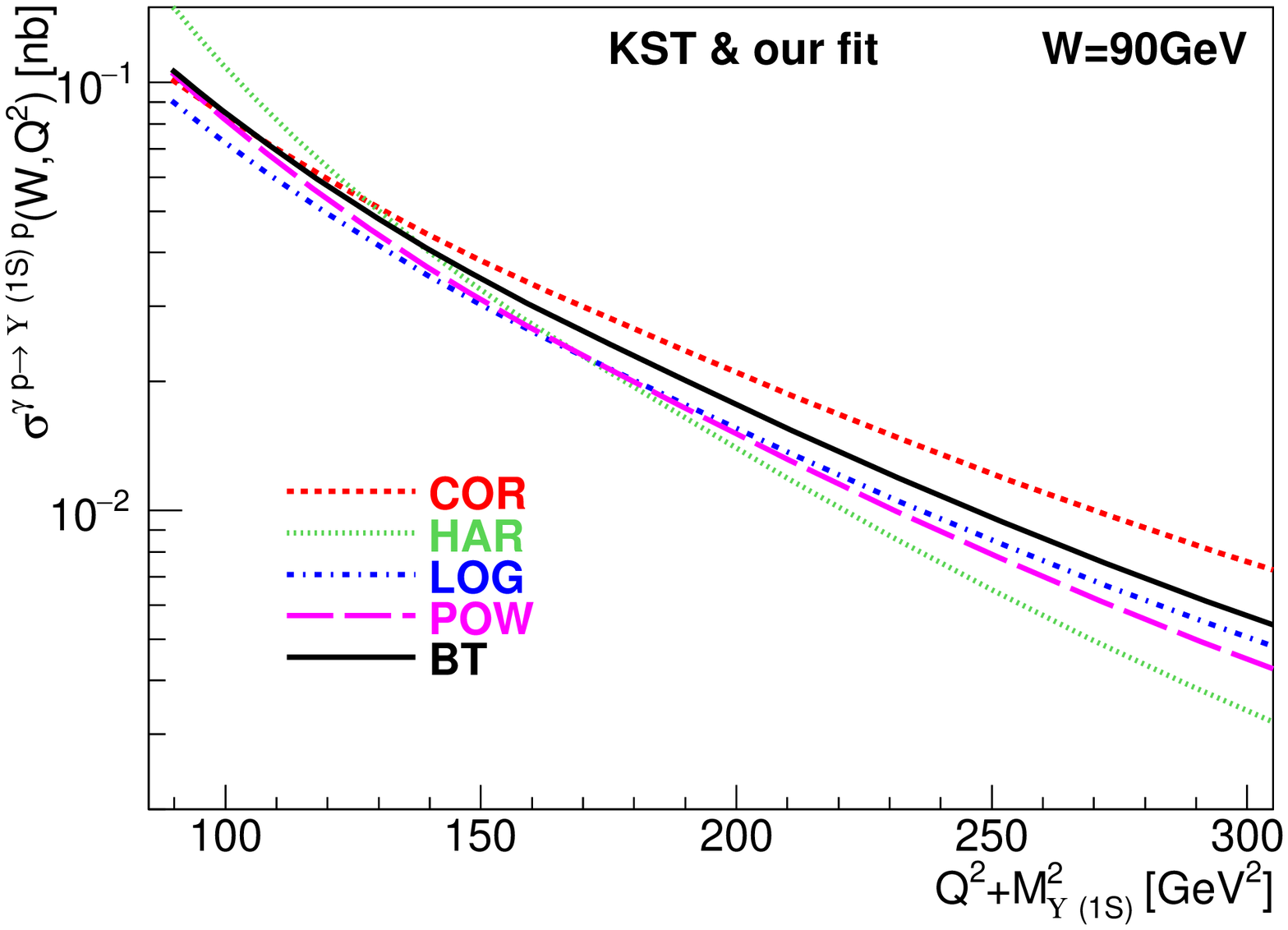}
  \caption{The same as Fig.~\ref{fig:upsi1S-Bcomp} but for different realistic
  $b-\bar b$ interaction potentials as described in Appendix~\ref{App:potentials}. }
%%%%%%%%%%%%%%%%%%%%%%%
  \label{fig:ups1S-pot}
%%%%%%%%%%%%%%%%%%%%%%%
 \end{center}
 \end{figure}
%==========================================================
%
%

%
%
%==========================================================
\subsection{Theoretical uncertainties caused by a shape of
the $c-\bar c$ ($b-\bar b$) interaction potential}
\label{Sec:psi1S-pot}
%==========================================================
%
%

Here we analyze how determination of the quarkonium wave functions generated by various interquark interaction potentials leads to a different behavior of the photo- and electroproduction cross sections. The results for $\Jpsi$ and $\Y$ are shown in Figs.~\ref{fig:psi1S-pot} and \ref{fig:ups1S-pot} in comparison with the available data. Our calculations were performed using the phenomenological KST parametrization for the dipole cross section and for the $1S$ quarkonium wave functions determined from the COR, HAR, LOG, POW and BT potentials described in Appendix~\ref{App:potentials}. Our observations are as follows:

(i)
The potentials labeled as HAR, BT and LOG well describe the photoproduction $\Jpsi$ data, whereas the potential POW slightly overestimates the data while the potential COR significantly underestimates them by a factor of about $2\div 2.5$.

(ii)
Such a different behavior originates from different charm quark masses used in various potentials. The potentials BT and LOG use $m_c=1.5\,\GeV$, while HAR adopts $m_c=1.4\,\GeV$, POW -- $m_c=1.3\,\GeV$ and COR takes $m_c=1.84\,\GeV$. Different potentials have only a small impact on the shape of wave functions for $1S$-state charmonium (see Fig.~\ref{fig:psir}). However, the photon wave function, Eq.~(\ref{gamma-wf}) is extremely sensitive to the value of $m_c$ that enters the argument of the Bessel function $K_0$. 

(iii)
The model predictions for the photoproduction cross section, quite expectedly, exhibit the following hierarchy: the smaller $c$-quark mass used in the realistic potential leads to higher values of the cross sections (see also Figs.~\ref{fig:psi1S-mq} and \ref{fig:ups1S-mq}).

(iv)
Dependence of the $\Jpsi$ electroproduction cross section (see the right panel of Fig.~\ref{fig:psi1S-pot} on the scaling variable $Q^2+M_{\Jpsi}^2$ follows from the structure of $\varepsilon^2$ in Eq.~(\ref{gamma-wf}). As was analyzed in Ref.~\cite{Nemchik:1996cw} the nonrelativistic approximation with $z=0.5$ can be safely used for production of charmonia and, especially, bottomonia. In this approximation, $\varepsilon^2$ takes the value $\propto Q^2+(2\,m_c)^2\approx Q^2+M_{\Jpsi}^2$.

(v)
Similarly to photoproduction of $1S$ charmonium, the right panel of Fig.~\ref{fig:psi1S-pot} shows a reasonable agreement of the data with our calculations using the COR, HAR, LOG and BT potentials. Differences in model predictions gradually decrease with $Q^2$ since the variation between the corresponding realistic potentials is weaker at smaller dipole transverse separations $r$ (see Fig.~\ref{fig:cb-V}). Only the HAR potential leads to much smaller values of the cross sections at large $Q^2$ due to a lack of the Coulomb-like behavior at small $r$. 

(vi)
Model predictions for the $\Y(1S)$ photoproduction cross section depicted in the left panel of Fig.~\ref{fig:ups1S-pot} exhibit a rather good description of the data with the use of all five realistic potentials considered in this work. Thus, we confirm the universality property of the quarkonia production cross sections as functions of the scaling variable $Q^2+M_V^2$. Here, due to such universality the theoretical uncertainty given by a spread between the results obtained with different interquark potentials (see the left panel of Fig.~\ref{fig:ups1S-pot}) directly corresponds to the results for $1S$ charmonia electroproduction at $Q^2\sim M_{\Y}^2$ (compare with the right panel of Fig.~\ref{fig:psi1S-pot}).

(vii) A small variance of the model predictions made also for $1S$ bottomonia electroproduction using different realistic potentials is demonstrated in the right panel of Fig.~\ref{fig:ups1S-pot}. However, this variance rises with $Q^2$ due to growing differences between the considered $b-\bar b$ interaction potentials at small $\tilde r\lesssim 0.1\,\fm$ (see the right panel of Fig.~\ref{fig:cb-V}).

%
%
%====================================
\subsection{Theoretical uncertainties caused by different parametrizations of the color dipole cross section}
\label{Sec:psi1S-dip}
%====================================
%
%

%
%               Fig.14
%====================================
\begin{figure}[!htbp]
\begin{center}
\includegraphics[width=0.47\textwidth]{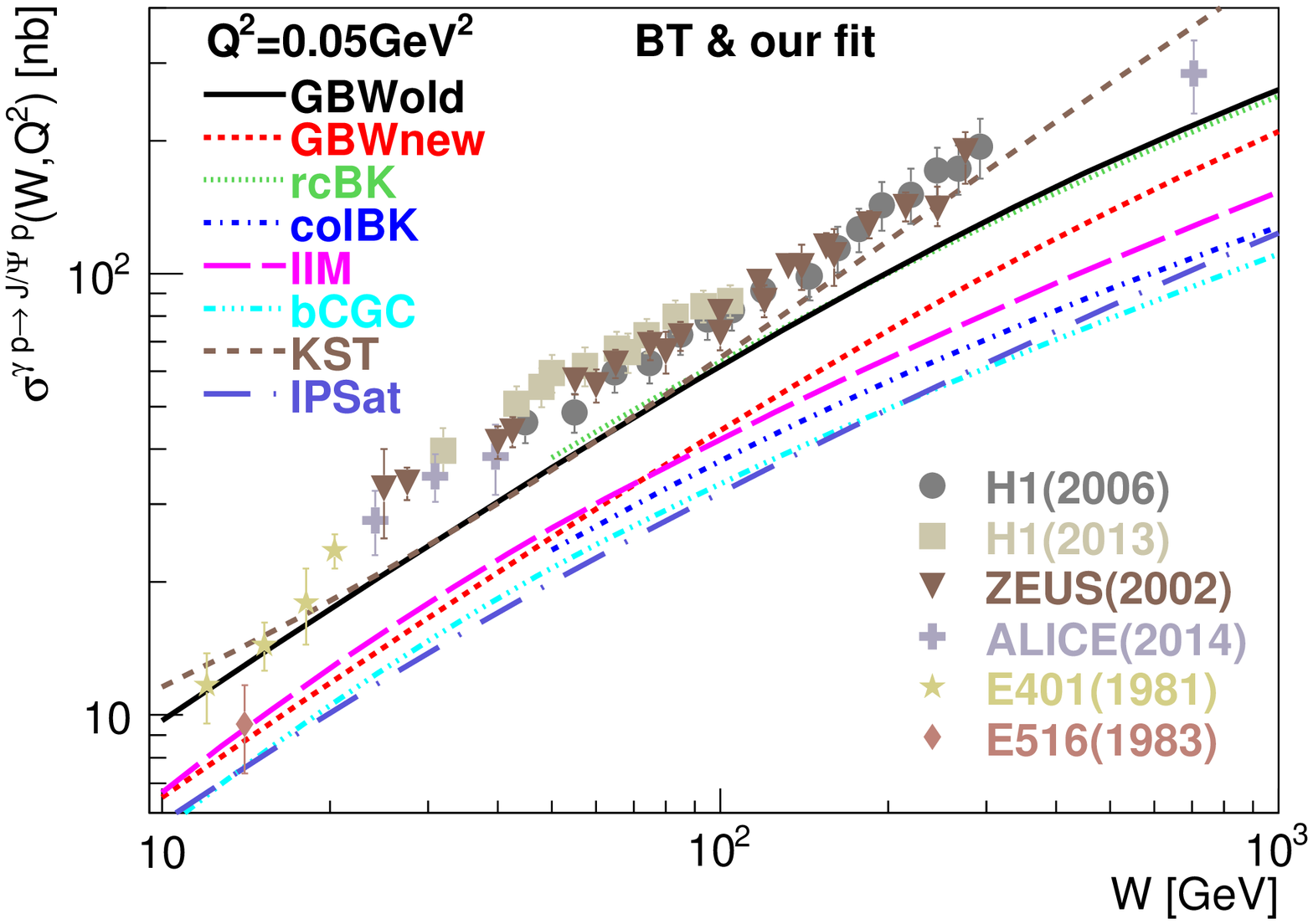}
\includegraphics[width=0.47\textwidth]{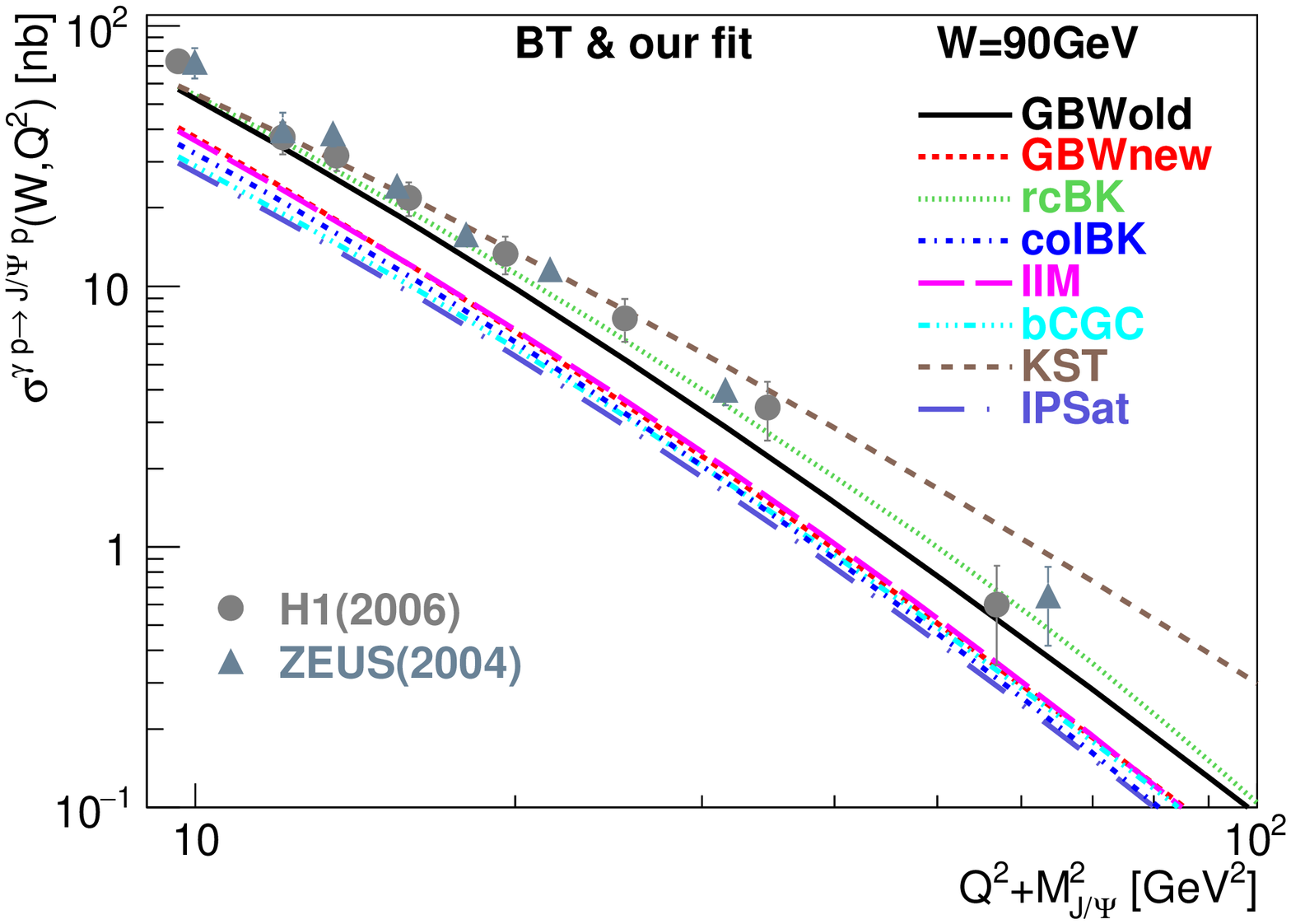}
  \caption{The same as Fig.~\ref{fig:psi1S-Bcomp} but for different phenomenological dipole cross sections described in Sect.~\ref{Sec:dipole-CS}.
}
%%%%%%%%%%%%%%%%%%%%%%%%%%%%%
  \label{fig:psi1S-sigcomp-1}
%%%%%%%%%%%%%%%%%%%%%%%%%%%%%
 \end{center}
 \end{figure}
%==================================== 
%
%

%
%               Fig.15
%====================================
\begin{figure}[!htbp]
\begin{center}
\includegraphics[width=0.47\textwidth]{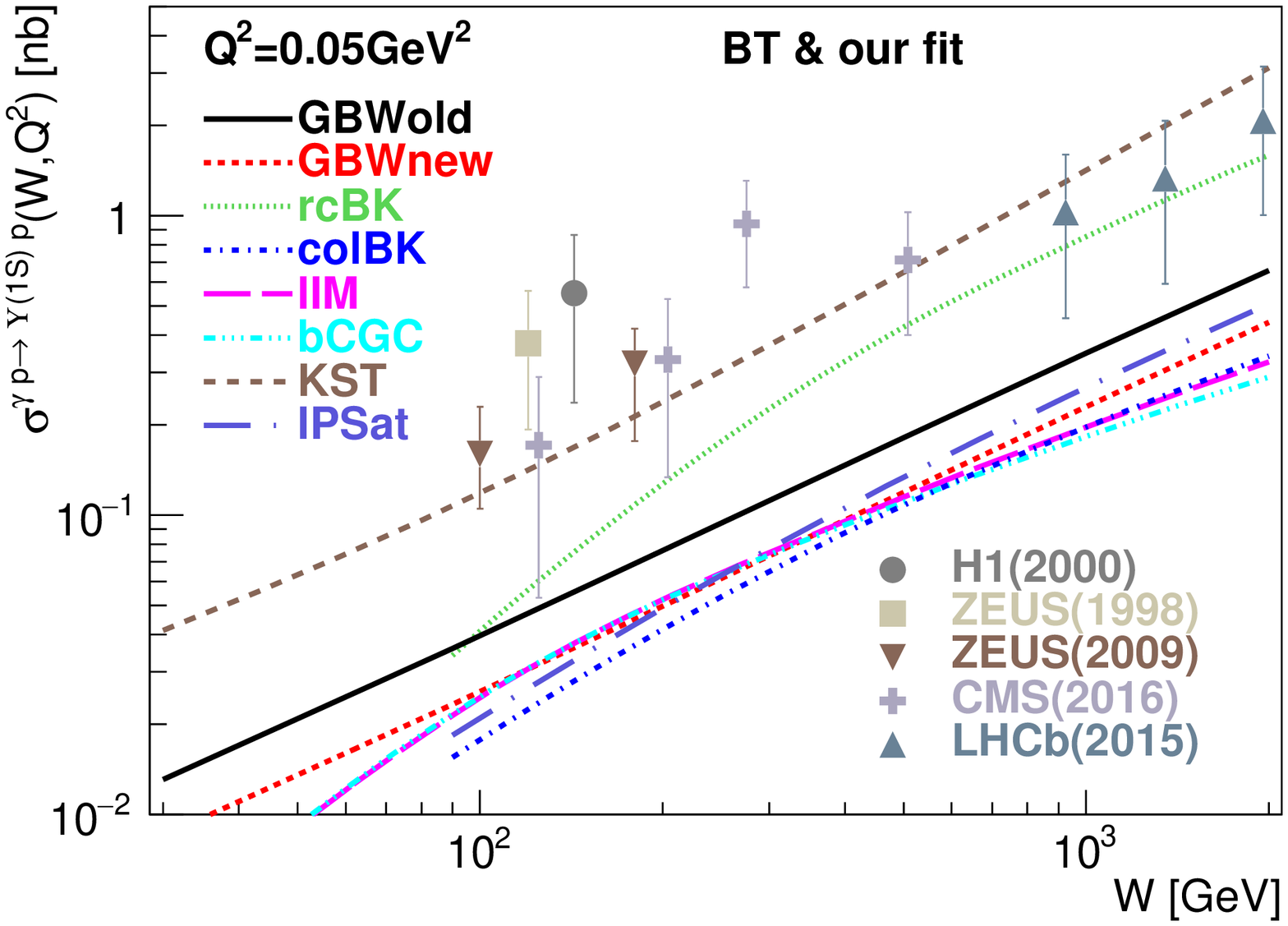}
\includegraphics[width=0.47\textwidth]{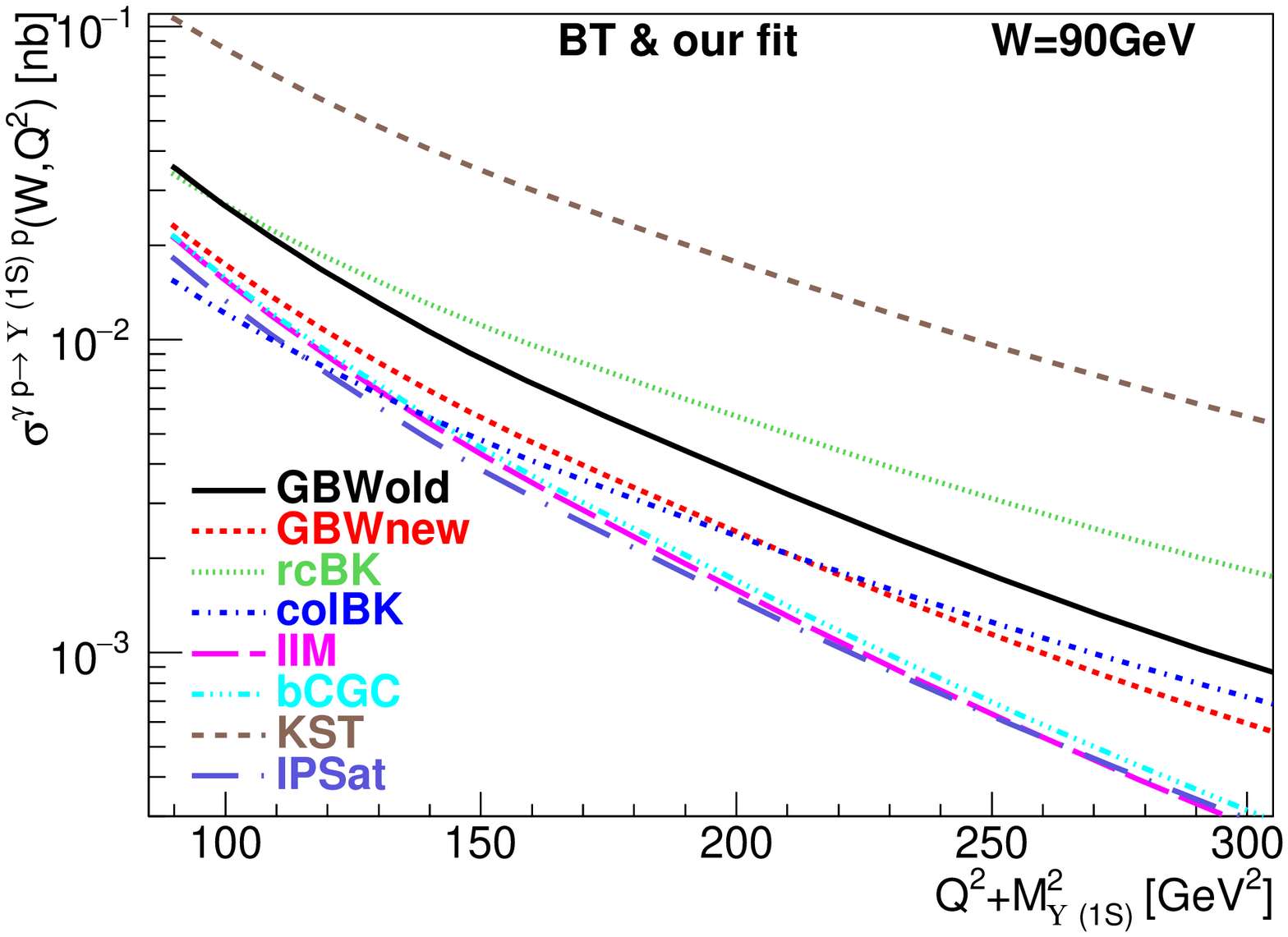}
  \caption{The same as Fig.~\ref{fig:upsi1S-Bcomp} but for different phenomenological dipole cross sections described in Sect.~\ref{Sec:dipole-CS}. }
%%%%%%%%%%%%%%%%%%%%%%%%%%%%%
  \label{fig:ups1S-sigcomp-1}
%%%%%%%%%%%%%%%%%%%%%%%%%%%%%
 \end{center}
 \end{figure}
%==================================== 
%
%

%
%               Fig.16
%====================================
\begin{figure}[!htbp]
\begin{center}
\includegraphics[width=0.95\textwidth]{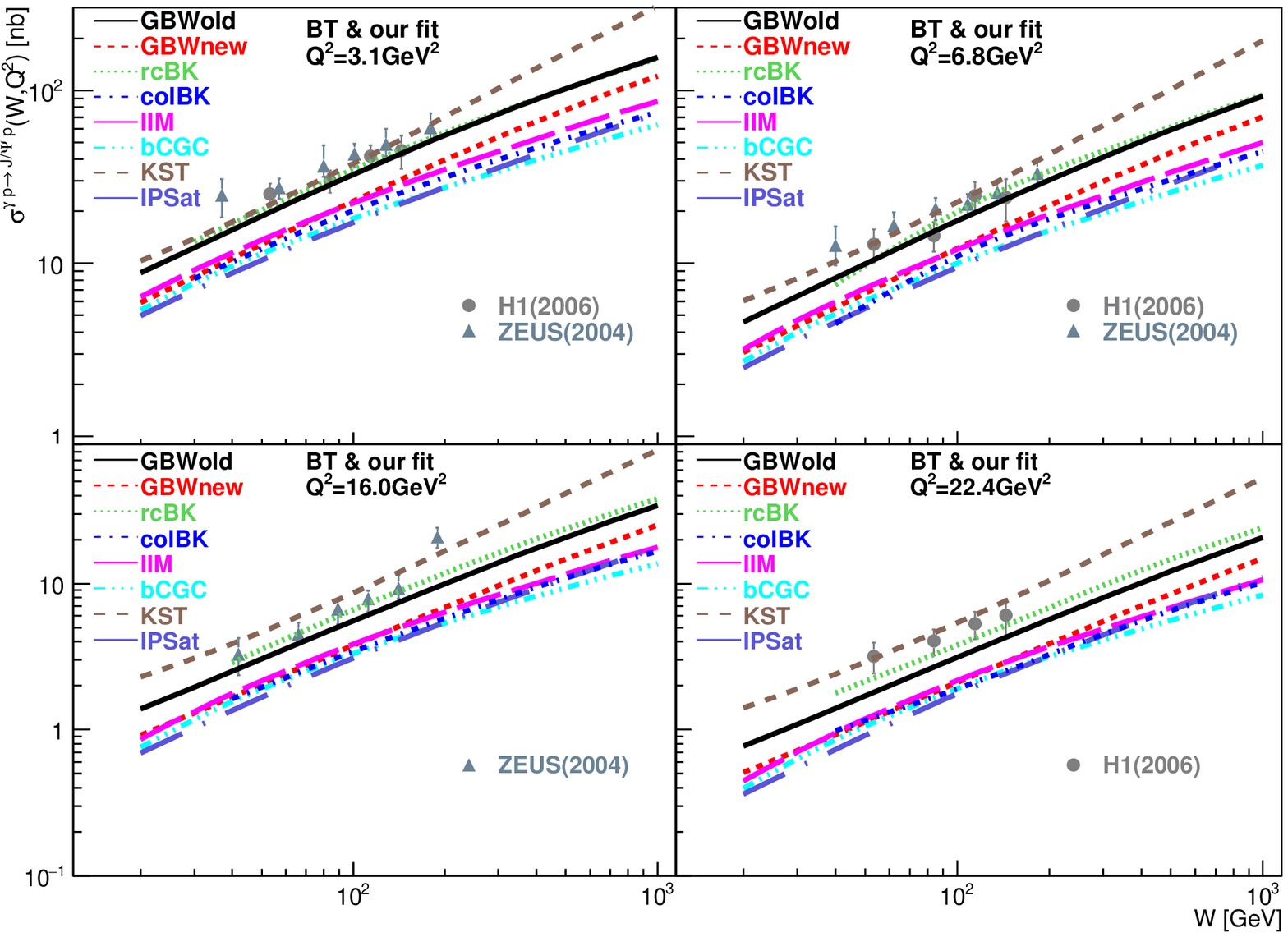}
  \caption{The exclusive $\Jpsi$ electroproduction cross section as a function of energy $W$ at several fixed values of $Q^2=3.1\,\GeV^2$ (top left panel), $Q^2=6.8\,\GeV^2$ (top right panel), $Q^2=16.0\,\GeV^2$ (bottom left panel), and $Q^2=22.4\,\GeV^2$ (bottom right panel). The model predictions, 
  including the Melosh spin rotation effects, were performed with the $\Jpsi$ wave function using the BT potential \cite{Buchmuller:1980su} and for different phenomenological dipole cross sections described in Sect.~\ref{Sec:dipole-CS}. The data are taken from H1 \cite{Aktas:2005xu} and ZEUS \cite{Chekanov:2004mw} Collaborations at HERA.}
%%%%%%%%%%%%%%%%%%%%%%%%%%%%%%
  \label{fig:psi1S-sigcomp-2}
%%%%%%%%%%%%%%%%%%%%%%%%%%%%%%
 \end{center}
 \end{figure}
%==================================== 
%
%

The calculations performed in the framework of color dipole approach are strongly correlated with the shape of the dipole cross section, $\sigma_{q\bar q}(r,x)$. In our predictions using the BT realistic potential for determination of the quarkonium wave functions, in Figs.~\ref{fig:psi1S-sigcomp-1}, \ref{fig:ups1S-sigcomp-1} and \ref{fig:psi1S-sigcomp-2} we test the eight main phenomenological parametrizations for $\sigma_{q\bar q}(r,x)$ found in the literature and discussed in Sect.~\ref{Sec:dipole-CS}. Here, the main observations are the following:

(i)
In the case of $1S$ charmonium photoproduction, the KST and GBWold dipole models give almost the same cross section at c.m. energies $W\lesssim 200\,\GeV$ describing the available data reasonably well. The other phenomenological parametrizations denoted as GBWnew, rcBK, coIBK, IM, bCGC and IPSat strongly underestimate the data by a factor of $2\div 3$ (see the left panel of Fig.~\ref{fig:psi1S-sigcomp-1}.

(ii)
In the electroproduction of $1S$ charmonia, the KST and GBWold dipole cross sections lead to the cross sections that differ from each other by a factor of $2\div 3$ at high $Q^2$ (see the right panel of Fig.~\ref{fig:psi1S-sigcomp-1}). Here, the KST parametrization provides the best description of $Q^2$-dependent data. Other six parametrizations used in our study grossly underestimate the data within the whole considered $Q^2$ interval.

(iii)
A similar conclusion as above can be made also from Fig.~\ref{fig:psi1S-sigcomp-2} where we studied the energy dependence of $1S$ charmonium electroproduction cross section at different fixed values of $Q^2$.

(iv)
In analogy to electroproduction of $1S$ charmonia, the model calculations using the KST phenomenological parametrization for the dipole cross section provide the best description of the available data on photoproduction of $1S$ bottomonia as shown in Fig.~\ref{fig:ups1S-sigcomp-1}. Except for the rcBK parametrization at large $W$, all other parametrizations lead to a significant underestimation of these data in the whole range of $W$. 

(v)
A huge variance of the model predictions for the photoproduction ($Q^2\to 0$) cross sections of $\Y(1S)$ using various parametrizations for the dipole cross section remains also in the case of electroproduction results shown in the right panel of Fig.~\ref{fig:ups1S-sigcomp-1}. Here, the spread between the results rises with $Q^2$ as a direct consequence of the growing differences between the dipole parametrizations at small transverse separations $r\lesssim 0.1\,\fm$. The latter is demonstrated by bottom panels of Fig.~\ref{fig:sigqq}.

%
%
%====================================
\subsection{Sensitivity of model predictions to quark mass}
\label{Sec:psi1S-mass}
%====================================
%
%

The quark mass has a strong impact on magnitudes of the model predictions as was presented and discussed earlier in Sect.~\ref{Sec:psi1S-pot}. Different realistic potentials (see Appendix~\ref{App:potentials}) used in our analysis of the quarkonium wave functions contain distinct values of quark masses ranging within the interval $m_c\in (1.3 - 1.84)\,\GeV$, for the charm quark, and $m_b\in (4.2 - 5.17)\,\GeV$, for the bottom quark. Here we test, taking the BT potential as a reference point with $m_c=1.48\,\GeV$ and $m_b=4.87\,\GeV$, how much our model predictions are modified by changing the quark mass from the minimal to maximal values corresponding to these intervals.
%
%               Fig.17
%===================================
\begin{figure}[!htbp]
\begin{center}
\includegraphics[width=0.47\textwidth]{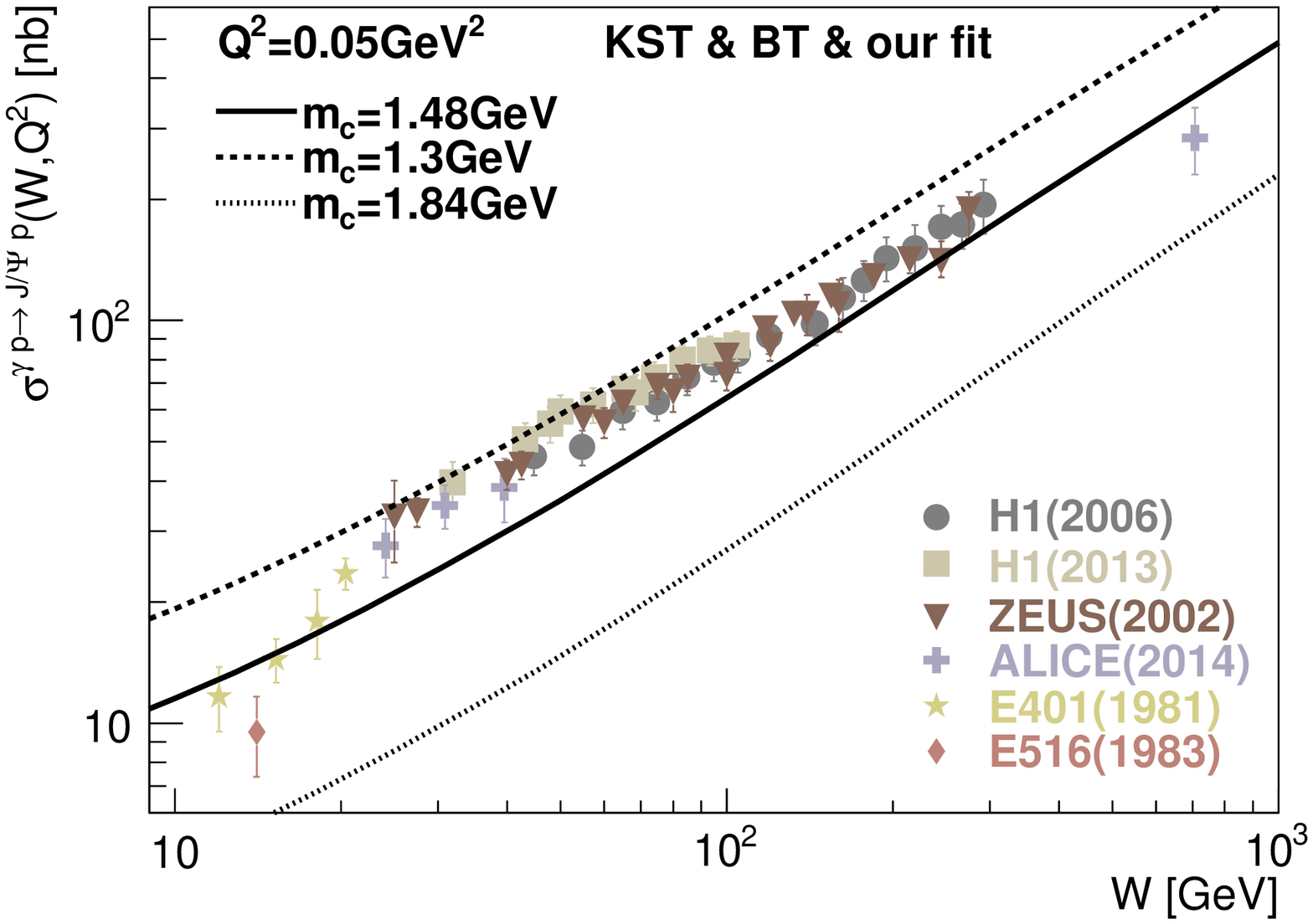}
\includegraphics[width=0.47\textwidth]{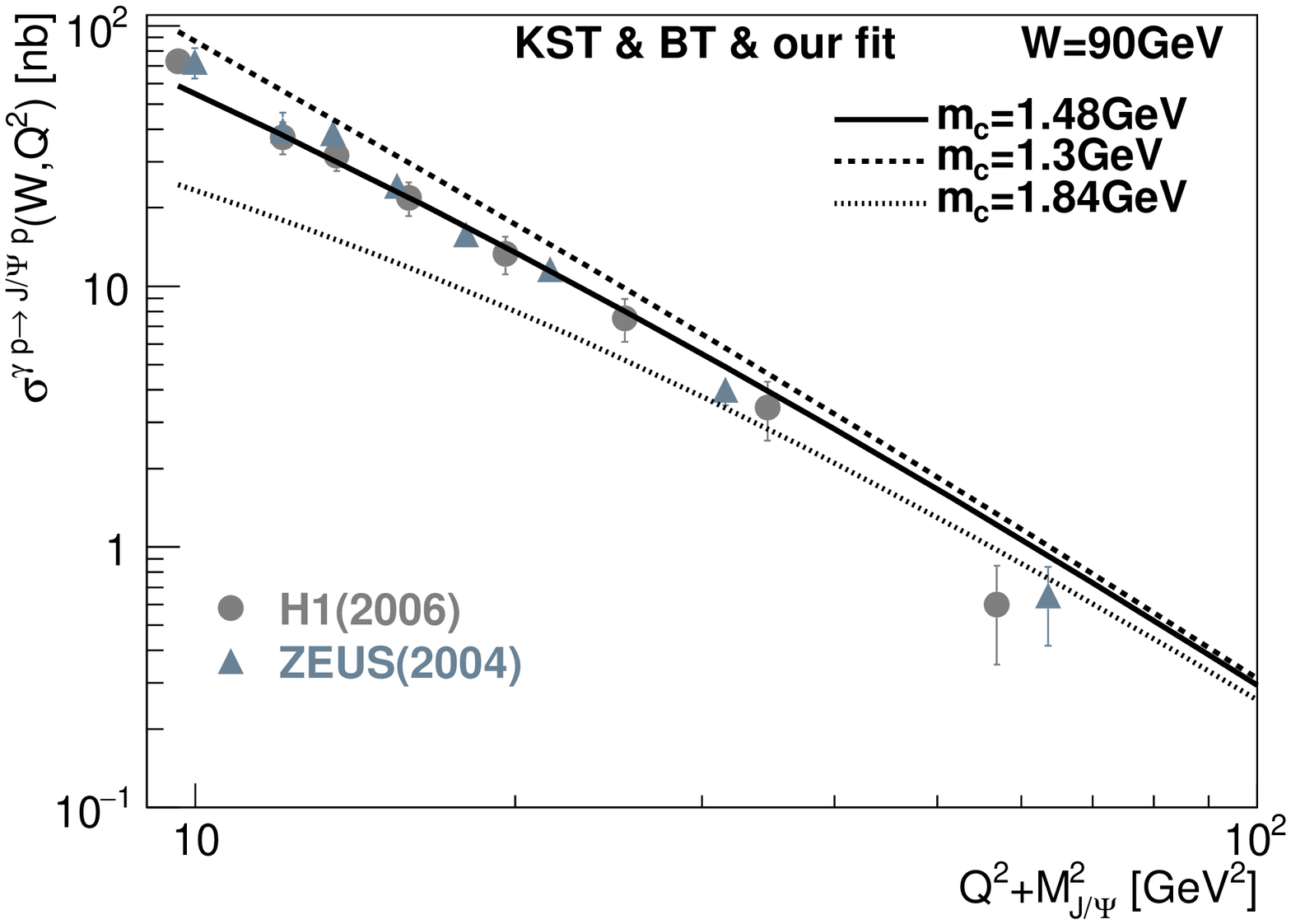}
  \caption{The same as Fig.~\ref{fig:psi1S-Bcomp} but for the test of sensitivity of the model predictions to the typical charm quark mass $m_c$ variations.}
%%%%%%%%%%%%%%%%%%%%%%%
  \label{fig:psi1S-mq}
%%%%%%%%%%%%%%%%%%%%%%%
 \end{center}
 \end{figure}
%=================================== 
%
%

Fig.~\ref{fig:psi1S-mq} clearly demonstrates the sensitivity of model results, taking the realistic BT potential and KST parametrization of the dipole cross section, to different quark mass values. Whereas our calculations, using the BT potential with $m_c=1.48\,\GeV$, lead to a reasonable description of the data, a modification of the charm quark mass to the lower ($m_c=1.3\,\GeV$) and higher ($m_c=1.84\,\GeV$) value causes a gross overestimation and underestimation of these data, respectively. Such a strong sensitivity to the value of the charm quark mass comes from the photon wave function, Eq.~(\ref{gamma-wf}), which contains $m_c$ in the argument of the Bessel function $K_0$. 

The sensitivity of model predictions to quark mass values gradually decreases with $Q^2$ since, in comparison to photoproduction limit $Q^2\to 0$, the quark mass scale plays a weaker role and can be neglected at large $Q^2\gg m_c^2$. Then, the model calculations naturally give very similar values for the $1S$ charmonium electroproduction cross section as is demonstrated in the right panel of Fig.~\ref{fig:psi1S-mq}. 
%
%             Fig.18
%=====================================
\begin{figure}[!htbp]
\begin{center}
\includegraphics[width=0.47\textwidth]{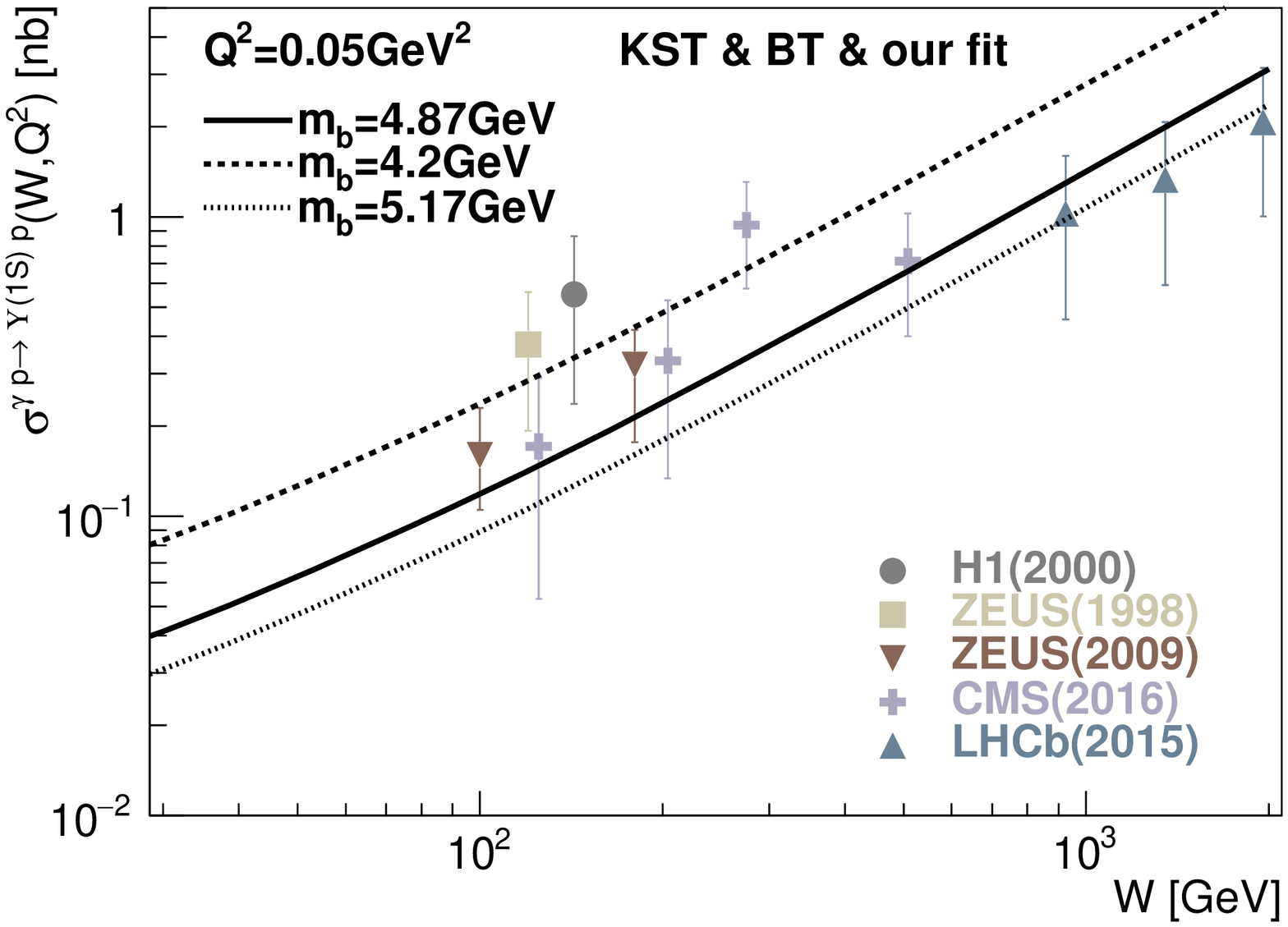}
\includegraphics[width=0.47\textwidth]{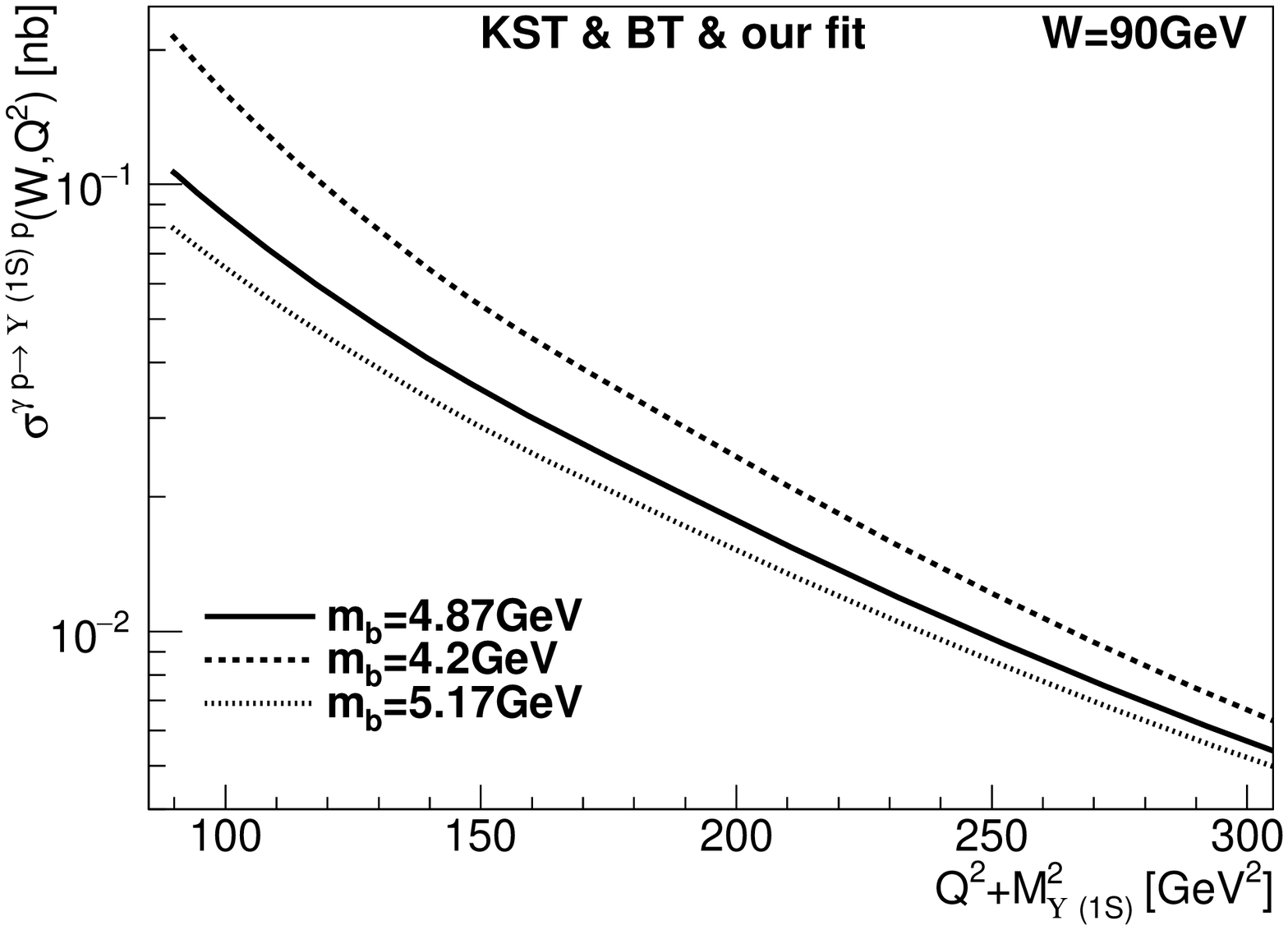}
\caption{The same as the left panel of Fig.~\ref{fig:upsi1S-Bcomp} but for the test of sensitivity of the model predictions to the bottom quark mass $m_b$.}
%%%%%%%%%%%%%%%%%%%%%%
  \label{fig:ups1S-mq}
%%%%%%%%%%%%%%%%%%%%%%
 \end{center}
 \end{figure}
%===================================== 
%
%

A variation of the model predictions with quark mass is presented in Fig.~\ref{fig:ups1S-mq} for the case of photo- and electroproduction of $\Y(1S)$. In comparison to charmonium production, here the sensitivity of the cross section to bottom quark mass variations is weaker due to a smaller relative change in $m_b$ and also gradually decreases with $Q^2$ at large $Q^2\gg m_b^2$ as expected.   

%
%
%====================================
\subsection{Spin rotation effects in electroproduction of $1S$ quarkonia}
\label{Sec:spin}
%====================================
%
%

%
%              Fig.19
%====================================
\begin{figure}[!htbp]
\begin{center}
\includegraphics[width=0.47\textwidth]{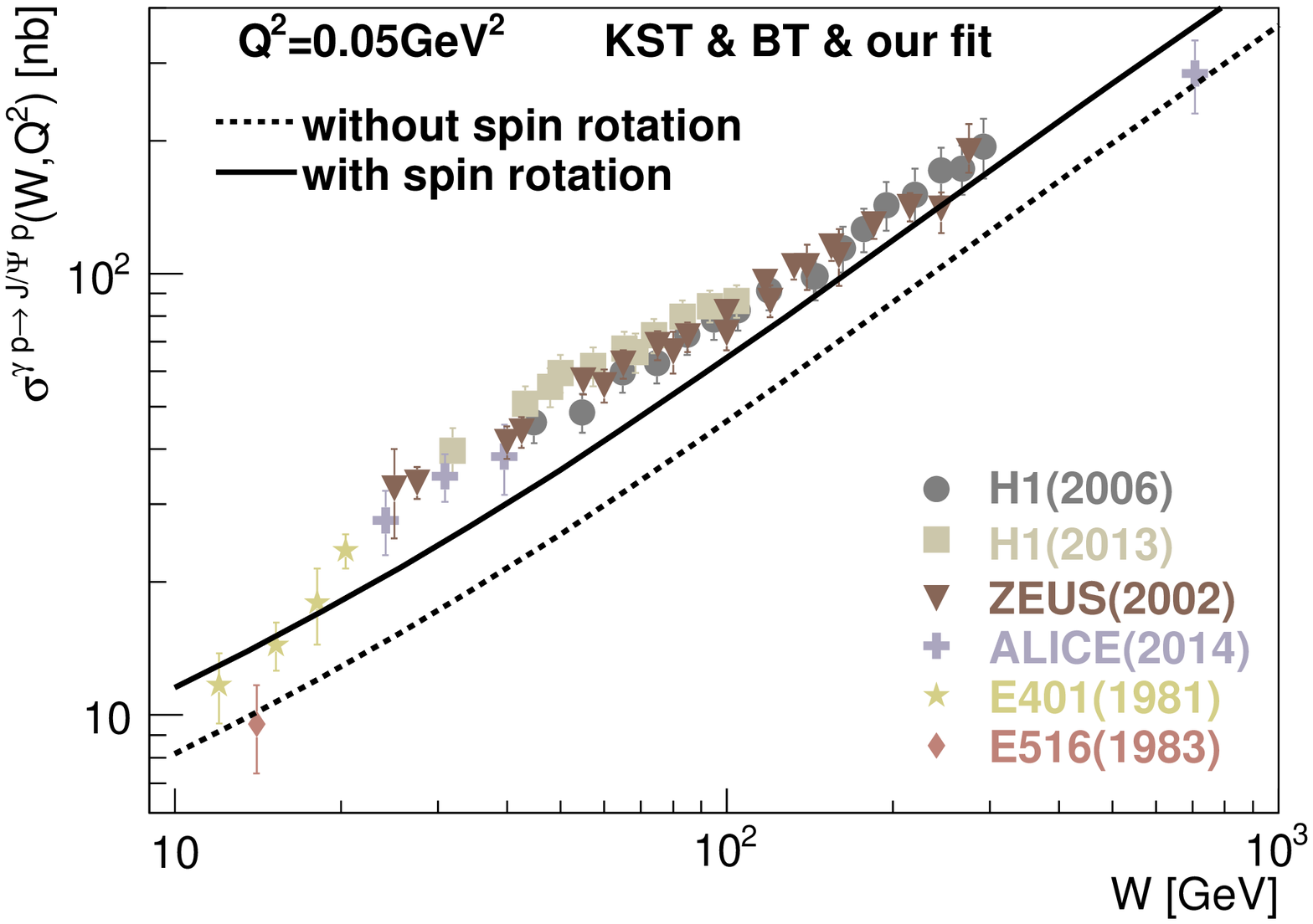}
\includegraphics[width=0.47\textwidth]{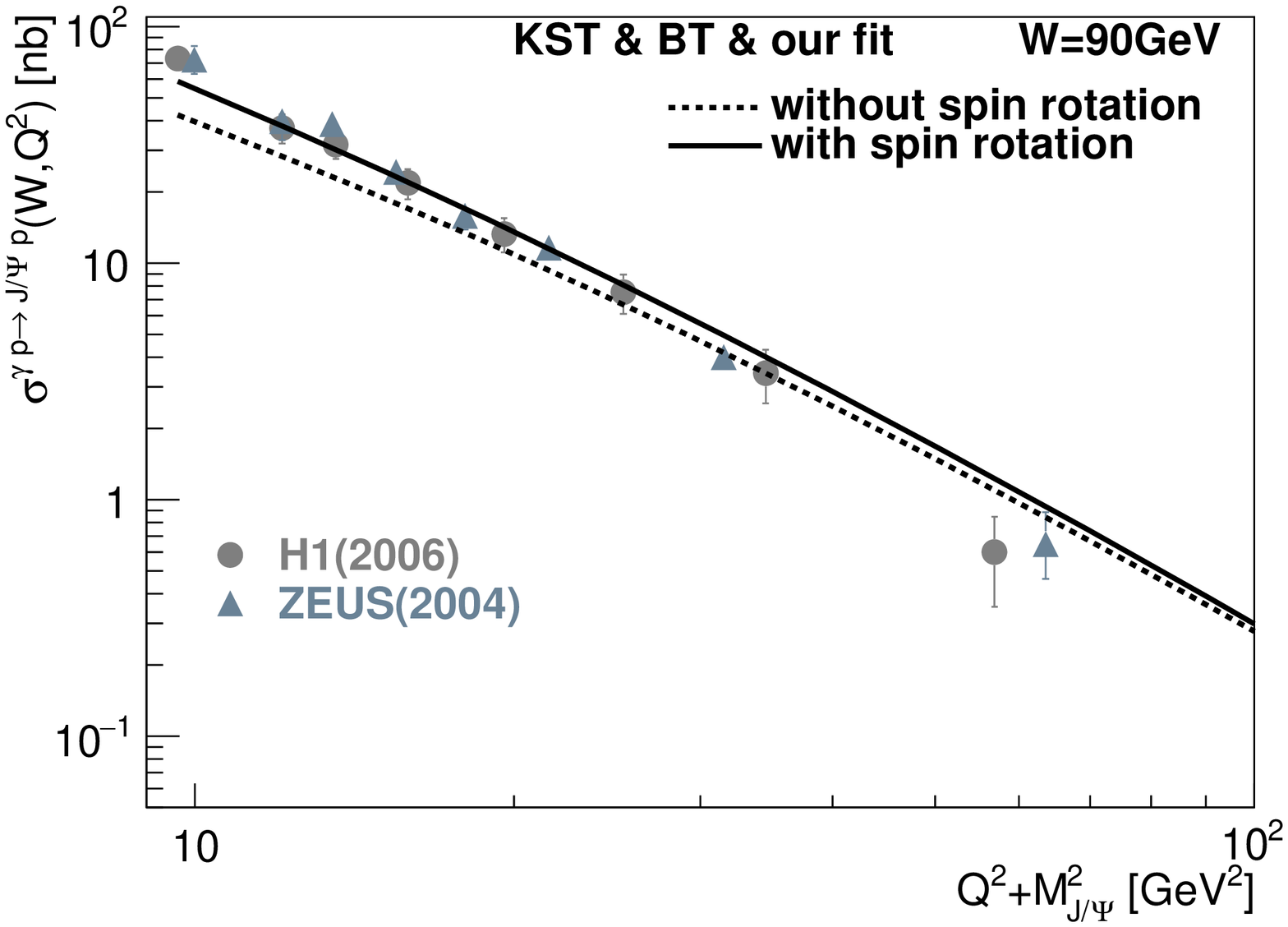}
  \caption{The same as Fig.~\ref{fig:psi1S-Bcomp} but showing the effect of the Melosh spin rotation in the exclusive $\Jpsi$ electroproduction cross section shown as a function of c.m. energy $W$ (left panel) and the scaling variable $Q^2+M_{\Jpsi}^2$ (right panel).}
%%%%%%%%%%%%%%%%%%%%%%%%%
  \label{fig:psi1S-spin}
%%%%%%%%%%%%%%%%%%%%%%%%%
 \end{center}
 \end{figure}
%====================================
%
%

%
%               Fig.20
%====================================
\begin{figure}[!htbp]
\begin{center}
\includegraphics[width=0.47\textwidth]{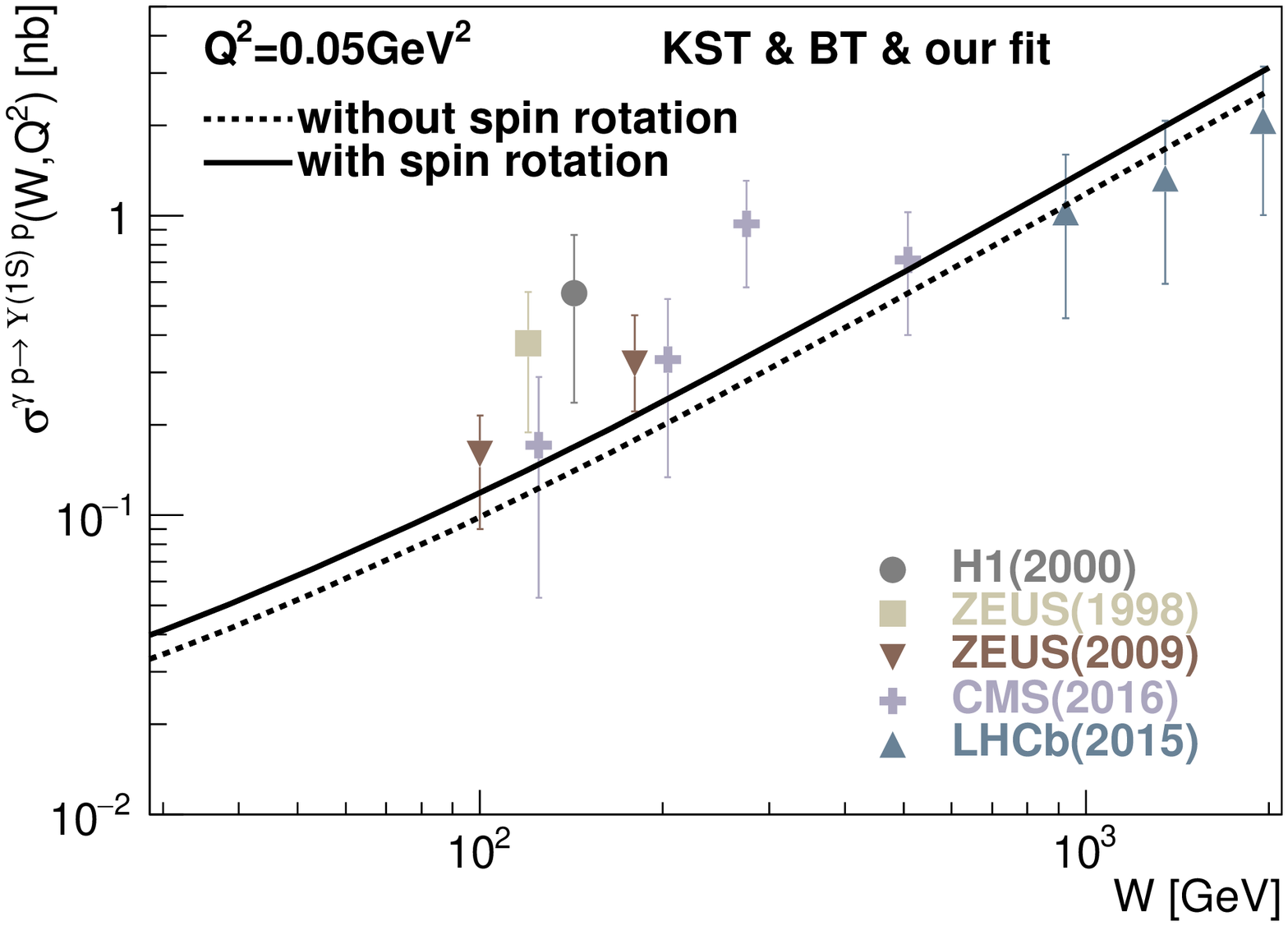}
\includegraphics[width=0.47\textwidth]{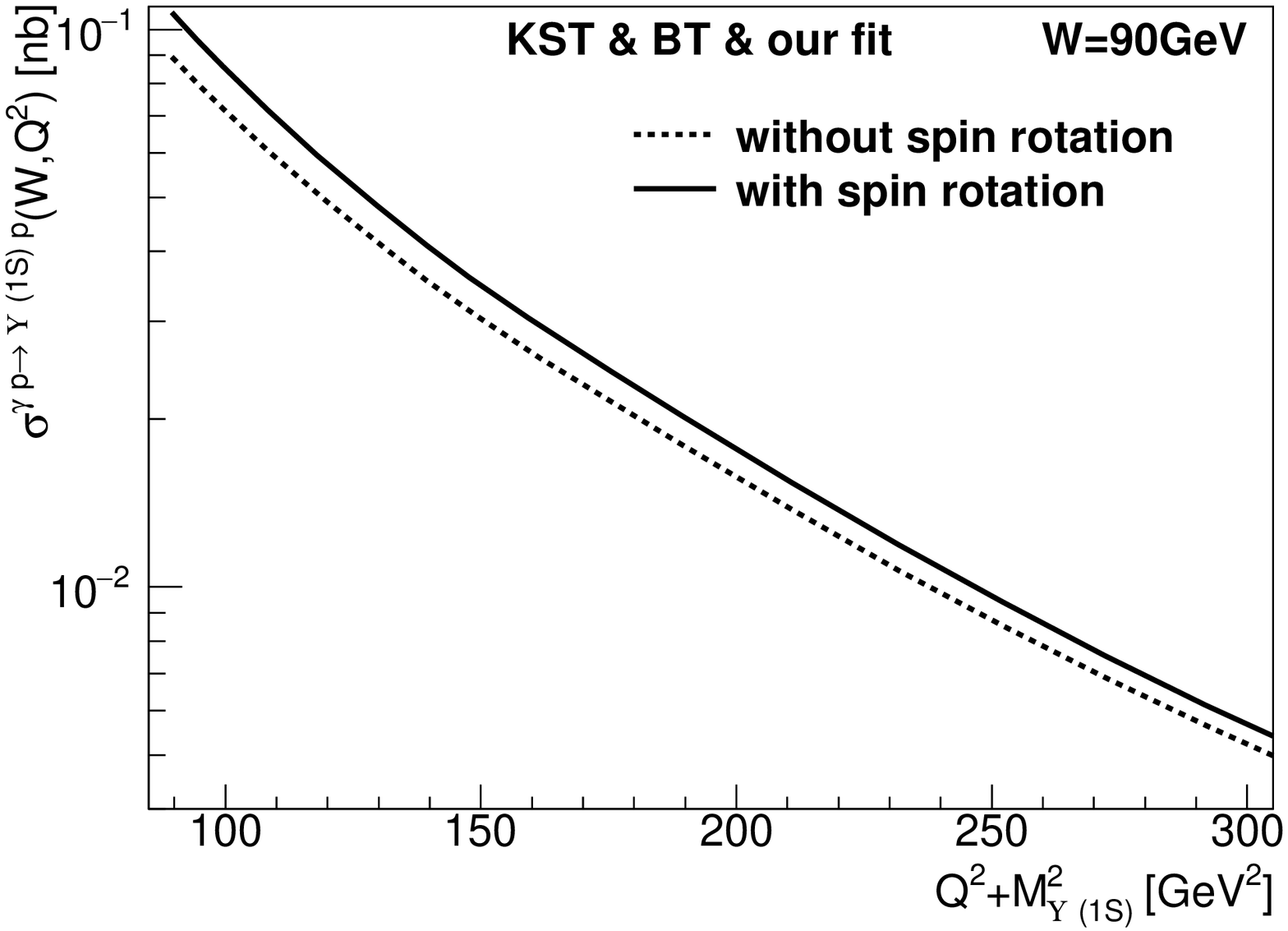}
\caption{The same as Figs.~\ref{fig:upsi1S-Bcomp} and \ref{fig:psi1S-spin} but for exclusive electroproduction of $\Y(1S)$ bottomonium.}
%%%%%%%%%%%%%%%%%%%%%%%%%
  \label{fig:ups1S-spin}
%%%%%%%%%%%%%%%%%%%%%%%%%
 \end{center}
 \end{figure}
%====================================
%
%

The effects of the Melosh spin rotation (see Eqs.~(\ref{AL}) and (\ref{AT}) in Appendix \ref{App:Amps}) in diffractive electroproduction of $S$-wave heavy quarkonia have been studied in detail in the framework of color dipole formalism in Ref.~\cite{jan-18}. For this reason, we present here only the main features of the spin rotation and demonstrate how much the spin effects can modify the corresponding photo- and electroproduction cross sections.

The onset of spin effects in photo- and electroproduction of $1S$ charmonium is presented in Fig.~\ref{fig:psi1S-spin}. It leads to an enhancement of the photoproduction cross section by $\approx 20\div 30\%$ leading a better agreement with the data (see the left panel of Fig.~\ref{fig:psi1S-spin}). This fact clearly supports an importance of the Melosh spin transformation, which is obviously neglected in many present studies of diffractive photo- and electroproduction of heavy quarkonia.

The right panel of Fig.~\ref{fig:psi1S-spin} demonstrates that the onset of spin effects gradually diminishes with the scaling variable $Q^2+M_{\Jpsi}^2$ and leads to a better description of the data, especially at small and medium $Q^2\lesssim 20\div 30\,\GeV^2$.

As was analyzed recently in Ref.~\cite{jan-18}, the universal properties in production of different vector mesons cause a similar onset of spin rotation effects in production of charmonia and bottomonia at the same fixed values of the scaling variable $Q^2+M_V^2$. For this reason, we predict a weak onset of these effects also in the photoproduction of $\Y(1S)$ state corresponding to electroproduction of $\Jpsi(1S)$ at $Q^2\sim M_{\Y}^2$ (compare the right panel of Fig.~\ref{fig:psi1S-spin} with the left panel of Fig.~\ref{fig:ups1S-spin}). The weak onset of the Melosh spin transformation in $\Y(1S)$ photoproduction decreases further with $Q^2$ as is demonstrated in the right panel of Fig.~\ref{fig:ups1S-spin}.

%
%
%====================================
\subsection{Theoretical uncertainties in predictions for the $\psip(2S)$-to-$\Jpsi(1S)$ 
and $\Yp(2S)$-to-$\Y(1S)$ ratios}
\label{Sec:2S-to-1S-ratio}
%====================================
%
%

%
%               Fig.21
%====================================
\begin{figure}[!htbp]
\begin{center}
\includegraphics[width=0.48\textwidth]{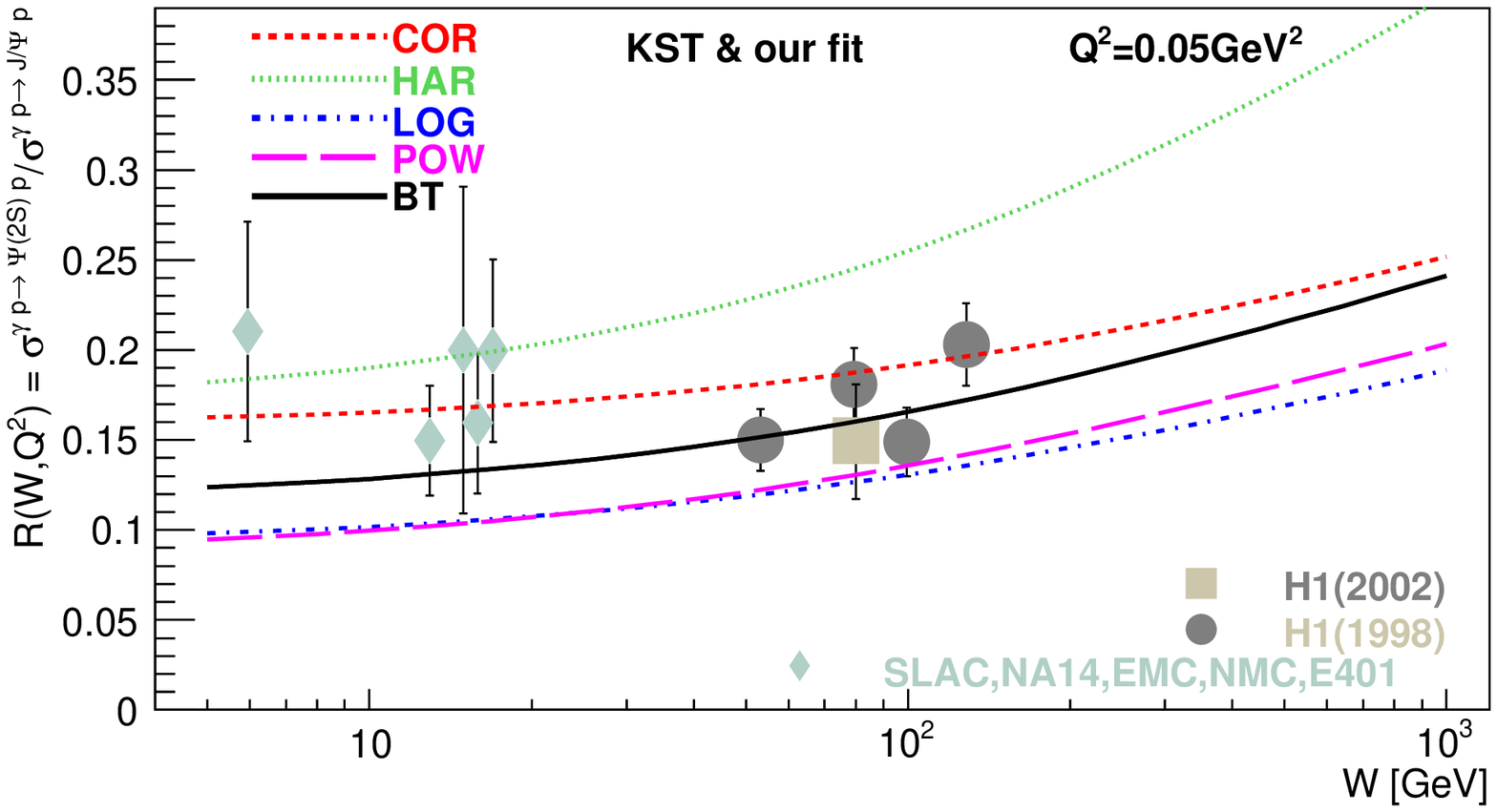}
\includegraphics[width=0.48\textwidth]{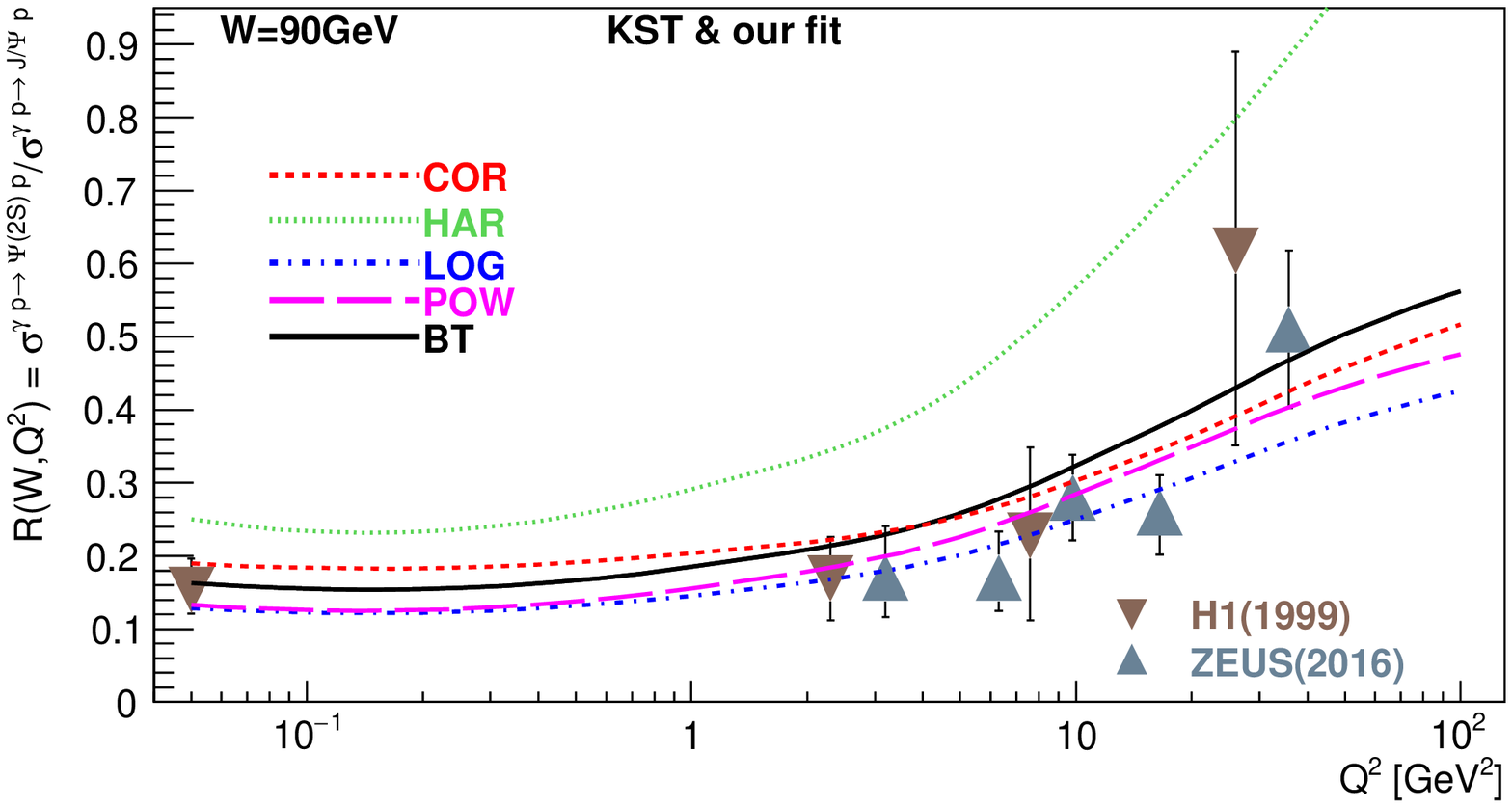}
\includegraphics[width=0.48\textwidth]{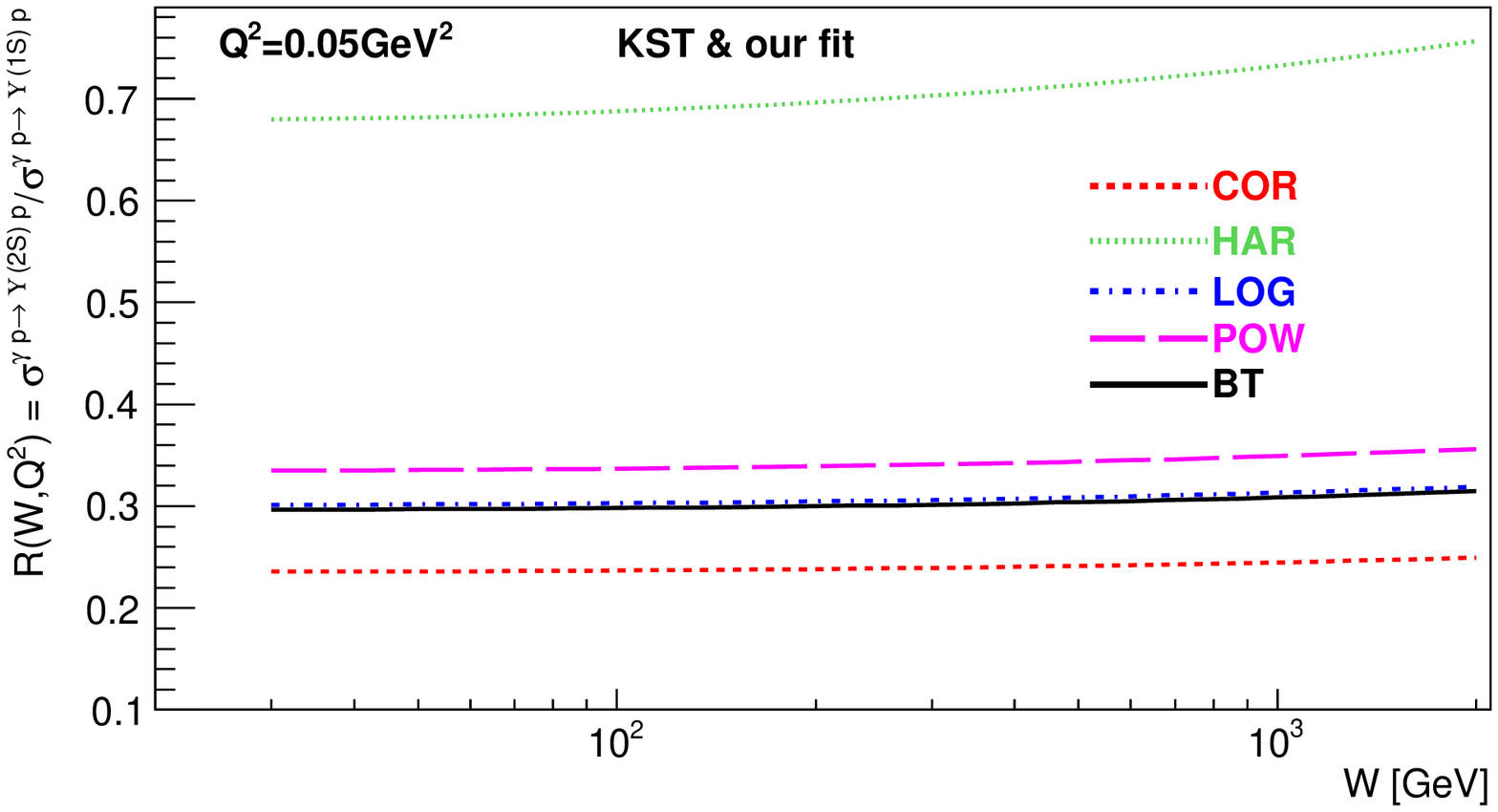}
\includegraphics[width=0.48\textwidth]{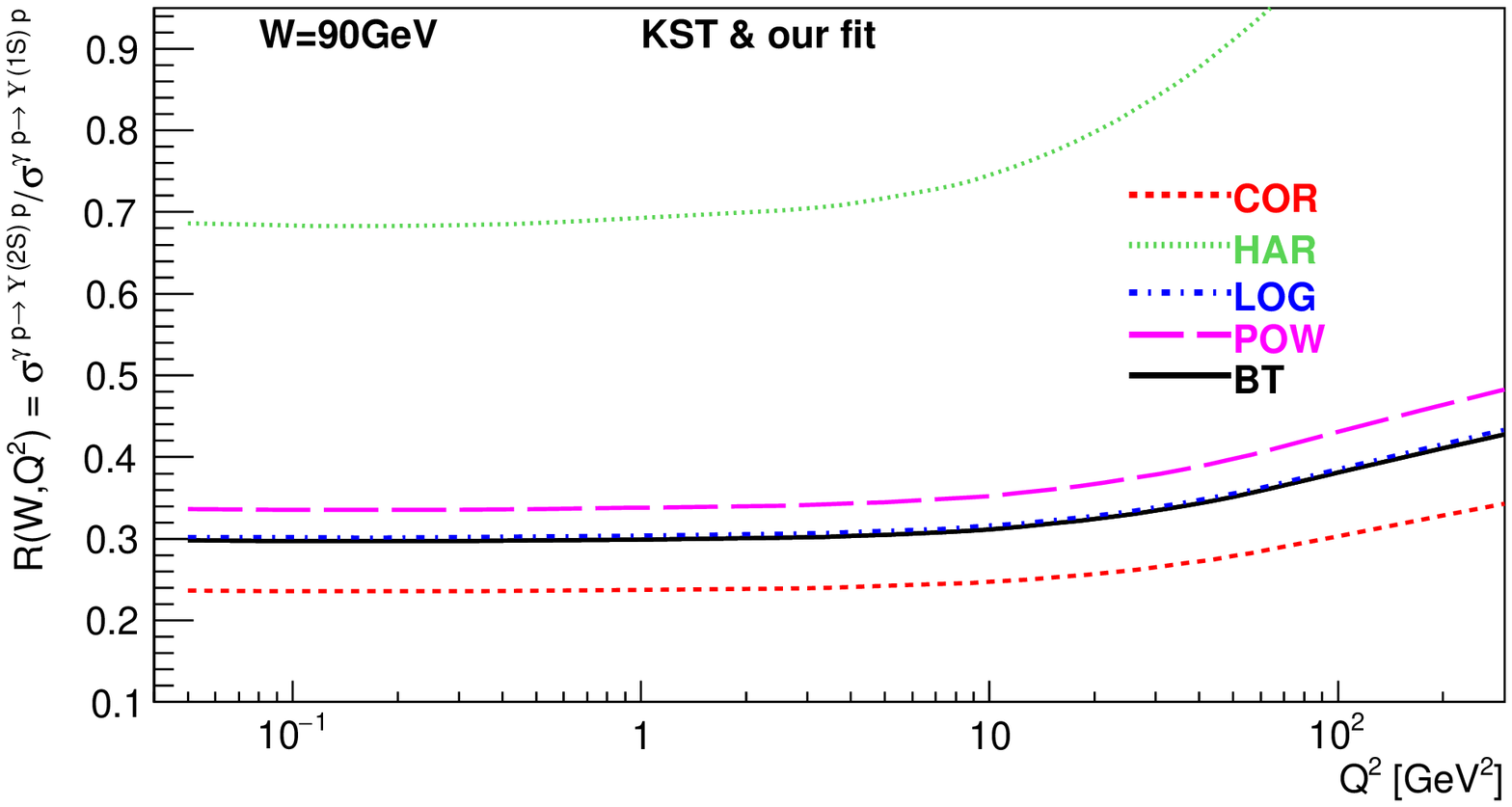}
  \caption{The color dipole model predictions for the $\psip(2S)$-to-$\Jpsi(1S)$ (top panels) and $\Yp(2S)$-to-$\Y(1S)$ (bottom panels) ratios of electroproduction cross sections as functions of c.m. energy $W$ at fixed $Q^2=0.05\,\GeV^2$ (left panels) and $Q^2$ at fixed $W=90\,\GeV$ (right panels) versus the data from H1 \cite{Adloff:1997yv,Adloff:2002re}, ZEUS \cite{Abramowicz:2016xls} and fixed target experiments \cite{Camerini:1975cy,Barate:1986fq,Binkley:1982yn,Aubert:1982tt,Amaudruz:1991sr}. The calculations were performed for the quarkonium wave functions generated by different realistic $c-\bar c$ and $b-\bar b$ interaction potentials as depicted in Appendix~\ref{App:potentials} and with the phenomenological KST dipole cross section \cite{Kopeliovich:1999am}. The results include the Melosh spin rotation.}
%%%%%%%%%%%%%%%%%%%%%%%%
  \label{fig:2S-1S-pot}
%%%%%%%%%%%%%%%%%%%%%%%%
 \end{center}
 \end{figure}
%============================= 
%
%

The theoretical uncertainties presented above in Sects.~\ref{Sec:psi1S-pot}, \ref{Sec:psi1S-dip}, \ref{Sec:psi1S-mass} and \ref{Sec:spin} can be tested by investigating also the ratios $R_{2S/1S}$ for charmonia $\psip(2S)$-to-$\Jpsi(1S)$ and bottomonia $\Yp(2S)$-to-$\Y(1S)$ photo- and electroproduction cross sections. Such a study enables us to minimize the uncertainties providing with more stable and accurate predictions, which can be verified by the future measurements.

%......................................................
\subsubsection{Dependence on $c-\bar c$ and $b-\bar b$ interaction potentials}
%......................................................

In Fig.~\ref{fig:2S-1S-pot}, using the KST phenomenological parametrization (see Sect.~\ref{Sec:dipole-CS}) for the dipole cross section, we test the sensitivity of model predictions for the $\psip(2S)$-to-$\Jpsi(1S)$ and $\Yp(2S)$-to-$\Y(1S)$ ratios with respect to the choice of interaction potentials which are employed in deriving the corresponding quarkonium LC wave functions. 

One can notice in top panels of Fig.~\ref{fig:2S-1S-pot} a good agreement of our calculations with the experimental data for all realistic potentials (COR, LOW, POW and BT), except for the HAR potential, which grossly overestimates the data at large $W$ and $Q^2$. It is caused by the lack of Coulomb-like behavior in the HAR potential, which amplifies the role of the node effect. This is based upon a stronger enhancement of the small-$r$ domain of the $\psip(2S)$ and $\Yp(2S)$ wave functions below the node position, therefore leading to a stronger reduction of the cancellation between low-$r$ and high-$r$ domains in the $2S$ production amplitude. Since the role of the Coulomb-like behavior increases in production of bottomonia, the bottom panels of Fig.~\ref{fig:2S-1S-pot} clearly demonstrate a huge difference in predictions between the HAR potential and all the other potentials. The latter generate only a small variance in the corresponding results for the $\Yp(2S)$-to-$\Y(1S)$ ratio.

%.......................................................................
\subsubsection{Dependence on the phenomenological dipole cross sections}
%........................................................................

In Sect.~\ref{Sec:psi1S-dip} we studied a correlation of the model predictions for photo- and electroproduction of $1S$ quarkonium with a shape of the color dipole cross section, $\sigma_{q\bar q}(r,x)$. We found a huge variance in the model predictions for the electroproduction cross section by using eight different popular parametrizations for $\sigma_{q\bar q}(r,x)$ discussed in Sect.~\ref{Sec:dipole-CS}. Here, we test how large is the theoretical uncertainty in the model predictions for the ratio $R_{2S/1S}(W,Q^2)$ caused by such a variety of different treatments of the target gluon density encoded in these parametrizations. 

\begin{figure}[!htbp]
\begin{center}
\includegraphics[width=0.48\textwidth]{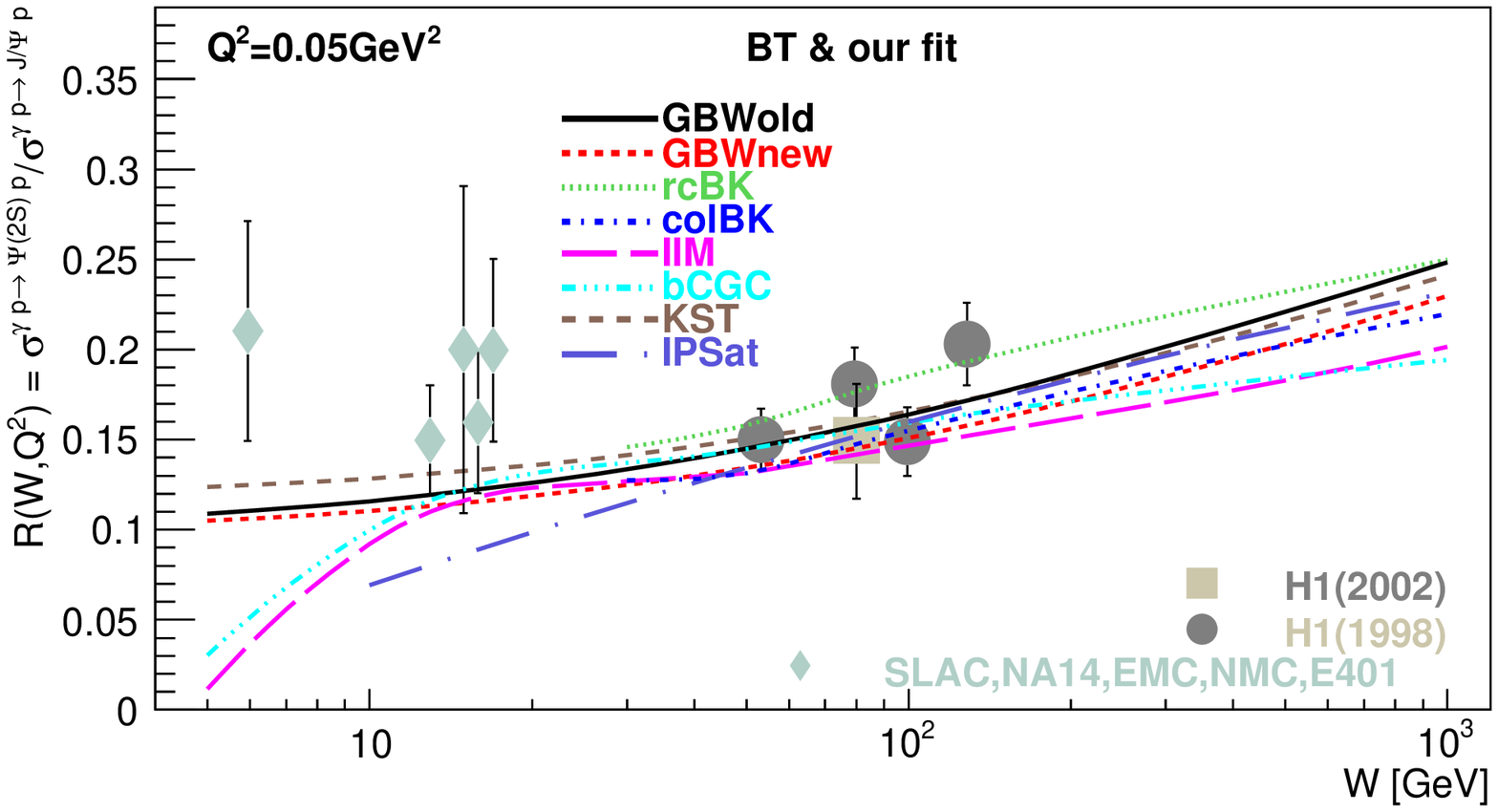}
\includegraphics[width=0.48\textwidth]{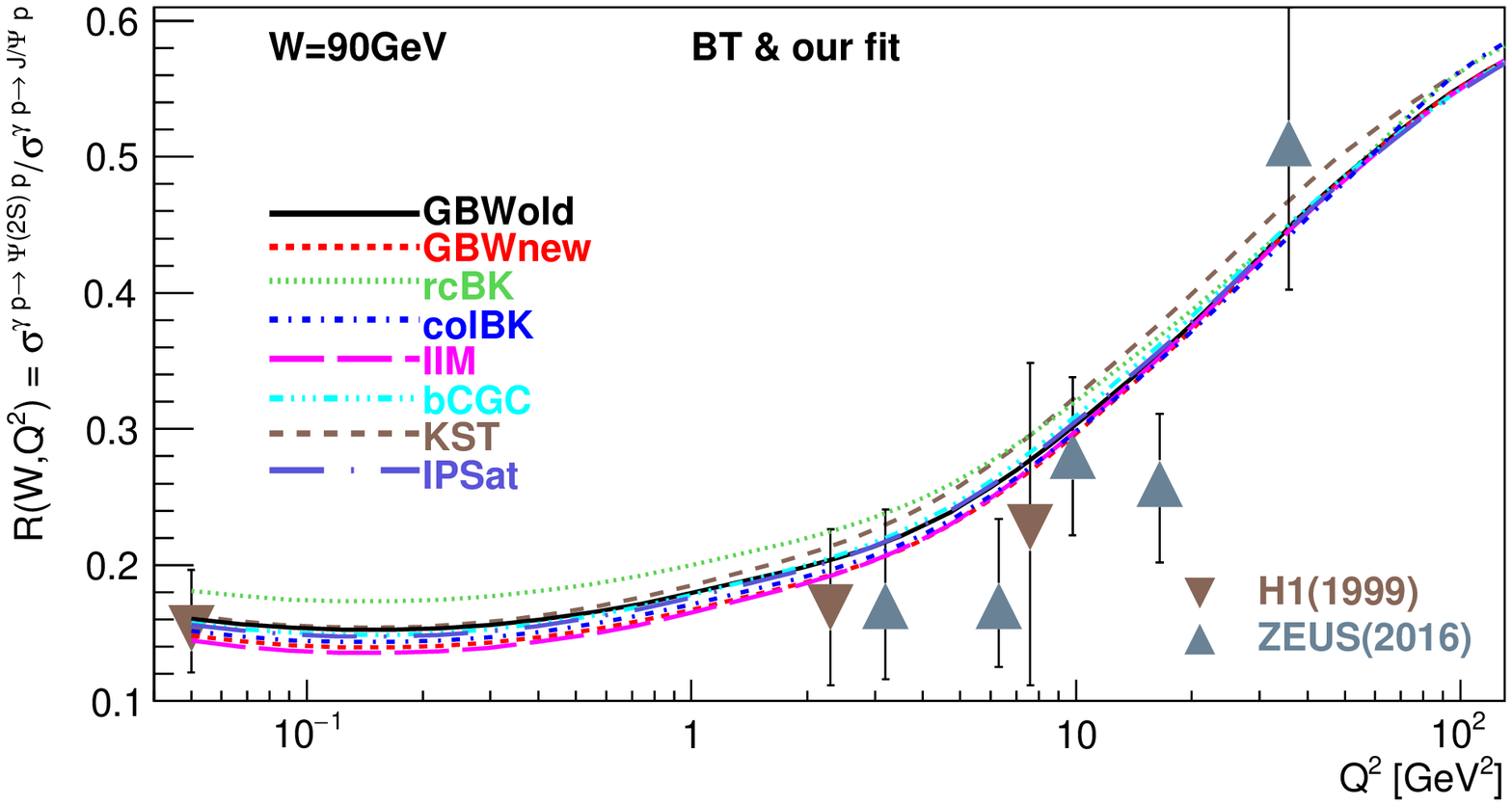}
\includegraphics[width=0.48\textwidth]{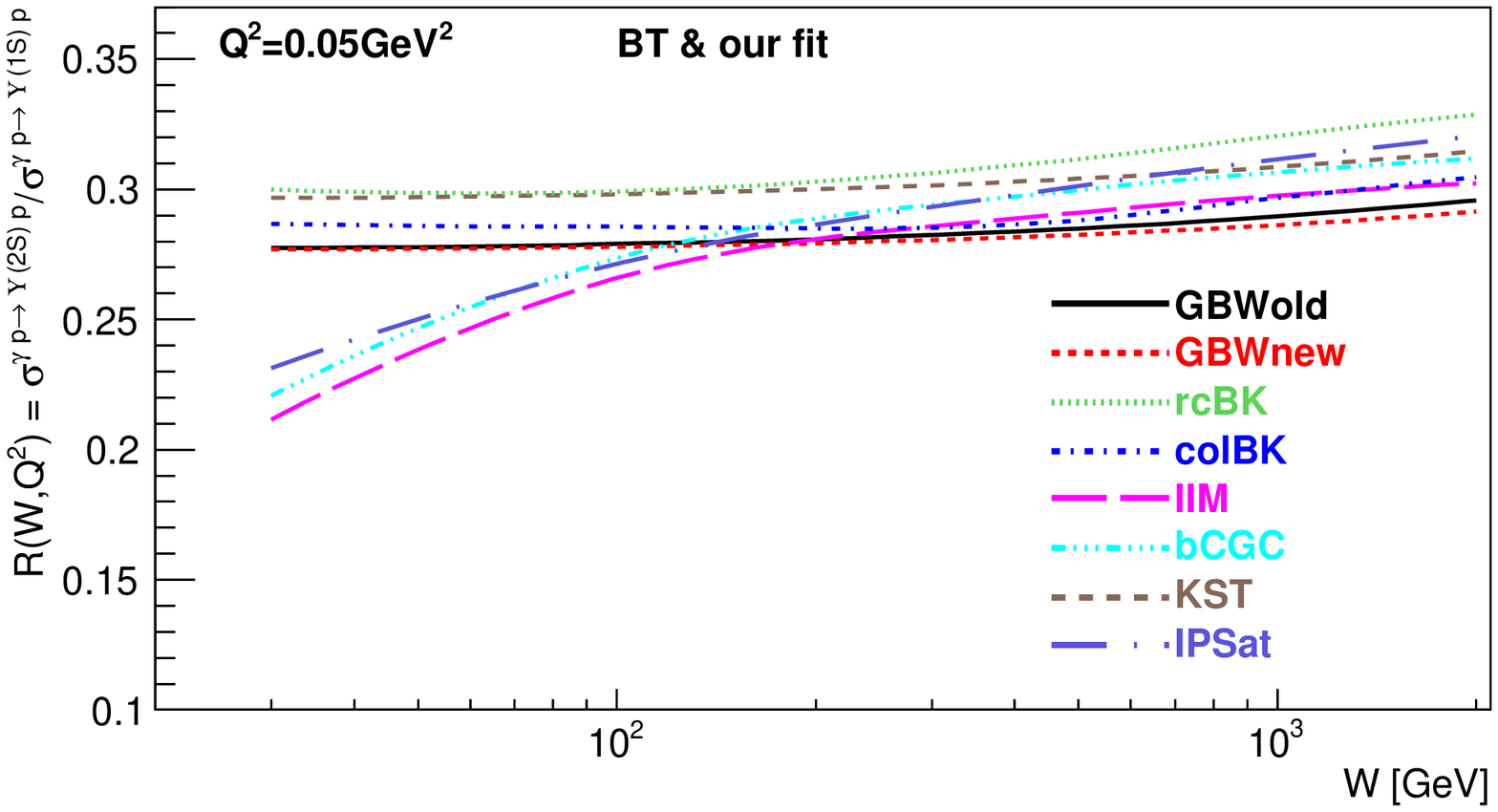}
\includegraphics[width=0.48\textwidth]{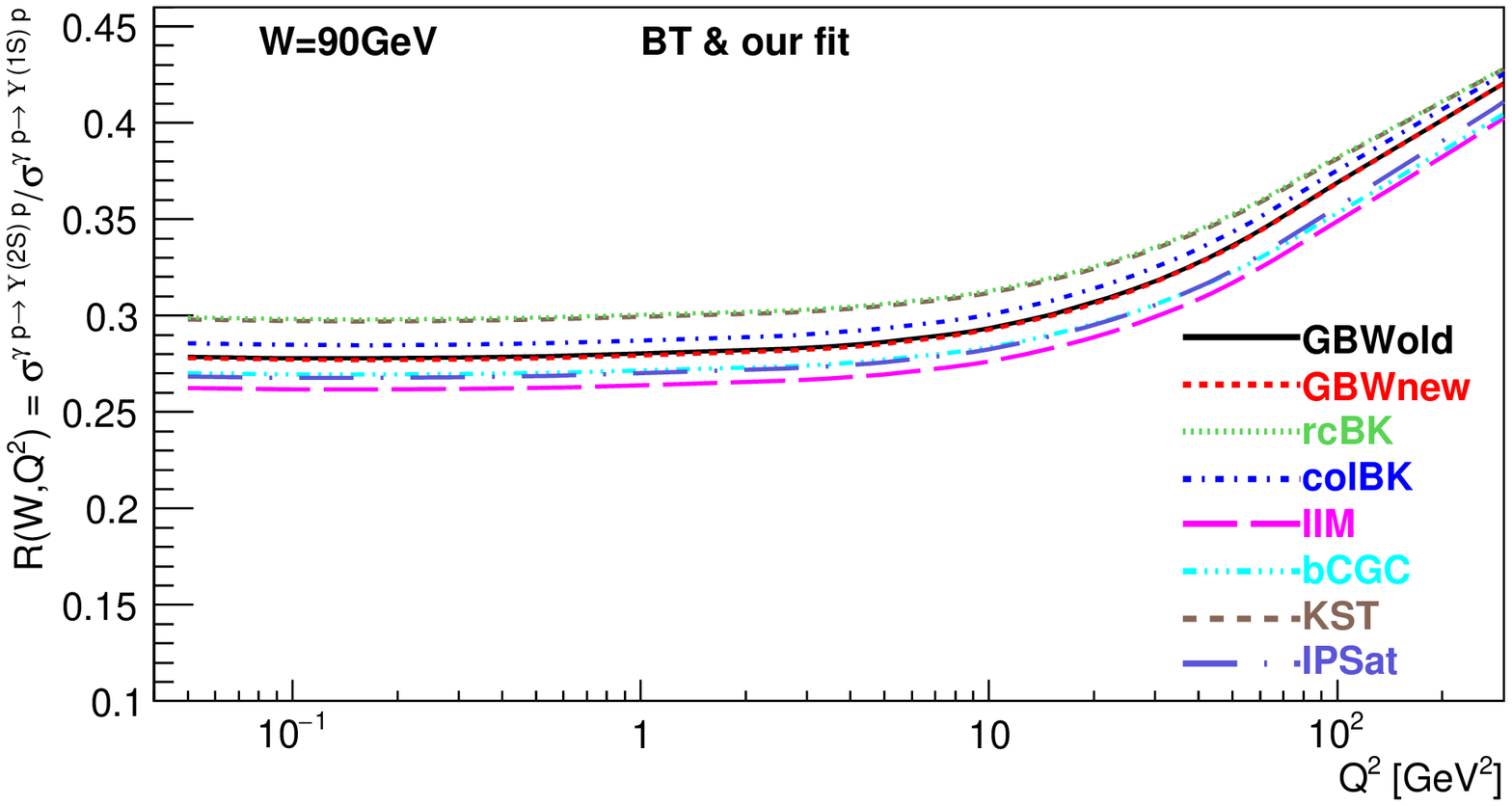}
  \caption{The same as Fig.~\ref{fig:2S-1S-pot} but for the quakonium wave functions generated by the realistic BT potential and for different dipole cross section parametrizations described in Sect.~\ref{Sec:dipole-CS}.}
%%%%%%%%%%%%%%%%%%%%%%%
  \label{fig:2S-1S-dip}
%%%%%%%%%%%%%%%%%%%%%%%
 \end{center}
 \end{figure}

The results of our calculations are depicted in Fig.~\ref{fig:2S-1S-dip}. One can see that, in comparison to the electroproduction cross section, the study of $R_{2S/1S}$ ratio (utilizing, for example, the realistic BT potential) allows to reduce substantially the uncertainty of our predictions stemming from different existing parametrizations for $\sigma_{q\bar q}(r,x)$ (compare Fig.~\ref{fig:2S-1S-dip} with the results of Sect.~\ref{Sec:psi1S-dip}). 

On the other hand, such a study makes it possible to analyze how the node effect manifests itself for different shapes of the color dipole cross section. The onset of the node effect is controlled by an increase of the ratio $R_{2S/1S}$ with energy $W$ and photon virtuality $Q^2$. The stronger is the cancellation in $2S$ production amplitude, the steeper is the rise of $R_{2S/1S}(W,Q^2)$ with a rate, which is slightly different for various dipole parametrizations. For production of $2S$ bottomonia the node effect is much weaker as one can see in the bottom panels of Fig.~\ref{fig:2S-1S-dip}.

Note, that the rise of such variations in the model predictions towards small energies can be influenced by a worse accuracy in dipole phenomenological parametrizations at the corresponding (large) values of Bjorken $x\gtrsim 0.1$. This is due to a natural limitation of the color dipole approach that is expected to fail at sufficiently large Bjorken $x$.
%
%                Fig.22
%==================--====================
%=======================================
%
%
%
%               Fig.23
%=======================================
\begin{figure}[!htbp]
\begin{center}
\includegraphics[width=0.48\textwidth]{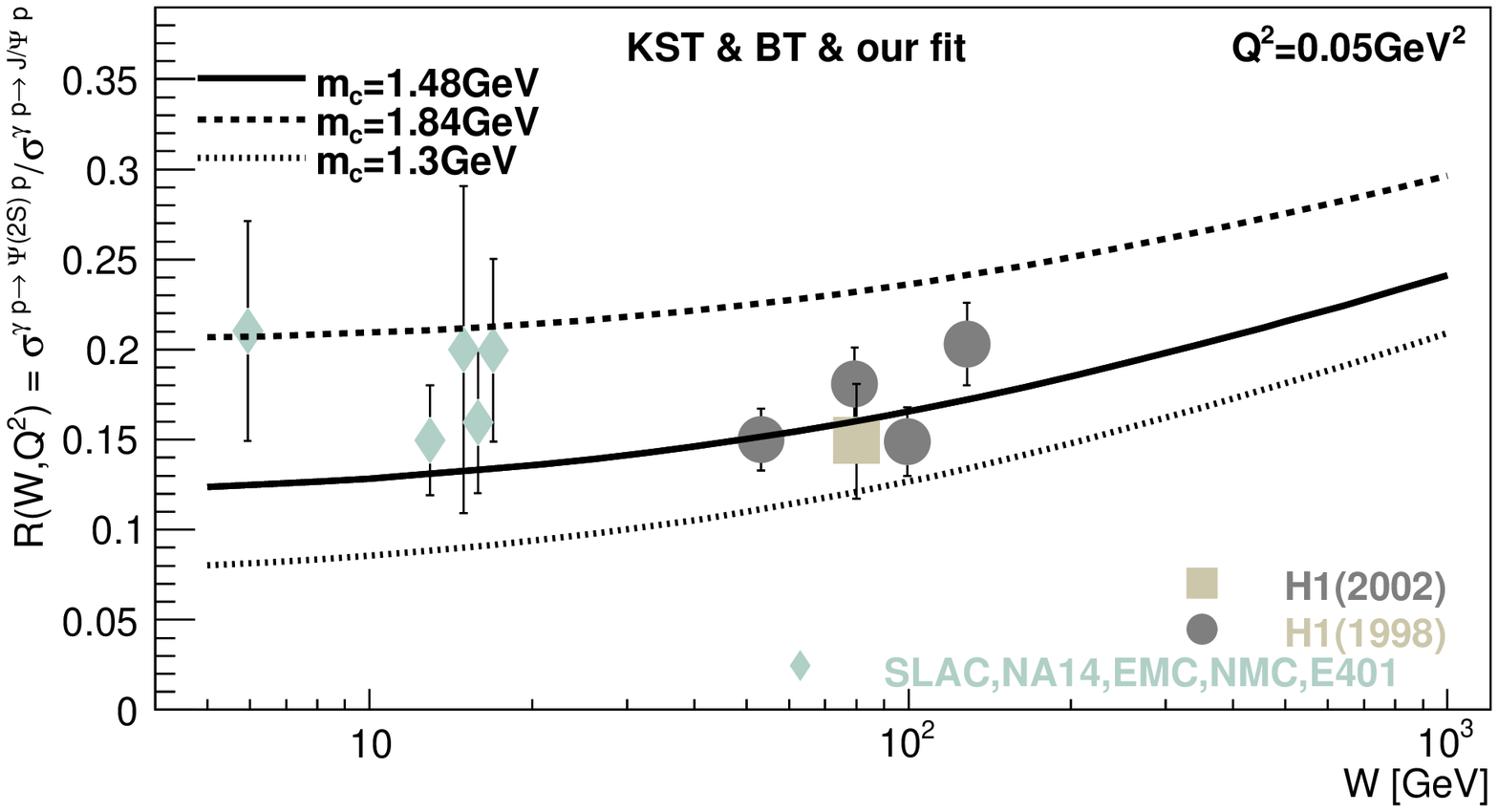}
\includegraphics[width=0.48\textwidth]{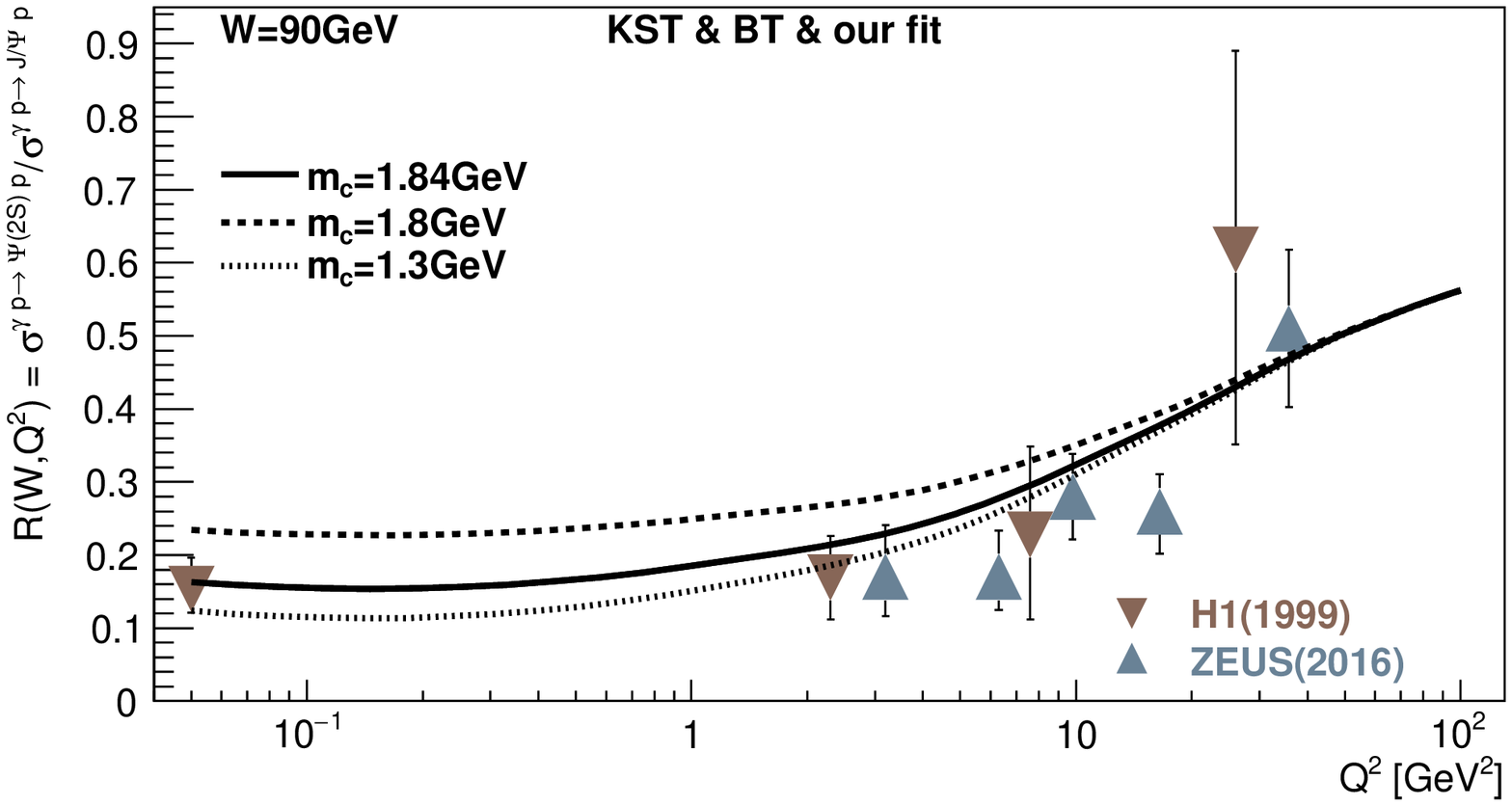}
\includegraphics[width=0.48\textwidth]{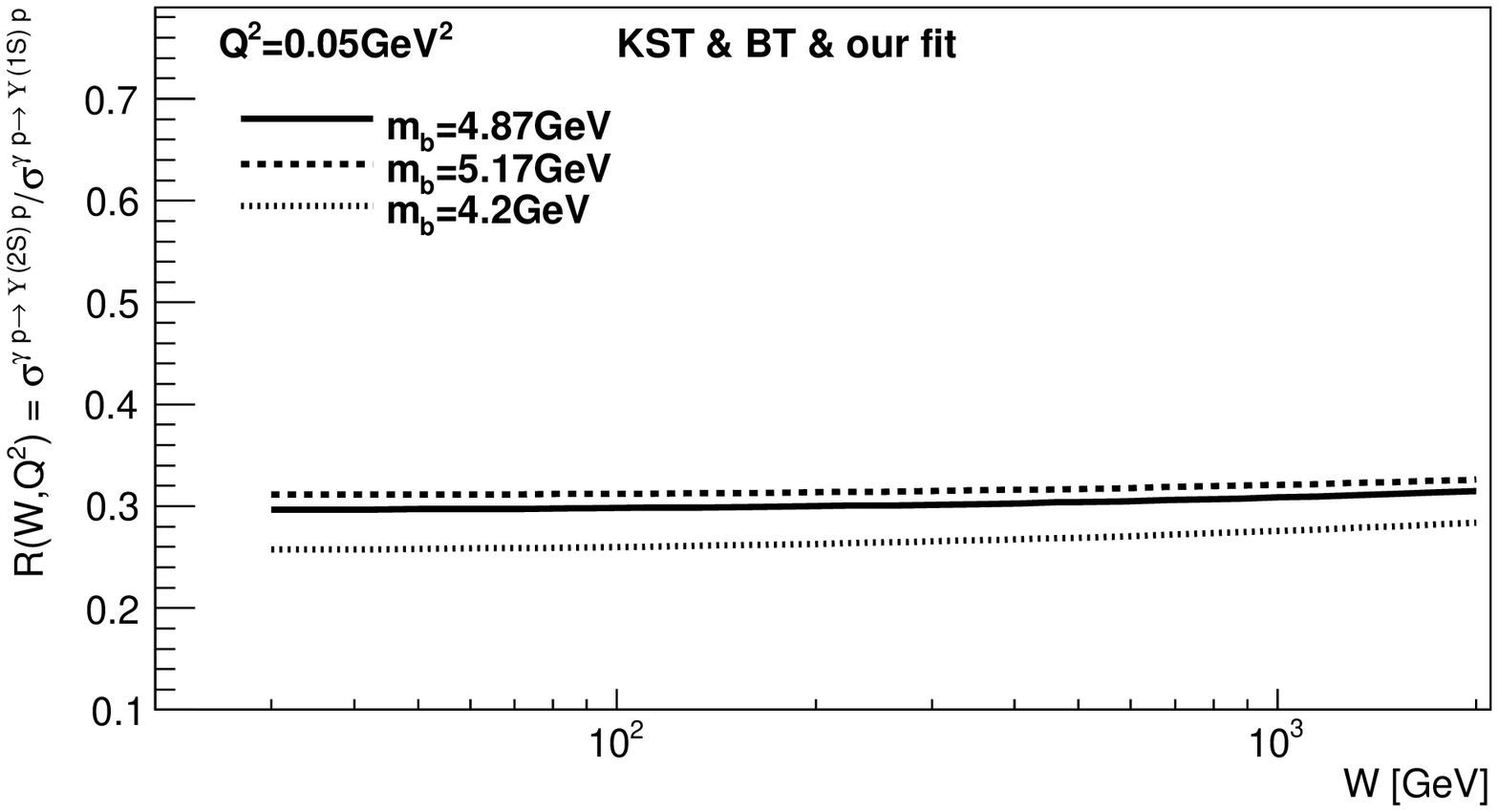}
\includegraphics[width=0.48\textwidth]{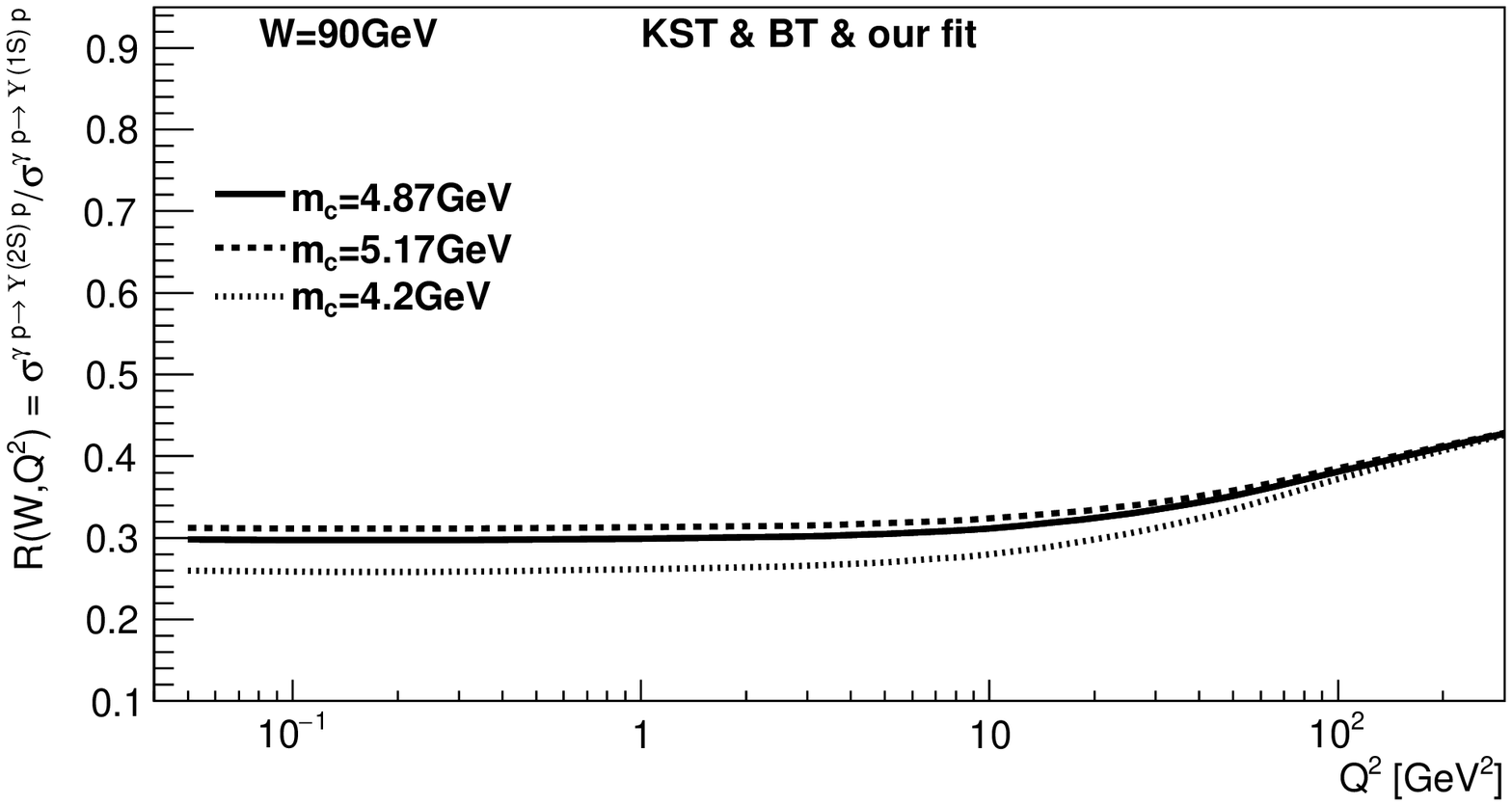}
  \caption{The same as Fig.~\ref{fig:2S-1S-pot} but for the test of sensitivity of the dipole model model predictions to the charm $m_c$ and bottom $m_b$ quark mass variations.}
%%%%%%%%%%%%%%%%%%%%%%%
  \label{fig:2S-1S-mq}
%%%%%%%%%%%%%%%%%%%%%%%
 \end{center}
 \end{figure}
%====================================== 
%
%

%...............................................................
\subsubsection{Dependence on the mass of charm and bottom quark}
%................................................................

The study of $R_{2S/1S}(W,Q^2)$ ratios in production of quarkonia also allows to minimize the underlined theoretical uncertainties in our knowledge of the corresponding quark mass value. In Fig.~\ref{fig:2S-1S-mq}, we test a variance in the model predictions taking values of $m_c$ and $m_b$ determined from the BT potential used in the calculations as well as the minimal and maximal $m_c$ and $m_b$ values occurring along all the other realistic potentials studied in this work as was described above in Sect.~\ref{Sec:psi1S-mass}. One can see that the sensitivity of $R_{2S/1S}$ to different values of $m_c$ and $m_b$ is much weaker in comparison to the results for the photo- and electroproduction cross sections (compare with Figs.~\ref{fig:psi1S-mq} and \ref{fig:ups1S-mq}). 

%.........................................
\subsubsection{Importance of spin effects}
%.........................................

\begin{figure}[!htbp]
\begin{center}
\includegraphics[width=0.48\textwidth]{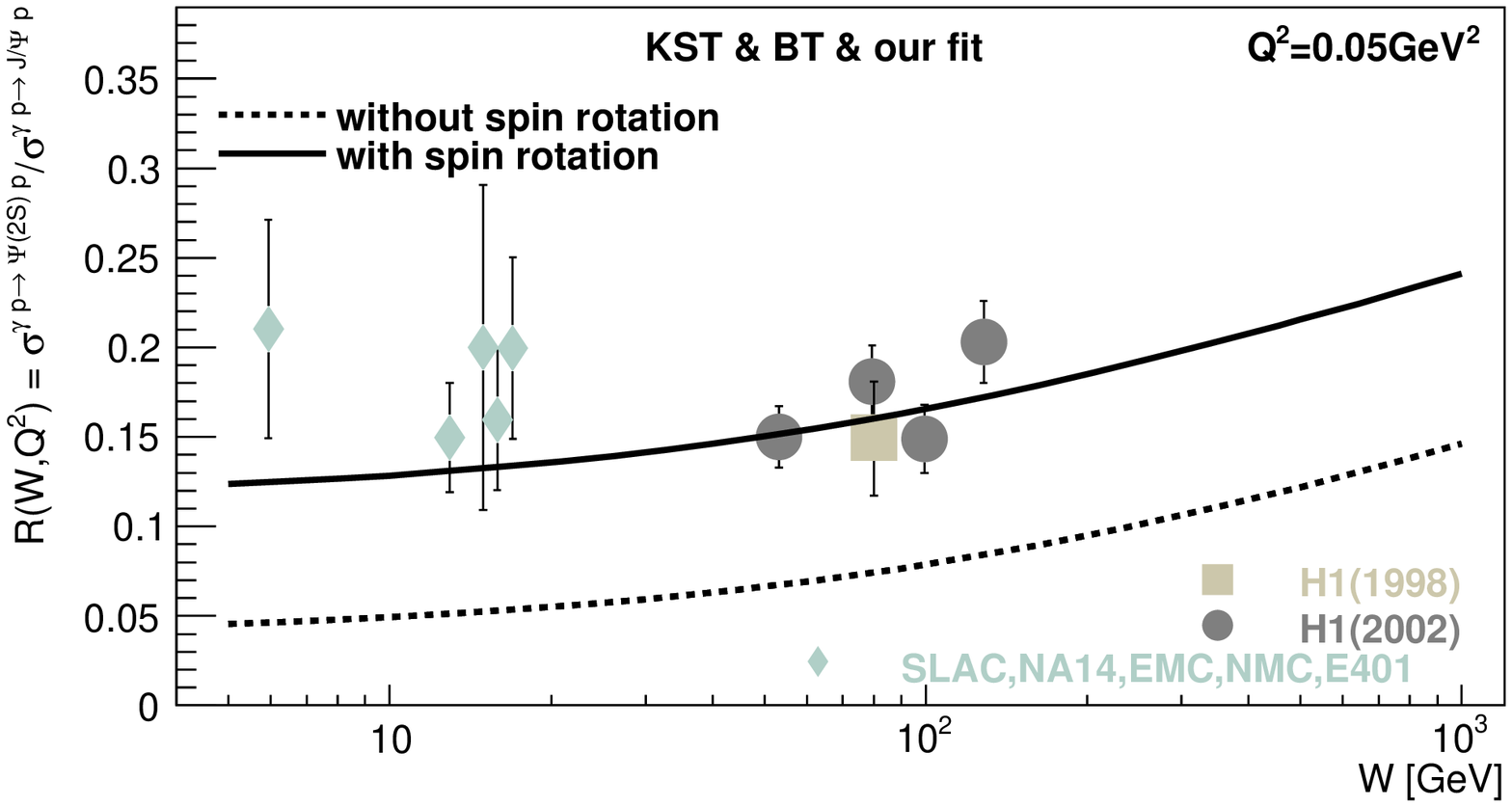}
\includegraphics[width=0.48\textwidth]{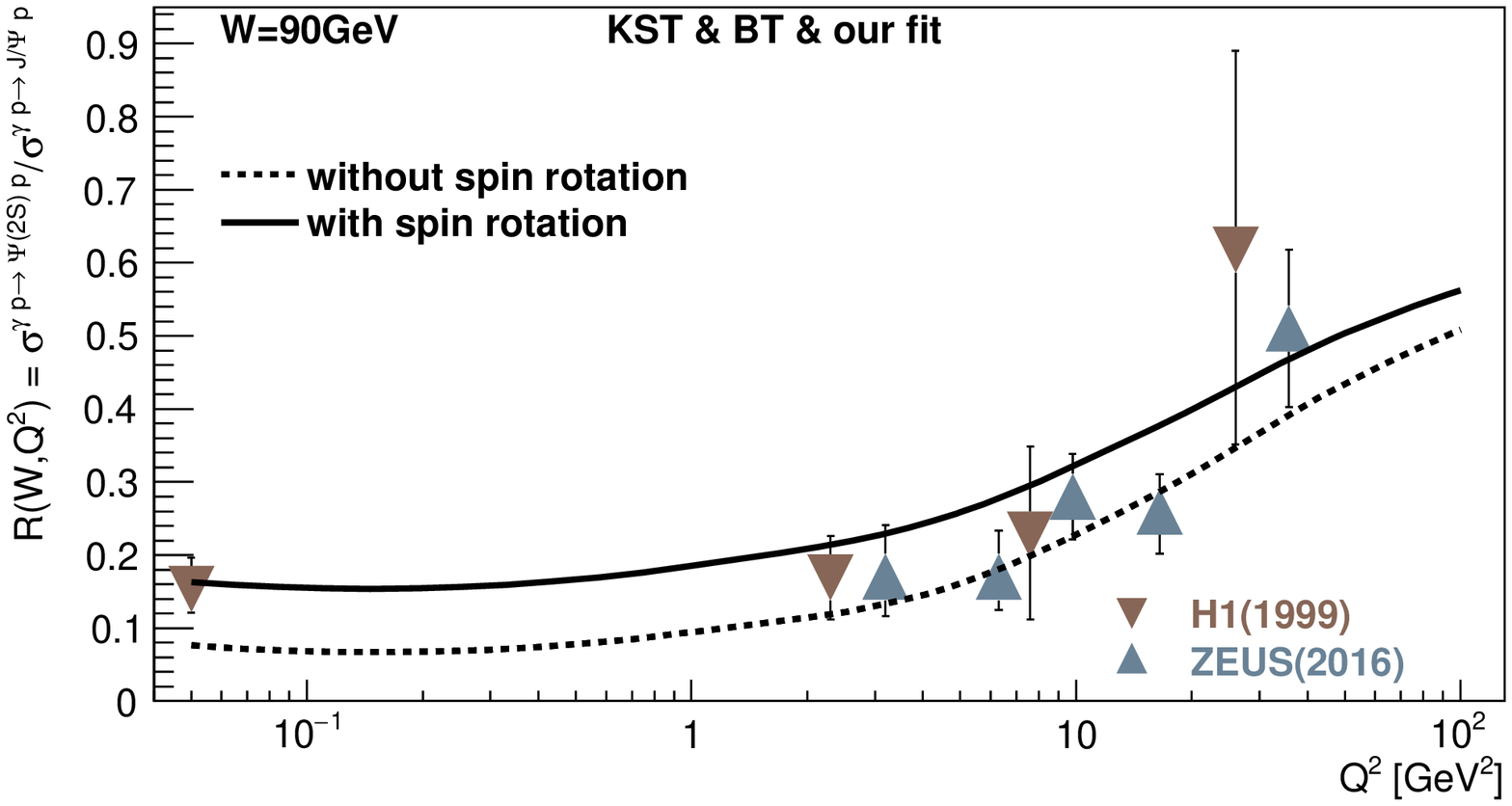}
\includegraphics[width=0.48\textwidth]{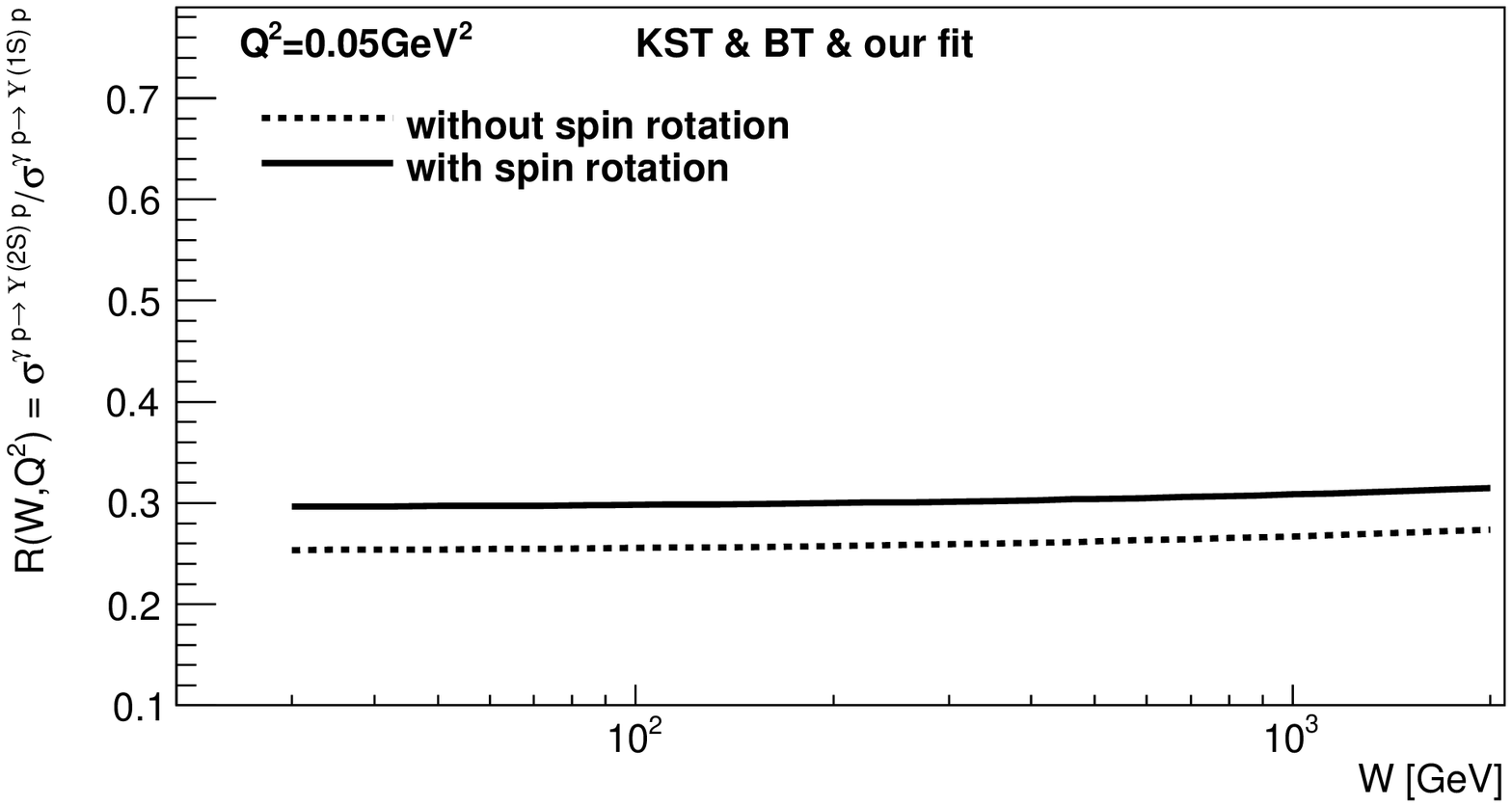}
\includegraphics[width=0.48\textwidth]{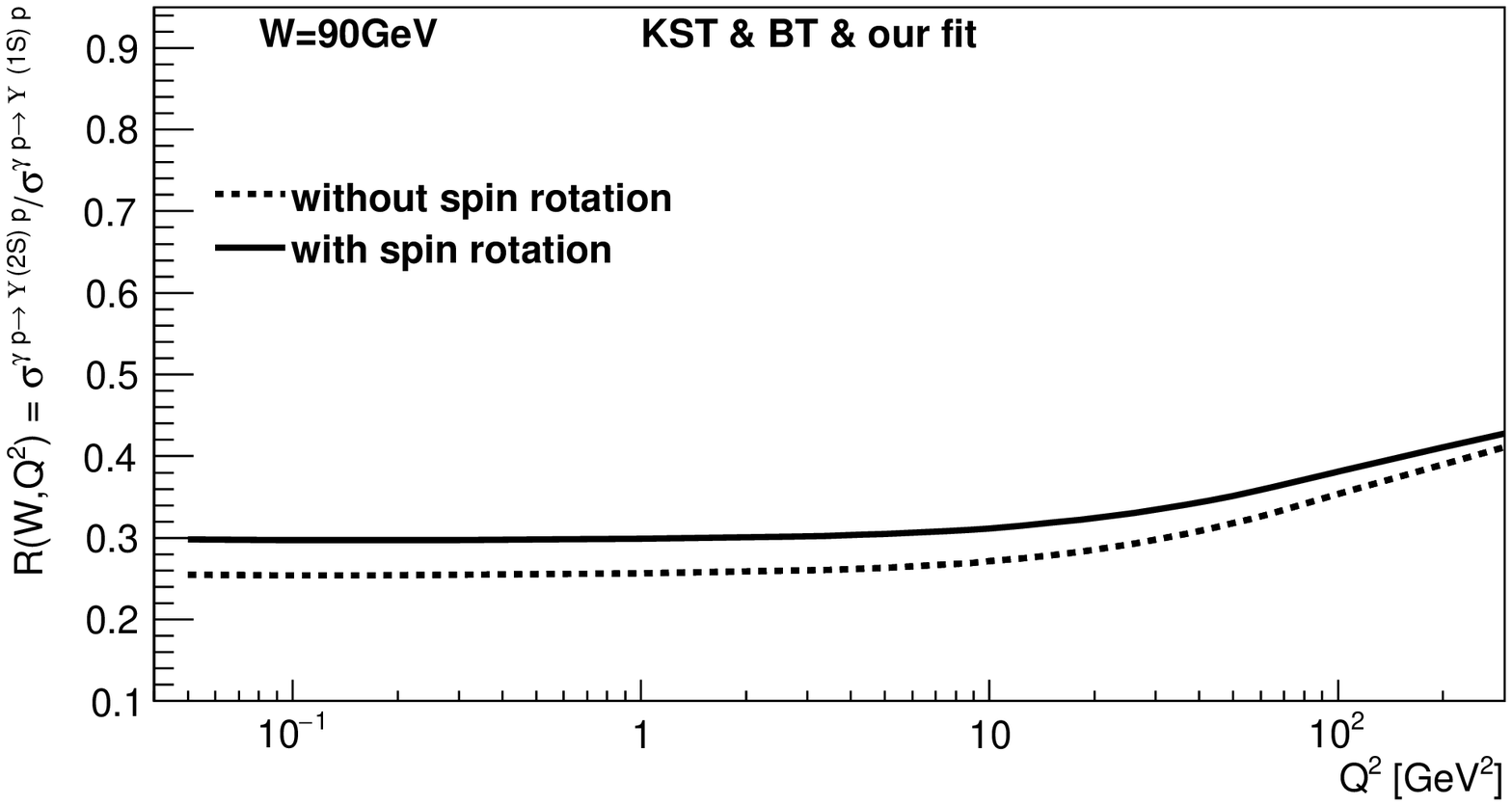}
\caption{The same as Fig.~\ref{fig:2S-1S-pot} but for demonstration of the importance of the Melosh spin rotation effects.}
%%%%%%%%%%%%%%%%%%%%%%%%%
  \label{fig:2S-1S-spin}
%%%%%%%%%%%%%%%%%%%%%%%%%
 \end{center}
 \end{figure}

In comparison to production of $1S$ quarkonia (see Sect.~\ref{Sec:spin}), as a consequence of the node effect leading to a cancellation in the production amplitude from regions below and above the node position, the onset of spin rotation effects is much stronger in proto- and electroproduction of radially-excited $\psip(2S)$, $\Yp(2S)$ and $\Ypp(3S)$ as was recently discussed in detail in Ref.~\cite{jan-18}. Here, we predict a dramatic effect of the Melosh spin transformation in charmonium electroproduction causing an increase of the $R_{2S/1S}(W,Q^2)$ ratio by a factor of $2\div 3$ as is demonstrated in the top panels of Fig.~\ref{fig:2S-1S-spin}.
%
%                Fig.24
%======================================
%====================================== 

One can see that such a substantial enhancement of $R_{2S/1S}$ due to the spin effects brings our predictions, using the KST dipole parametrization and the realistic BT potential, to the values close to the experimental data. Here, the rise of $R_{2S/1S}(W,Q^2)$ with c.m. energy $W$ and with $Q^2$ is yet another manifestation of the node effect as was discussed in Ref.~\cite{jan-18}.

Due to a weaker node effect at larger $Q^2$, we predict that the spin rotation effects gradually diminish with $Q^2$ as is demonstrated in the right panels of Fig.~\ref{fig:2S-1S-spin} for charmonium and bottomonium ratios $R_{2S/1S}$. Since the same values of the scaling variable $Q^2+M_V^2$ lead to a similar onset of various effects in production of different quarkonia, we predict a weak onset of spin effects also in photo- and electroproduction of bottomonia (see also Ref.~\cite{jan-18}) at the corresponding photon virtuality  $Q^2(\Y)\approx Q^2(\Jpsi) + M_{\Jpsi}^2$ as is shown in the bottom panels of Fig.~\ref{fig:2S-1S-spin}. 

%
%
%====================================
\subsection{Theoretical uncertainties in predictions for the ratio $\sigma_L^{\gamma^*\,p\to\Jpsi(\Y)\,p}/\sigma_T^{\gamma^*\,p\to\Jpsi(\Y)\,p}$}
\label{Sec:psi1S-LT}
%====================================
%
%

\begin{figure}[!htbp]
\begin{center}
\includegraphics[width=0.45\textwidth]{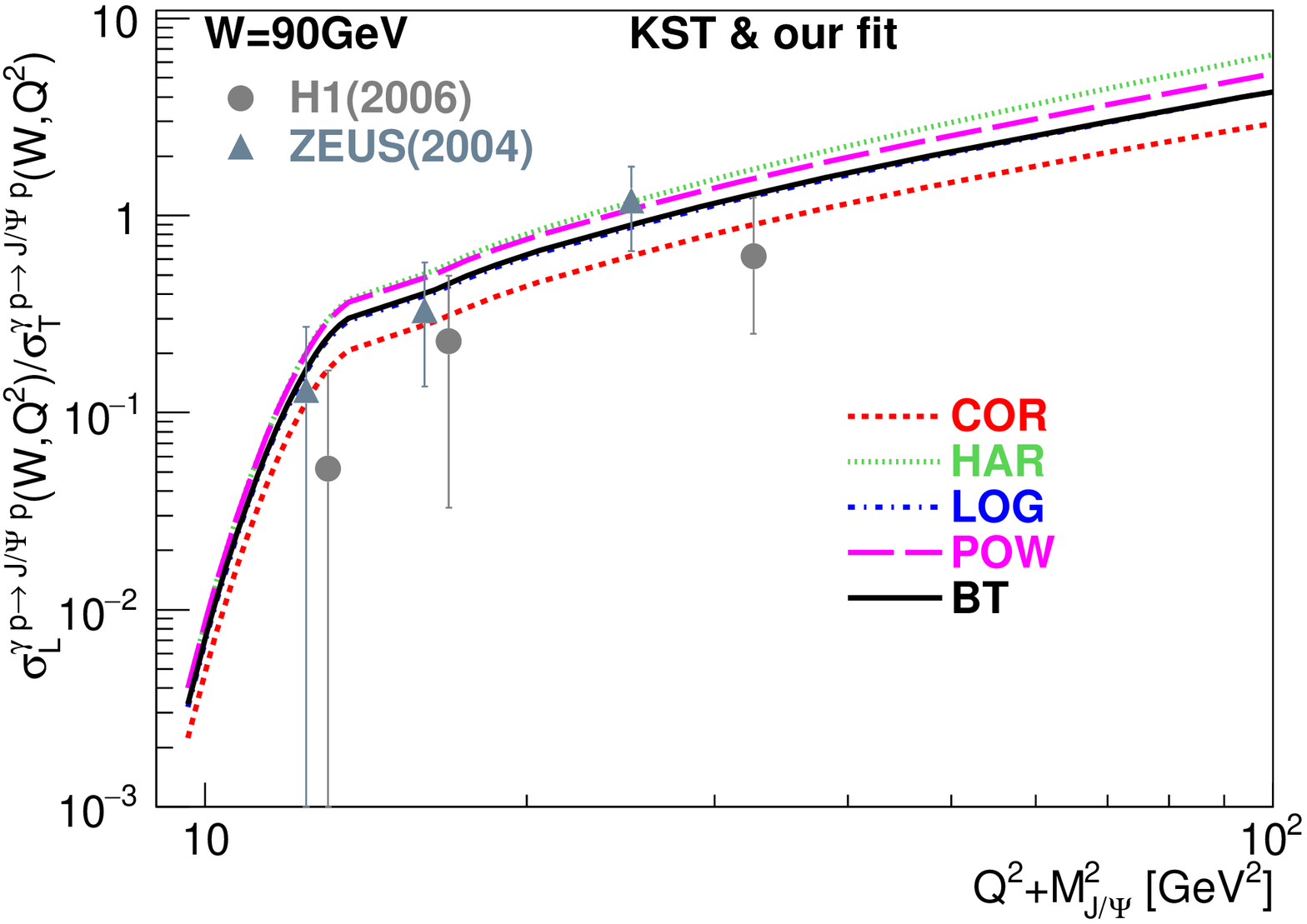}
\includegraphics[width=0.45\textwidth]{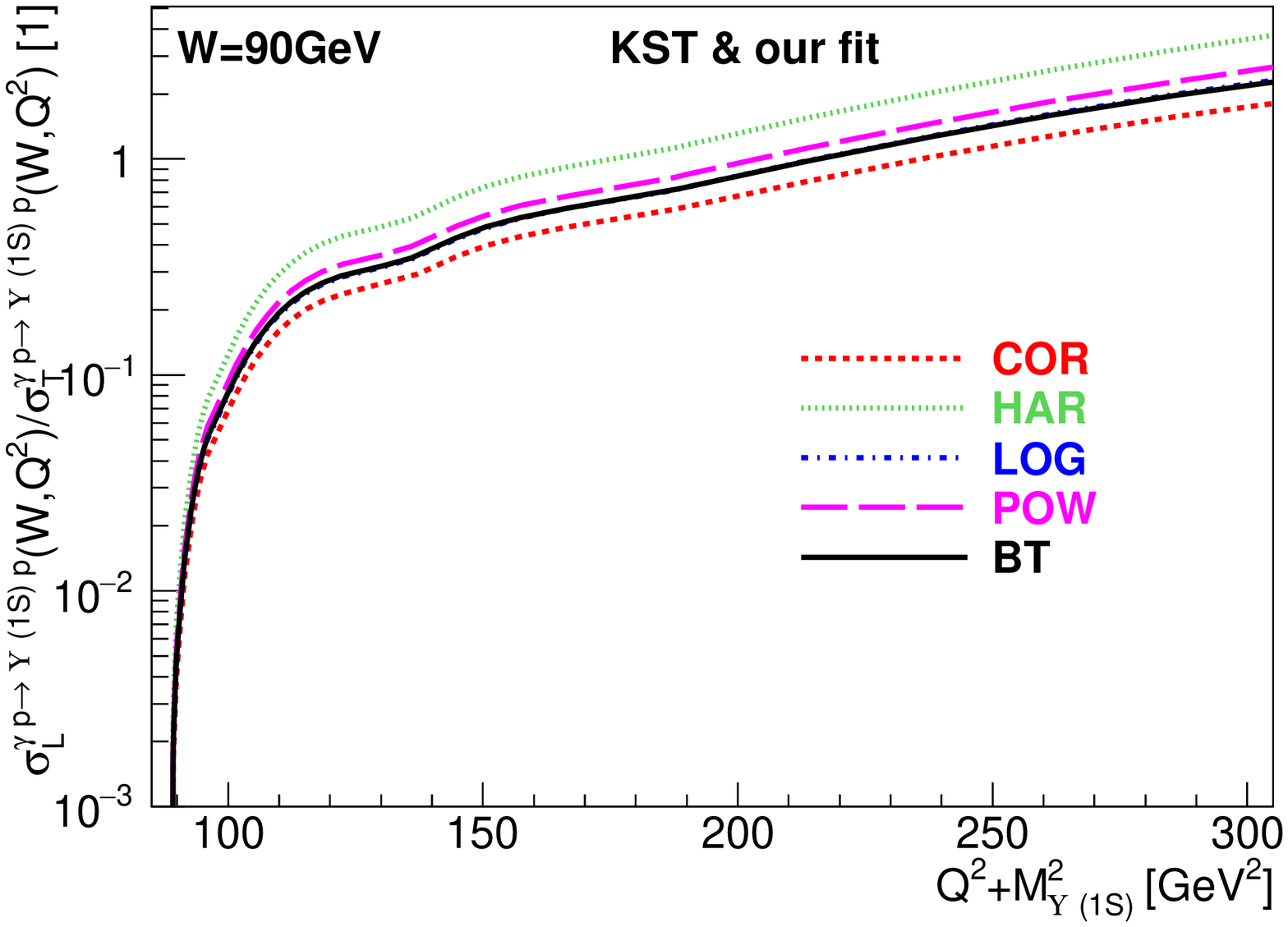}
\caption{The ratio of integrated cross sections for elastic electroproduction of longitudinally ($L$) and transversely ($T$) polarized $\Jpsi(1S)$ (left panel) and $\Y(1S)$ (right panel) as a function of the scaling variable $Q^2+M_{\Jpsi}^2$ (left panel) and $Q^2+M_{\Y}^2$ (right panel) at fixed c.m. energy $W=90\,\GeV$. The data are taken from H1 \cite{Aktas:2005xu} and ZEUS \cite{Chekanov:2004mw} collaborations. The results, including also the Melosh spin rotation effects, were obtained using the phenomenological KST dipole cross section \cite{Kopeliovich:1999am} and five different $c-\bar c$ and $b-\bar b$ interaction potentials described in Appendix~\ref{App:potentials}.}
%%%%%%%%%%%%%%%%%%%%%%%%%
 \label{fig:psi1S-LTcomp}
%%%%%%%%%%%%%%%%%%%%%%%%%
 \end{center}
 \end{figure}

The theoretical uncertainties in predictions, presented above in Sects.~\ref{Sec:psi1S-pot}, \ref{Sec:psi1S-dip}, \ref{Sec:psi1S-mass} and \ref{Sec:spin}, can be eliminated to a large extent by investigating the ratio of the elastic electroproduction cross sections of longitudinally and transversely polarized quarkonia. In Fig.~\ref{fig:psi1S-LTcomp} we present our results for such ratios $R_{LT}^{\Jpsi} = \sigma_L^{\gamma^*\,p\to\Jpsi\,p} / \sigma_T^{\gamma^*\,p\to\Jpsi\,p}$ and $R_{LT}^{\Y} = \sigma_L^{\gamma^*\,p\to\Y\,p} / \sigma_T^{\gamma^*\,p\to\Y\,p}$ as functions of the scaling variables $Q^2+M_{\Jpsi}^2$ and $Q^2+M_{\Y}^2$, respectively. One can see a rather good agreement of $R_{LT}^{\Jpsi}$ with the existing data for all considered $c-\bar c$ potentials. Our predictions for the ratio $R_{LT}^{\Y}(Q^2)$ can be tested by future measurements.
%
%                 Fig.25
%========================================
%=======================================
%
%

Here we would like to emphasize that the variation in model predictions for the ratio $R_{LT}$ using different quarkonium wave functions generated by distinct potentials is much less pronounced than that observed in Sect.~\ref{Sec:psi1S-pot} for the standard photo- and electroproduction cross sections (see Figs.~\ref{fig:psi1S-pot} and \ref{fig:ups1S-pot}).

%
%
%====================================
\subsection{The skewness effect in electroproduction of quarkonia}
\label{Sec:psi1S-skew}
%====================================
%
%

The skewness correction is frequently interpreted in the literature as an effect when the gluons attached
to the $Q\bar Q$ fluctuation of the photon carry (very) different light-front fractions $x$ and $x'$ of the proton momentum \cite{Shuvaev:1999ce, Martin:1999wb}. The corresponding expression for the correction factor $R_g$ has the following form \cite{Shuvaev:1999ce},
%
%**************************************************
\beq
R_g(\lambda_{T,L})
=
\frac{2^{2\lambda_{T,L} +3}}{\sqrt{\pi}}\frac{\Gamma(\lambda_{T,L} + 5/2)}{\Gamma(\lambda_{T,L} + 4)}\, ,
%%%%%%%%%%
\label{rg}
%%%%%%%%%%
\eeq
%***************************************************
%
with $\lambda_{T,L}$ determined by Eq.~(\ref{lambda}) for both the photon polarizations $T$ and $L$. Consequently, the skewness correction is then accounted for by multiplying the cross section $\sigma_{T,L}^{\gamma^*\,p\to V\,p}(x,Q^2)$, Eq.~(\ref{total-cs}), by a factor of $R_g^2(\lambda_{T,L})$.

However, the shape of the correction factor in Eq.~(\ref{rg}) has been derived in Ref.~\cite{Shuvaev:1999ce} within the next-to-leading order approximation assuming the strong inequalities $x'\ll x\ll 1$ in the small-$t$ region and for the specific power-law form of the diagonal gluon density of the target. Here, such a small-$x$ shape of the gluon density is not fully probed within kinematic regions studied in the present paper, and consequently, may not be fully consistent with those extracted from different dipole models used in our calculations. 

The statement from Ref.~\cite{Rezaeian:2013tka} that the skewness correction given by Eq.~(\ref{rg}) can be incorporated into the bCGC dipole model is not fully consistent for the case of electroproduction of heavy quarkonia. The dipole amplitude is related to the gluon structure function of the target only at sufficiently large $Q^2\sim\Lambda/r^2$ (see also Eq.~(\ref{dip-pdf})), where the dipole sizes $r\lesssim r_0\sim 0.3\,\fm$ ($r_0$ is the gluon propagation radius \cite{spots,drops}) and the large numerical factor $\Lambda\approx 10$ have been estimated in Ref.~\cite{Nikolaev:1994cn}. In the case of quarkonium production, this condition requires rather large values of the saturation scale squared corresponding to the bCGC dipole model, $Q_s^2(x)$. This leads to rather small values of the Bjorken variable $x\lesssim 10^{-5}\div 10^{-6}$ necessary for justification of Eq.~(\ref{rg}) for the skewness correction. Such small $x$-values correspond to way too large c.m. energies $W\gtrsim 10^3\,\GeV$, which are far beyond the energy range studied in the present paper. The same conclusion concerns also the other dipole models since the corresponding saturation scales are similar to that in the bCGC parametrization. 
%
%                Fig.26
%=======================================
\begin{figure}[!htbp]
\begin{center}
\includegraphics[width=0.47\textwidth]{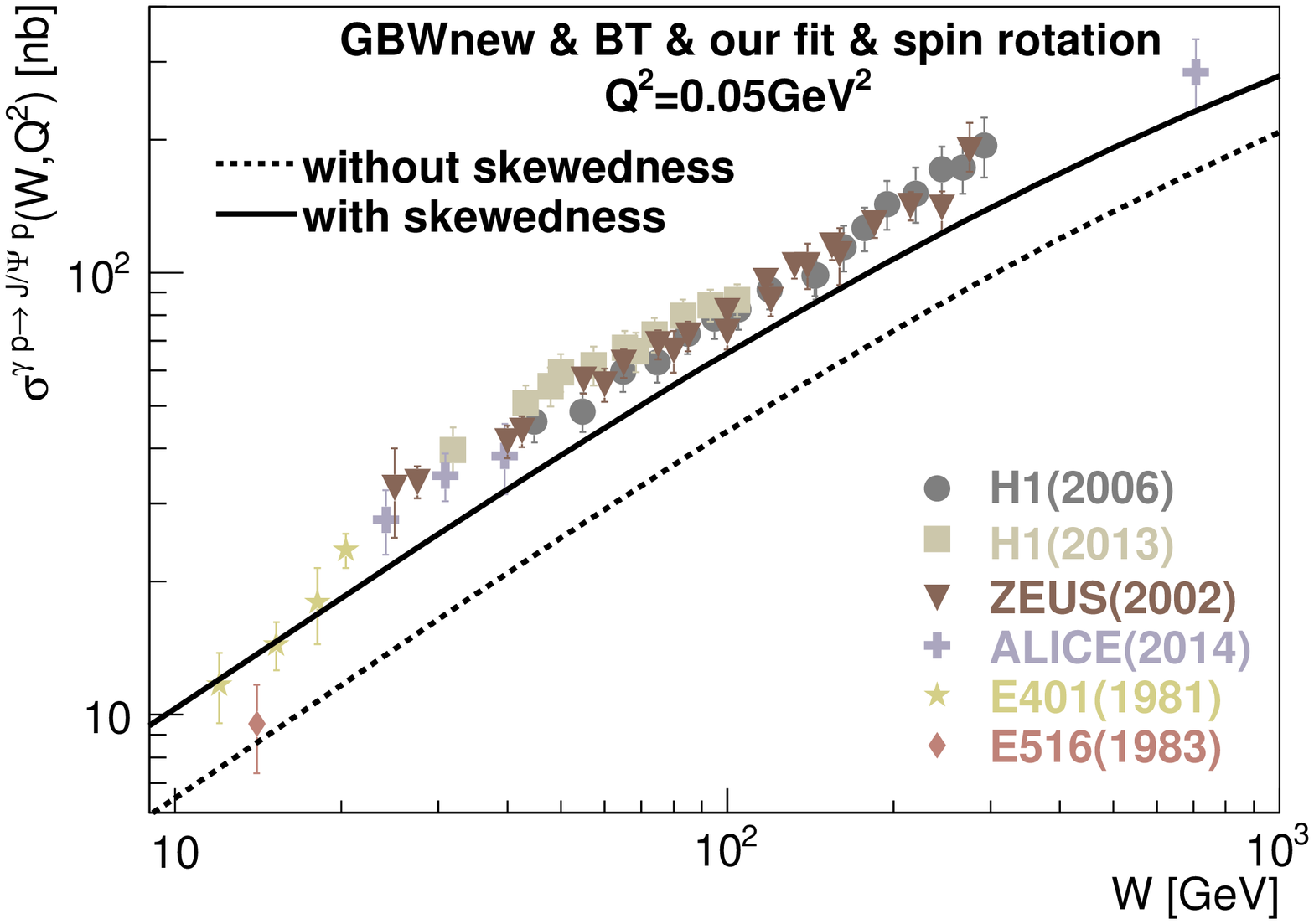}
\includegraphics[width=0.47\textwidth]{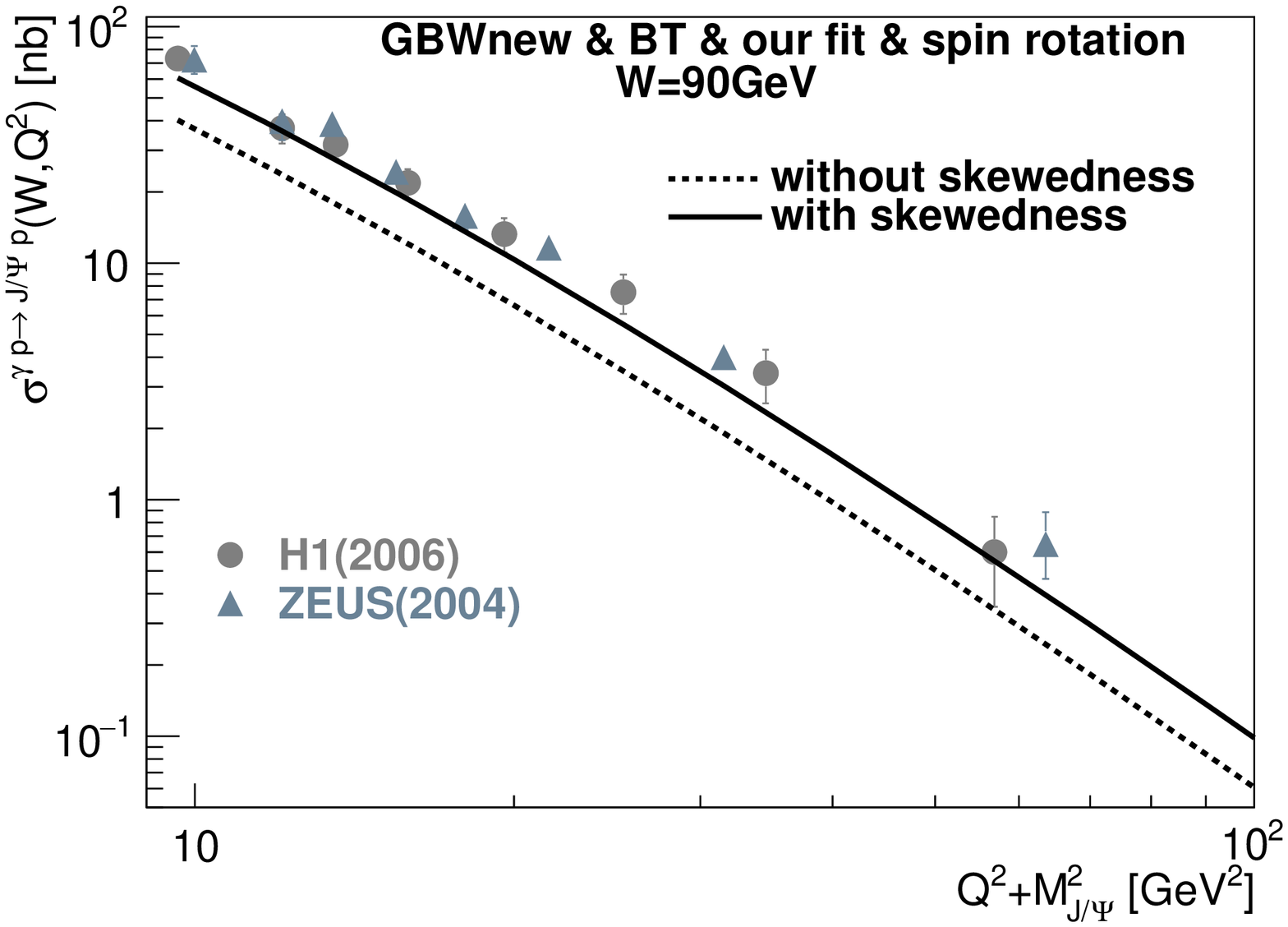}
  \caption{The same as Fig.~\ref{fig:psi1S-Bcomp} but for illustration of the onset of the skewness effect
  as a function of c.m. energy $W$ and the scaling variable $Q^2+M_{\Jpsi}^2$. The model calculations were performed with the $\Jpsi$ wave function generated by the BT potential \cite{Buchmuller:1980su} and with the
  phenomenological GBWnew dipole cross section  \cite{Kowalski:2006hc} including the Melosh spin rotation.}
%%%%%%%%%%%%%%%%%%%%%%%%%
  \label{fig:psi1S-skew}
%%%%%%%%%%%%%%%%%%%%%%%%%
 \end{center}
 \end{figure}
%========================================= 
%
%

\begin{figure}[!htbp]
\begin{center}
\includegraphics[width=0.47\textwidth]{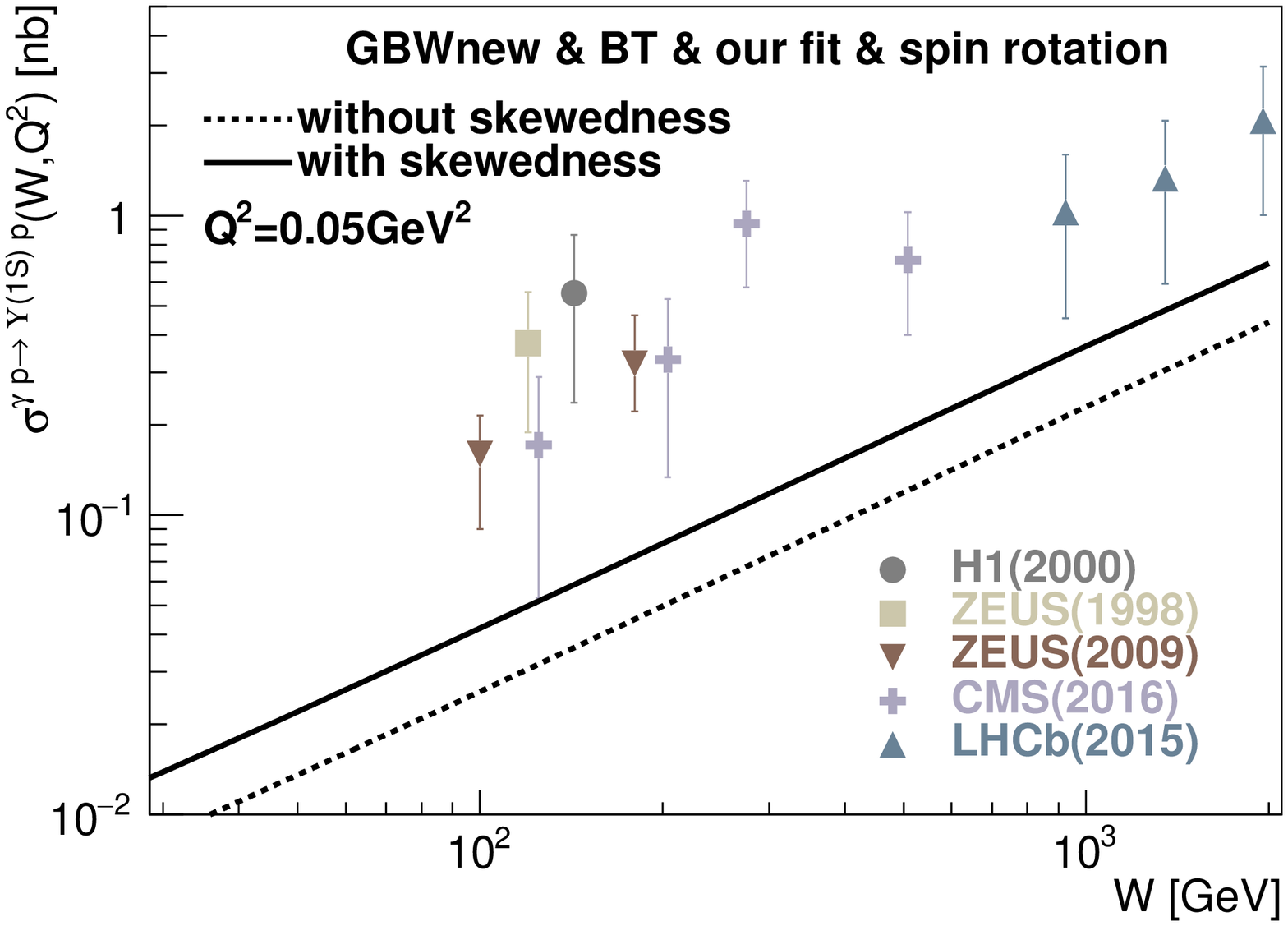}
\includegraphics[width=0.47\textwidth]{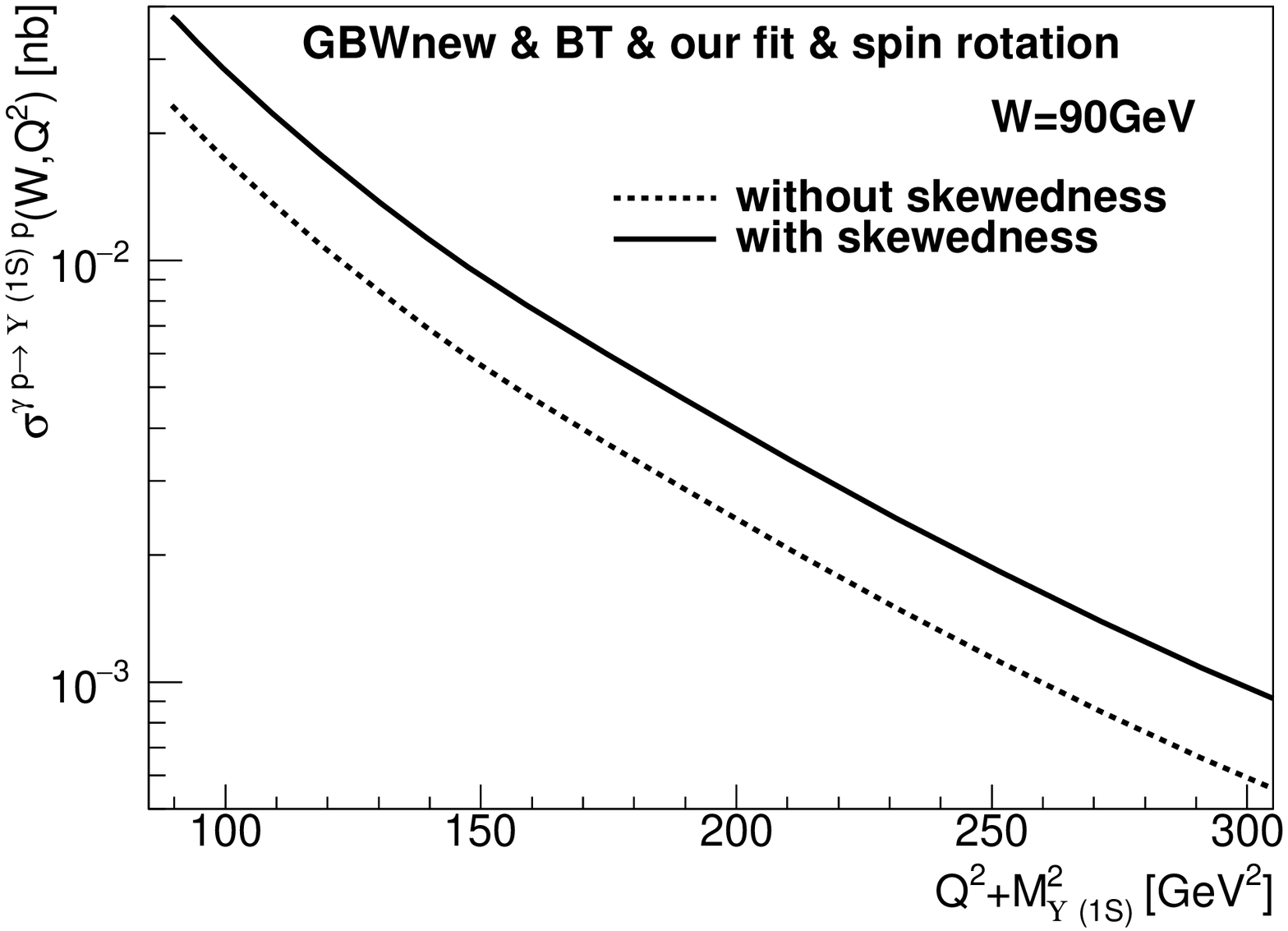}
  \caption{The same as Fig.~\ref{fig:psi1S-skew} but for exclusive electroproduction of $1S$ bottomonia. The data are taken from from H1 \cite{Adloff:2000vm}, ZEUS \cite{Breitweg:1998ki,Chekanov:2009zz}, CMS \cite{CMS:2016nct} and LHCb \cite{Aaij:2015kea} Collaborations.}
%%%%%%%%%%%%%%%%%%%%%%%%%
  \label{fig:ups1S-skew}
%%%%%%%%%%%%%%%%%%%%%%%%%
 \end{center}
 \end{figure}

Since the exact analytical expression for $R_g$ is not available in the literature, we present here only a phenomenological estimation of the onset of the skewness effect in electroproduction of heavy quarkonia relying on the known approximate relation, Eq.~(\ref{rg}). The results are depicted in Figs.~\ref{fig:psi1S-skew} and \ref{fig:ups1S-skew} for the case of electroproduction of charmonia and bottomonia in the ground state, respectively.

The model calculations have been performed, as an example, with the phenomenological GBWnew dipole cross section  \cite{Kowalski:2006hc} and with the quankonium wave functions generated by the realistic BT potential
\cite{Buchmuller:1980su}. One can see from Figs.~\ref{fig:psi1S-skew} and \ref{fig:ups1S-skew} that the skewness correction increases the photo- and electroproduction cross section of quarkonia by a factor of $\sim 1.5\div 1.6$. As was analyzed in Sect.~\ref{Sec:psi1S-dip}, neglecting the skewness correction, only the KST and GBWold dipole parametrizations lead to the best description of the available data on quarkonium electroprodution, whereas other phenomenological dipole cross sections grossly underestimate these data. Consequently, one can expect that the onset of the factor $R_g$ in our calculations should cause a slight overestimation of data for the KST but would lead to an improvement of the data description using not only GBWnew but also other dipole parametrizations. Namely, such an effort to obtain a better agreement with the data typically generates the main reason to include formally the skewness effects adopting only an approximate relation (\ref{rg}) based on assumptions, which can not be naturally adopted or justified for an arbitrary process. This is the basic motivation for us not to include the skewness factor in the rest of the calculations in the previous sections, instead, showing the more justified color dipole model predictions and estimates for the underlined theoretical uncertainties.
%
%               Fig.27
%========================================
%========================================= 
%
%

%
%
%
%========================================
\section{Conclusions}
\label{Sec:conclusions}
%========================================
%
%
%

We have presented an exploratory and comprehensive study of elastic photo- and electroproduction of heavy quarkonia within the color dipole formalism. The main motivation is based on a growing interest in this topic, mainly in connection with an extensive ongoing investigation of quarkonium production processes in ultra-peripheral collisions at the RHIC and LHC facilities. Although the color dipole approach is well-known already of about thirty years and a wealth of research has been done, it is frequently used in the literature without a deeper understanding of the underlined theoretical uncertainties in predictions caused by various effects and properties of particular ingredients entering into the production amplitudes. Consequently, in order to obtain a better agreement with the data, this leads to an ongoing effort to include some additional new phenomena or additional ingredients instead of a better understanding the corresponding uncertainties or performing more accurate calculations. For this reason, in this paper we try to describe and analyze various sources of theoretical uncertainties and study their impact on the magnitude of the corresponding electroproduction cross sections for a large variety of quarkonia states and physics inputs.

In the color dipole formalism the production amplitude, given by the factorized light-cone expression (\ref{prod-amp}), has the following ingredients: (i) the perturbative light-cone wave functions for the heavy $Q\bar Q$ fluctuation of the photon, (ii) the light-cone wave functions for the $S$-wave quarkonia states, and (iii) a phenomenological dipole cross section $\sigma_{q\bar q}(r,x)$ describing the interaction of the $Q\bar Q$ fluctuation with a proton target.

A description of the photon wave function is well-known and quite well understood, so it should not cause major uncertainties in calculations of the production amplitude. On the other hand, the determination of the quarkonium light-cone wave functions remains rather uncertain. Here, we adopted the frequently used prescription from Ref.~\cite{Terentev:1976jk} for the transition from the $Q\bar Q$ rest frame to the infinite momentum one. The corresponding quarkonium wave functions in the $Q\bar Q$ rest frame have been obtained by solving the Schr\"odinger equation for various $Q-\bar Q$ interaction potentials. Such an ambiguity in determination of quarkonium wave functions represents one of more relevant sources of theoretical uncertainties.

The essential ingredient in our calculations of the photo- and electroproduction cross sections of heavy quarkonia is the dipole cross section $\sigma_{q\bar q}(r,x)$. Here, we adopted the total of eight main phenomenological parametrizations for $\sigma_{q\bar q}(r,x)$ found in the literature that exhibit a saturated form at large transverse separations (dipole sizes) $r$ as well as roughly satisfy the characteristic small-$r$ behavior, $\sigma_{q\bar q}(r,x)\propto r^2$ for $r\to 0$ (color transparency). The differences in the corresponding parametrizations for $\sigma_{q\bar q}(r,x)$ represent another source of theoretical uncertainties in calculations of dipole amplitudes and, subsequently, of the corresponding electroproduction cross sections.

In order to avoid a double counting, the effect of higher Fock states, $Q\bar QG$, $Q\bar QGG$, ..., containing gluons in the photon wave function can be reabsorbed into the energy (Bjorken $x$) dependence of $\sigma_{q\bar q}(r,x)$. On the other hand, the dipole cross section has a steeper rise with energy at smaller dipole sizes due to more intensive gluon radiation. Here all cross sections at different dipole sizes are expected to follow the universal asymptotic properties at very large energies controlled by the Froissart bound.

The model predictions for the exclusive quarkonium electroproduction cross sections depend on the magnitude of the diffraction slope $B$ (see Eq.~(\ref{total-cs})) for the corresponding elastic process $\gamma^*\,p\to V\,p$, where the vector meson $V = \Jpsi(1S), \psip(2S), \Y(1S), \Yp(2S), \Ypp(3S)$, etc. The energy dependence of $B(W)$ has been obtained by the fit to the available data at HERA (see Eq.~\ref{BW} and Table~\ref{Tab:slope-pars}). Since the data on the $Q^2$ behavior of the slope parameter are very scarce, we adopted a phenomenological model from Ref.~\cite{jan-98} leading to an empirical parametrization (\ref{BQ2}), which gives the values of $B(Q^2)$ in a reasonable agreement with the data. Within the same model, we have included also the differences in slope parameters $B(1S)-B(2S)$, corresponding to production of the $1S$-ground state and $2S$-radially excited quarkonia. These differences come as a direct manifestation of the node effect in the quankonium wave functions, in particular, leading to a cancellation in the production amplitude coming from regions in $r$ below and above the node position. We have verified that different parametrizations of the energy evolution, with modelled $Q^2$ behavior of the slope parameter, cause only rather small uncertainties in the model predictions using various combinations of the quarkonium wave functions and phenomenological dipole parametrizations for $\sigma_{q\bar q}(r,x)$. 

Another source of uncertainties studied in this work refers to the effect of the Melosh spin rotation, which is often neglected in the literature. We found that such spin effects are very important, especially in elastic photoproduction of quarkonia. They lead to a $\approx 20\div 30\,\%$ rise of the $\Jpsi(1S)$ photoproduction cross section contributing to a better agreement of the model predictions with the data. However, they cause even more dramatic effect in $\psip(2S)$ photoproduction substantially increasing the corresponding cross sections, as well as the $\psip(2S)$-to-$\Jpsi(1S)$ ratio, by a factor of $2\div 3$ (see also Ref.~\cite{jan-18}).

We have also presented and discussed a large sensitivity of the model predictions to the value of heavy quark mass $m_Q$ which is caused by the photon wave function, Eq.~(\ref{gamma-wf}) containing $m_Q$ in the argument of the Bessel function $K_0$.

Although the skewness correction is frequently used in calculations of the quarkonium photo- and electroproduction cross sections, only an approximate relation, Eq.~(\ref{rg}), is known for the corresponding correction factor $R_g$. Since the exact analytical formula for $R_g$ is not available in the literature, we estimated a magnitude of this effect relying on the known expression (\ref{rg}) and found that the skewness correction increaces the quarkonium electroproduction cross section by a factor of $\sim 1.5\div 1.6$. However, it is questionable to what extent and with what accuracy the approximate relation, Eq.~(\ref{rg}), can be applied to quarkonium electroproduction within the kinematic ranges studied in the present paper.

Finally, we have found that all these sources of theoretical uncertainties can be reduced to a large extent when investigating the ratios of the cross sections such as $R_{2S/1S}(W,Q^2) = \sigma^{\gamma^*\,p\to \psip(2S) (\Yp(2S))\,p}(W,Q^2) / \sigma^{\gamma^*\,p\to \Jpsi(1S) (\Y(1S))\,p}(W,Q^2)$, as well as $R_{L/T}(W,Q^2) = \sigma_L^{\gamma^*\,p\to \Jpsi (\Y)\,p}(W,Q^2) / \sigma_T^{\gamma^*\,p\to \Jpsi (\Y)\,p}(W,Q^2)$. We have demonstrated that, in comparison to the standard quarkonium electroproduction cross sections, the ratios $R_{2S/1S}$ and $R_{L/T}$ exhibit much smaller variations generated by these uncertainties and thus produce more stable and accurate results, which can be tested by the future experiments.

To summarize, in our current analysis performed within the color dipole formalism we have used for the first time a combination of several new ingredients simultaneously, such as the proper light-cone wave functions of heavy quarkonia generated by realistic interquark interaction potentials, together with the Melosh spin rotation and the most recent models for the saturated dipole cross section. We have successfully described the existing $\Jpsi$, $\psip$ and $\Y$ photo- and electroproduction data off the nucleon target. This encourages us to extend consequently such an analysis, going beyond the NRQCD approximation, also for nuclear targets and verify our predictions for vector meson photoproduction by comparing with the recent data obtained from ultra-peripheral heavy-ion collisions at RHIC and LHC. The corresponding new predictions can be tested then by the future (e.g. LHeC) measurements. 

Finally, we would like to emphasize that the most of the results presented in the current paper can also be obtained interactively on our webpage \href{https://hep.fjfi.cvut.cz/vm.php}{https://hep.fjfi.cvut.cz/vm.php}, where the model predictions for the photo- and electroproduction cross sections can be readily computed for various combinations of the quarkonium wave functions with particular dipole parametrizations for $\sigma_{q\bar q}(r,x)$ including or neglecting the Melosh spin rotation effects. Such an online tool is expected to be very useful for QCD practitioners and experimentalists working in the research areas connected to quarkonia physics.

%
%
%
%==========================================
\section*{Acknowledgements}
%==========================================
%
%
%

J.C.~is supported by the grant 17-04505S of the Czech Science Foundation (GACR).
J.N.~is partially supported by grants LTC17038 and LTT18002 of the Ministry of Education, Youth and Sports of the Czech Republic, by projects of the European Regional Development Fund CZ02.1.01/0.0/0.0/16\_013/0001569 and CZ02.1.01/0.0/0.0/16\_019/0000778, and by the Slovak Funding Agency, Grant 2/0007/18.
M.K.~is supported in part by the Conicyt Fondecyt grant Postdoctorado N$^\circ$3180085 (Chile) and by the grant LTC17038 of the Ministry of Education, Youth and Sports of the Czech Republic.
R.P.~is supported in part by the Swedish Research Council grants, contract numbers 621-2013-4287 and 2016-05996, by CONICYT grant MEC80170112, by the Ministry of Education, Youth and Sports of the Czech Republic, project LTC17018, as well as by the European Research Council (ERC) under the European Union's Horizon 2020 research and innovation programme (grant agreement No 668679). The work has been performed in the framework of COST Action CA15213 ``Theory of hot matter and relativistic heavy-ion collisions'' (THOR).

%%%%%%%%%%%%%%%%%%%%%%%%%%%%%%%%%%%%%%%%%%
%%%%%%%%%%%%%%%%%%%%%%%%%%%%%%%%%%%%%%%%%%
%%%%%%%%%%%%%%%%%%%%%%%%%%%%%%%%%%%%%%%%%%
\appendix
%%%%%%%%%%%%%%%%%%%%%%%%%%%%%%%%%%%%%%%%%%
%%%%%%%%%%%%%%%%%%%%%%%%%%%%%%%%%%%%%%%%%%
%%%%%%%%%%%%%%%%%%%%%%%%%%%%%%%%%%%%%%%%%%

%
%
%
%=========================================
\section{Quarkonia potentials}
\label{App:potentials}
%=========================================
%
%
%

In order to compute the quarkonium wave function, one needs to specify an interaction potential between heavy quarks. Here, we provide the details of several distinct models for interquark potentials used in our numerical analysis.

%
%
%=========================================
\subsection{Harmonic oscillator}
%=========================================
%
%

The potential for harmonic oscillator (denoted as \textbf{HAR})
%
%*****************************************
\begin{eqnarray}
V(\r)
=
\frac{1}{2}\,m_Q\,\omega^{2}\,\r^{2} \,, \qquad \omega=\frac{1}{2}(M_{2S}-M_{1S}) \,,
\end{eqnarray}
%*****************************************
%
is the simplest and the most common choice that leads to the Gaussian shape of the wave function. The masses of charm $c$ and bottom $b$ quarks are taken to be $m_c=1.4$ GeV and $m_b=4.2$ GeV, respectively. The parameter $\omega$ is fixed to $0.3$ GeV, for charmonia, and to $0.28$ GeV, for bottomonia. The Schr\"odinger equation with this potential has an analytic solution
%
%*****************************************
\begin{eqnarray}
u(\r)
=
\exp \biggl [ -\frac{1}{4}\,m_Q\,\omega\, \r^2\biggr ] \,,
\end{eqnarray}
%*****************************************
%
however, we obtain a solution of the Schr\"odinger equation for the harmonic oscillator numerically. 

%
%
%=========================================
\subsection{Cornell potential}
%=========================================
%
%

The Cornell potential (\textbf{COR}) given by
%
%*****************************************
\begin{eqnarray}
V(\r)
=
- \frac{k}{\r}+\frac{\r}{a^{2}}\quad k=0.52\quad a=2.34\,\mathrm{GeV}^{-1} \,,
\end{eqnarray}
%*****************************************
%
with $m_c=1.84$ GeV and $m_b=5.17$ GeV, was initially proposed in Refs.~\cite{Eichten:1978tg,Eichten:1979ms} and 
was also used in quarkonia photoproduction studies in Refs.~\cite{Hufner:2000jb,Kowalski:2003hm}. 

%
%
%=========================================
\subsection{Logarithmic potential}
%=========================================
%
%

The logarithmic potential ({\textbf{LOG}) given by
%
%*****************************************
\begin{eqnarray}
V(\r)
=
- 0.6635\,\GeV + (0.733\,\GeV)\,
\log\bigl (\r\cdot 1\,\GeV\bigr ) \,,
\end{eqnarray}
%*****************************************
%
with $m_c=1.5\,\GeV$ and $m_b=5.0\,\GeV$, is motivated by Ref.~\cite{Quigg:1977dd} and
was also used in quarkonia photoproduction studies in Ref.~\cite{Hufner:2000jb}. 

%
%
%=========================================
\subsection{Power-law potential}
%=========================================

The effective power-law potential (\textbf{POW}) is given by
%
%*****************************************
\begin{eqnarray}
V(\r)
=
- 6.41\,\GeV + (6.08\,\GeV)\,(\r\cdot 1\,\GeV)^{0.106} \,,
\end{eqnarray}
%*****************************************
%
with $m_c=1.334\,\GeV$ and $m_b=4.721\,\GeV$, is motivated by Ref.~\cite{Martin:1980jx,Martin:1980xh} and the values were taken from Ref.~\cite{Barik:1980ai}.

%
%
%=========================================
\subsection{Buchm\"uller-Tye potential}
%=========================================
%
%

The Buchm\"uller-Tye potential (\textbf{BT}) \cite{Buchmuller:1980su} has a Coulomb-like behaviour at small $\r$ and a string-like behaviour at large $\r$. 
Its structure is similar to the Cornell potential but with additional corrections, particularly effective at small $\r$. Namely,
%
%*****************************************
\begin{eqnarray}
V(\r)
=
\frac{k}{\r}-\frac{8\pi}{27}\frac{v(\lambda \r)}{\r} \,,
\end{eqnarray}
%****************************************
%
for $\r\geq 0.01\,\fm$, and
%
%****************************************
\begin{eqnarray}
V(\r)
=
-\frac{16\pi}{25}\frac{1}{\r\,\ln\bigl(w(\r)\bigr)}
\left(1 + 2\left(\gamma_E+\frac{53}{75}\right)
\frac{1}{\ln\bigl(w(\r)\bigr)} -
\frac{462}{625}\,\,
\frac{\ln\Bigl( \ln\bigl( w(\r) \bigr)\Bigr)}
{\ln\bigl( w(\r)\bigr)}\right) \,,
\end{eqnarray}
%****************************************
%
for $\r < 0.01\,\fm$. Here,
%
%****************************************
\begin{eqnarray}
w(\r) 
= 
\frac{1}{\lambda^2_{\rm MS}\,\r^2}\,,\qquad
\lambda_{\rm MS}=0.509\,\GeV\,,\qquad 
k=0.153\,\GeV^2\,,\qquad
\lambda=0.406\,\GeV\,,
\end{eqnarray}
%****************************************
%
$\gamma_E=0.5772$ is the Euler constant, and the function $v(x)$ is provided numerically in Ref.~\cite{Buchmuller:1980su}. This potential uses the following quark mass values: $m_c=1.48\,\GeV$ and $m_b=4.87\,\GeV$. 

%
%
%
%===============================================
\section{Spatial quarkonium wave function in the $Q\bar Q$ rest frame}
\label{App:Schrodinger}
%===============================================
%
%
%

The spatial part of the quarkonium wave function satisfies the Schr\"odinger equation \cite{Hufner:2000jb}
%
%****************************************
\begin{eqnarray}
\left(-\frac{\Delta}{2\mu}+V(\r)\right)\Psi_{nlm}(\vec\r)=E_{nl}\Psi_{nlm}(\vec\r) \,, \qquad
\mu=\frac{m_Q}{2} \,,
\label{schr}
\end{eqnarray}
%****************************************
%
where $\mu$ is the reduced mass of the $Q\bar Q$ pair, and the operator $\Delta$ acts on the coordinate $\r$ and has the following form
%
%****************************************
\begin{eqnarray}
\Delta
=
\sum\limits_{i=1}^{3}\frac{\partial^{2}}{\partial x_i^{2}}
=
\frac{1}{\r^{2}}\frac{\partial}{\partial \r}\left(\r^{2}
\frac{\partial}{\partial \r}\right)
+
\frac{1}{\r^{2}\sin\theta}\,
\frac{\partial}{\partial\theta}\left(\sin\theta\,
\frac{\partial}{\partial\theta}\right)
+
\frac{1}{\r^{2}\sin^{2}\theta}\,\frac{\partial^{2}}{\partial\varphi^{2}} \,.
\label{laplace}
\end{eqnarray}
%****************************************
%
Factorizing the spatial wave function into the radial and angular parts,
%
%****************************************
\begin{eqnarray}
\Psi_{nlm}(\vec\r)=\psi_{nl}(\r)\, Y_{lm}(\theta,\varphi)
\end{eqnarray}
%****************************************
%
the Schr\"odinger equation (\ref{schr}) with (\ref{laplace}) can be expressed as the following two equations,
%
%****************************************
\begin{eqnarray}
\frac{1}{\r}\frac{\partial^{2}}{\partial \r^{2}}\bigl(\r\psi(\r)\bigr)
+
m_Q\bigl(E-V(\r)\bigr)\psi(\r)
&=&
\frac{l(l+1)}{\r^{2}}\psi(\r)
\nonumber\\
\frac{1}{\sin\theta}\,\frac{\partial}{\partial\theta}
\left(\sin\theta\,\frac{\partial Y(\theta,\varphi)}{\partial\theta}\right)
+
\frac{1}{\sin^{2}\theta}\,\frac{\partial^{2} Y(\theta,\varphi)}
{\partial\varphi^{2}}
&=&
- l(l+1) Y(\theta,\varphi)
\end{eqnarray}
%*****************************************
%
with $l=0$ for $S$-wave states, $l=1$ for $P$-waves, etc. 
The first differential equation for the radial wave function $\psi(\r)$ in the $Q\bar Q$ rest frame can be rewritten in a more convenient form
%
%*****************************************
\begin{eqnarray}
%%%%%%%%%%%%%%%
\label{res-eq}
%%%%%%%%%%%%%%%
\frac{\partial^{2}u(\r)}{\partial \r^{2}}
=
(V_{\rm eff}\bigl(\r)-\epsilon_Q\bigr)u(\r)\,, \qquad 
V_{\rm eff}(\r)
=
m_QV(\r)+\frac{l(l+1)}{\r^{2}}\,,\qquad 
\epsilon_Q
=
m_Q\,E\,.
\end{eqnarray}
%****************************************
%
where the new radial wave function $u(\r)$ is related to $\psi(\r)$ satisfying the following normalization,
%
%****************************************
\begin{eqnarray}
u(\r)
=
\sqrt{4\pi}\,\r\psi(\r) \,, \qquad 
\int\limits_0^{\infty}|u(\r)|^2\dfr \r=1\,, \qquad 
\int|\psi(\r)|^2\dfr^3 \r=1 \,.
%%%%%%%%%%%%%%%%
\label{VMwave_r}
%%%%%%%%%%%%%%%%
\end{eqnarray}
%****************************************
%==============================================
\begin{figure}[!htbp]
\begin{center}
\includegraphics[width=0.47\textwidth]{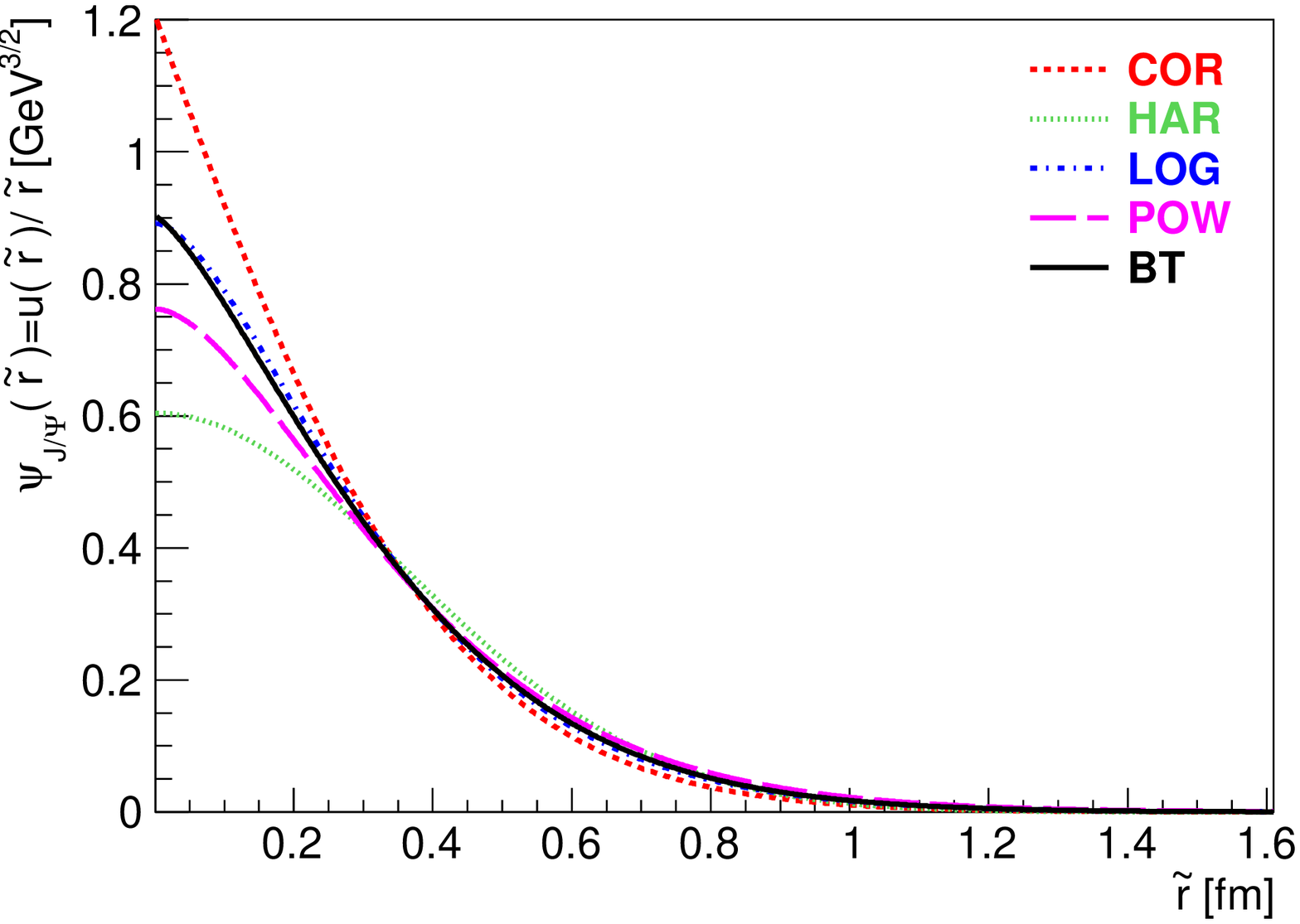}
\includegraphics[width=0.47\textwidth]{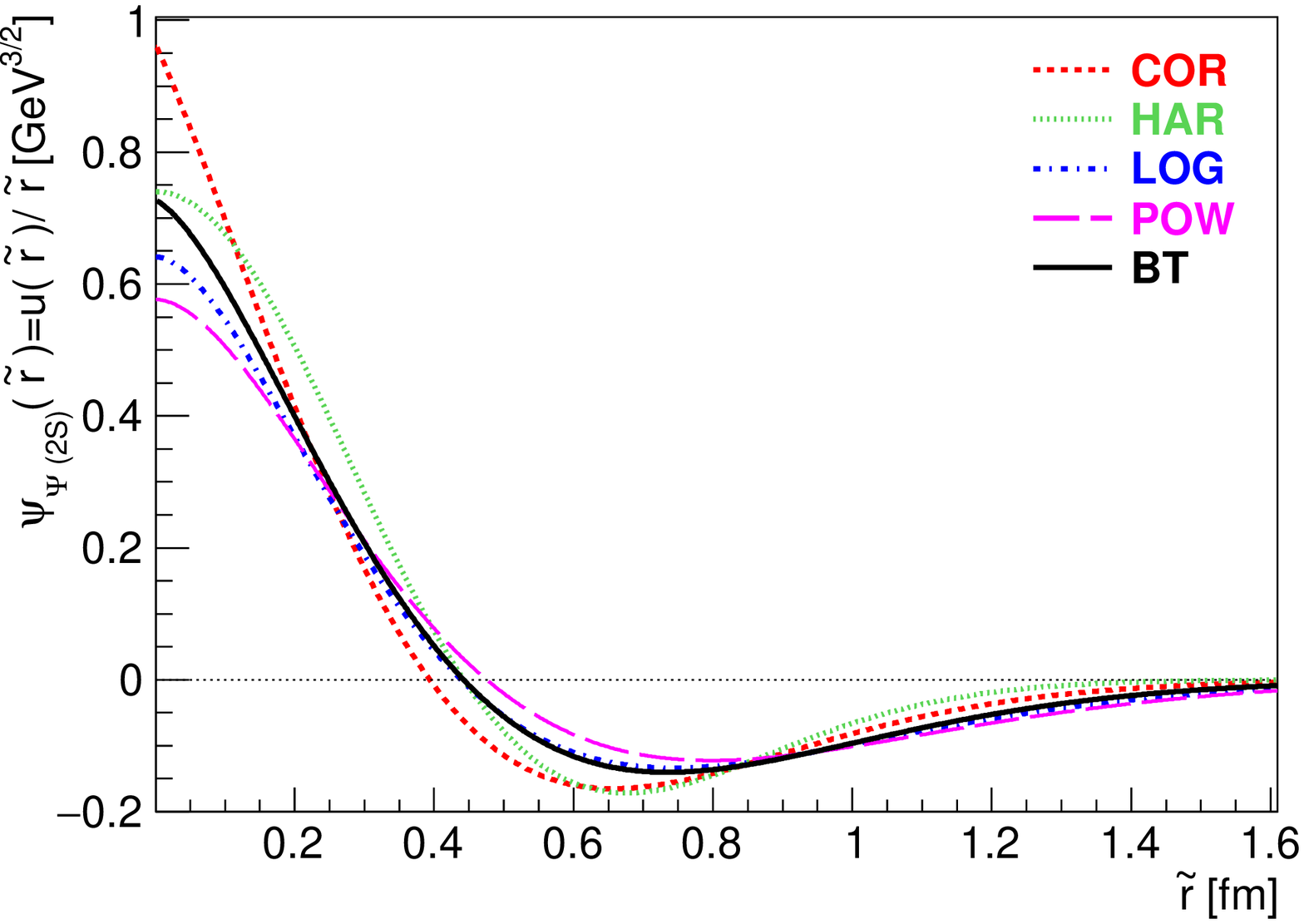}
  \caption{The radial part of the wave function $\psi(\r)$ for the $\Jpsi(1S)$ (left panel) and $\psip(2S)$ (right panel) mesons as a solution of the Schr\"odinger equation for five distinct $c-\bar c$ interaction potentials described in the Appendix~\ref{App:potentials}.}
%%%%%%%%%%%%%%%%%%
  \label{fig:psir}
%%%%%%%%%%%%%%%%%%
 \end{center}
 \end{figure}
%==============================================
%==============================================
 \begin{figure}[!hbtp]
\begin{center}
\includegraphics[width=0.47\textwidth]{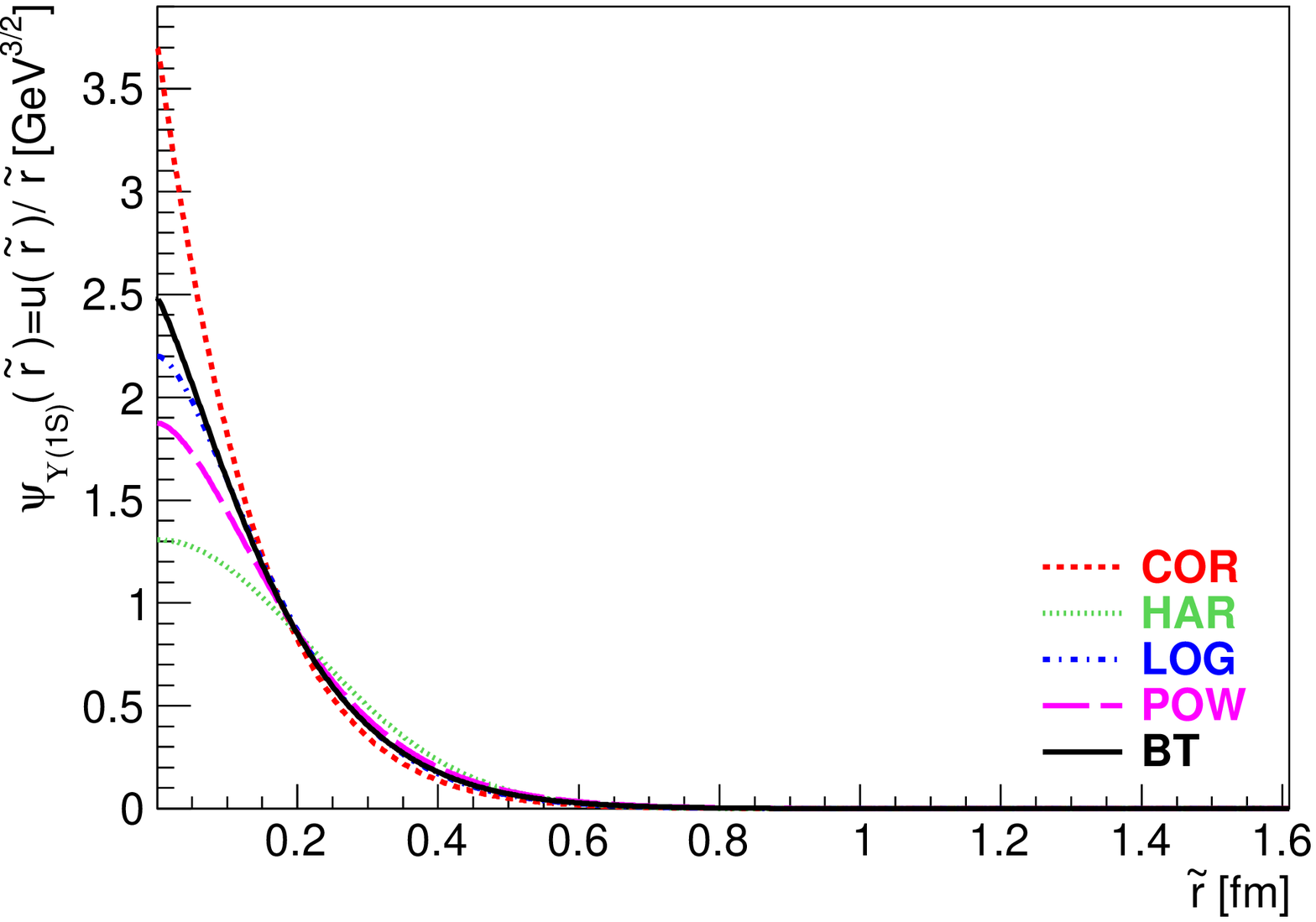}
\includegraphics[width=0.47\textwidth]{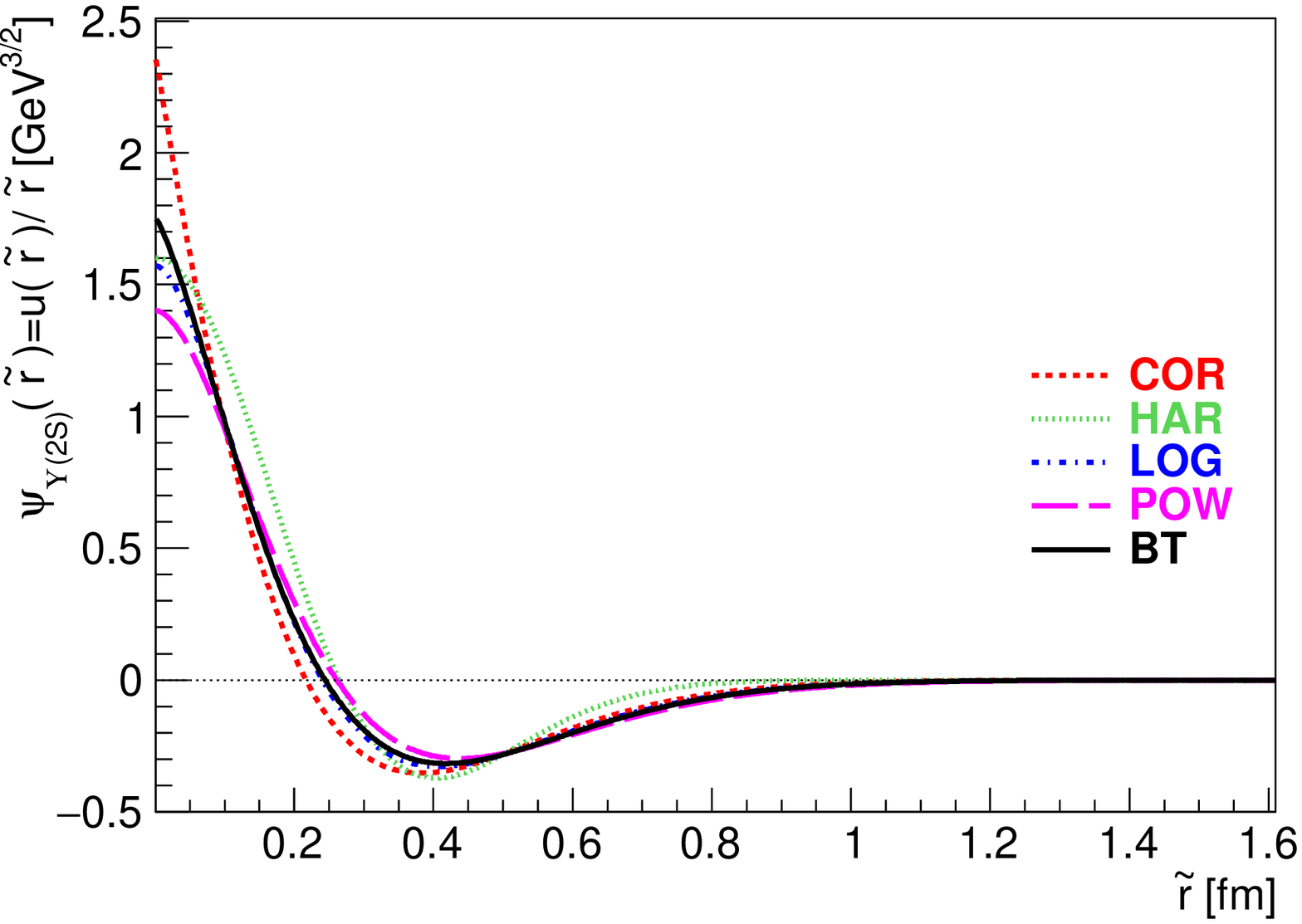}
  \caption{The radial part of the wave function $\psi(\r)$ for the $\Y(1S)$ (left panel) and $\Yp(2S)$ (right panel) mesons as a solution of the Schr\"odinger equation for five distinct $b-\bar b$ interaction potentials described in the Appendix~\ref{App:potentials}.}
%%%%%%%%%%%%%%%%%%
  \label{fig:upsr}
%%%%%%%%%%%%%%%%%%
 \end{center}
 \end{figure}
%==============================================
 
The Schr\"odinger equation (\ref{res-eq}) can be solved numerically, e.g. as a special case of the second-order differential equation by means of the Numerov method \cite{thijssen2007} or converting this equation into a set of the first-order differential equations by means of the Runge-Kutta method \cite{Lucha:1998xc}, for each of the five distinct $Q-\bar Q$ interaction potentials discussed in Appendix~\ref{App:potentials}. The numerical results for the radial wave function $\psi(\r)$ generated by various $c-\bar c$ interaction potentials are shown in Fig.~\ref{fig:psir} for the $\Jpsi(1S)$ (left panel) and $\psip(2S)$ (right panel) states. The corresponding results for the $\Y(1S)$ and $\Yp(2S)$ radial wave functions are depicted in Fig.~\ref{fig:upsr}. 

One can see that the variation in the results for $\psi(\r)$, using various interaction potentials, increases towards small $\r$ in the region where a Coulomb-like behavior of potentials becomes important. The enhanced sensitivity of numerical results to the choice of the $Q-\bar Q$ interaction potential appears especially for the $2S$ radially-excited charmonium state due to the nodal structure of the corresponding radial wave function.

%
%
%
%===============================================
\section{Expressions for amplitudes $\mathcal{A}_{L,R}$}
\label{App:Amps}
%===============================================
%
%
%

The resulting expressions for the amplitudes of quarkonia photo- and electroproduction in the polarised photon-nucleon scattering read \cite{Hufner:2000jb}
%
%********************************************
\begin{eqnarray}
&& \mathrm{Im}\mathcal{A}_{L}(x,Q^2)=\int\limits_0^1\dfr z\int\dfr^2r\,\Sigma_{L}(z,r;Q^2)\,\sigma_{q\bar q}(x,r) \,,  
%%%%%%%%%%%%
\label{AL} \\
%%%%%%%%%%%%
&& \Sigma_{L} = Z_q\,\frac{\sqrt{N_c\alpha_{em}}}{2\pi\sqrt{2}}\,4Qz(1-z)K_0(\varepsilon r)\int p_T\dfr p_TJ_0(p_Tr)
\Psi_{V}(z,p_T)\frac{m_Tm_L+m_Q^2}{m_Q(m_T+m_L)} \,, \nonumber
\end{eqnarray}
%********************************************
%
for a longitudinally polarised photon\footnote{Here, we have found an additional factor of $\sqrt{2}$ which was not included in similar calculations of Ref.~\cite{Hufner:2000jb}.}, and
%
%********************************************
\begin{eqnarray}
&& \mathrm{Im}\mathcal{A}_{T}(x,Q^2)=\int\limits_0^1\dfr z\int\dfr^2r \left[\Sigma^{(1)}_{T}(z, r;Q^2)\sigma_{q\bar q}(x,r) +
\Sigma^{(2)}_{T}(z, r;Q^2)\frac{\dfr \sigma_{q\bar q}(x,r)}{\dfr r}\right]\,, %%%%%%%%%%%%%%
\label{AT}  \\
%%%%%%%%%%%%%%
&& \Sigma^{(1)}_{T} = Z_q\,\frac{\sqrt{N_c\alpha_{em}}}{2\pi\sqrt{2}}\,2K_0(\varepsilon r)\int \dfr p_TJ_0(p_Tr)\Psi_{V}(z,p_T)
p_T\,\frac{m_T^2+m_Tm_L-2p_T^2z(1-z)}{m_T+m_L} \,, \nonumber \\
&& \Sigma^{(2)}_{T} = Z_q\,\frac{\sqrt{N_c\alpha_{em}}}{2\pi\sqrt{2}}\,2K_0(\varepsilon r)\int \dfr p_T J_1(p_Tr)
\Psi_{V}(z,p_T)\frac{p_T^2}{2}\,\frac{m_T+m_L+m_T(1-2z)^2}{m_T(m_T+m_L)} \,, \nonumber
\end{eqnarray}
%********************************************
%
for a transversely polarised photon. In the above formulas,
%
%*********************************************
\begin{eqnarray}
m_T^2 = m_Q^2 + p_T^2 \,, \qquad m_L^2 = 4m_Q^2\,z(1-z) \,,
\end{eqnarray}
%**********************************************
%
such that the meson mass squared reads
%
%**********************************************
\begin{eqnarray}
M_V^2 = \frac{m_T^2}{z(1-z)} \,.
\end{eqnarray}
%**********************************************

%
%
%
%======================================================
\bibliographystyle{unsrt}

\begin{thebibliography}{99}
%=======================================================
%
%
%

\bibitem{Brambilla:2010cs} 
   N.~Brambilla {\it et al.};
%  {\it Heavy quarkonium: progress, puzzles, and opportunities},
   Eur. Phys. J. C\textbf{71}, 1534 (2011).

\bibitem{Ivanov:2004ax} 
  I.P.~Ivanov, N.N.~Nikolaev and A.A.~Savin;
  %``Diffractive vector meson production at HERA: From soft to hard QCD,''
  Phys.\ Part.\ Nucl.\ \textbf{37}, 1 (2006).

\bibitem{Brambilla:2004wf} 
  N.~Brambilla {\it et al.} [Quarkonium Working Group];
  %``Heavy quarkonium physics,''
  \textbf{hep-ph/0412158}.
  %%CITATION = HEP-PH/0412158;%%

\bibitem{satz} 
   T.~Matsui and H.~Satz;
   Phys. Lett. B\textbf{178}, 416 (1986).

\bibitem{Nemchik:1996cw} 
   J.~Nemchik, N.N.~Nikolaev, E.~Predazzi and B.G.~Zakharov;
%  {\it Color dipole phenomenology of diffractive electroproduction of light vector mesons at HERA},
   Z. Phys. C\textbf{75}, 71 (1997).

\bibitem{Frankfurt:1995jw} 
  L.~Frankfurt, W.~Koepf and M.~Strikman;
  %``Hard diffractive electroproduction of vector mesons in QCD,''
  Phys.\ Rev.\ D\textbf{54}, 3194 (1996).

\bibitem{Melosh:1974cu} 
  H.J.~Melosh;
  %``Quarks: Currents and constituents,''
  Phys.\ Rev.\ D\textbf{9}, 1095 (1974).

\bibitem{Hufner:2000jb} 
  J.~Hufner, Y.P.~Ivanov, B.Z.~Kopeliovich and A.V.~Tarasov;
  %``Photoproduction of charmonia and total charmonium proton cross-sections,''
  Phys.\ Rev.\ D\textbf{62}, 094022 (2000).

\bibitem{jan-18} 
  M.~Krelina, J.~Nemchik, R.~Pasechnik and J.~Cepila;
  %``Spin rotation effects in diffractive electroproduction of heavy quarkonia,''
  \textbf{arXiv:1812.03001 [hep-ph]}.
  %%CITATION = ARXIV:1812.03001;%%

\bibitem{Nemchik:1994fp} 
   J.~Nemchik, N.N.~Nikolaev, B.G.~Zakharov;
%  {\it Scanning the BFKL pomeron in elastic production of vector mesons at HERA},
   Phys. Lett. B\textbf{341}, 228 (1994).

\bibitem{Kopeliovich:1991pu} 
  B.Z.~Kopeliovich and B.G.~Zakharov;
  %``Quantum effects and color transparency in charmonium photoproduction on nuclei,''
  Phys.\ Rev.\ D\textbf{44}, 3466 (1991).

\bibitem{Kopeliovich:1993pw} 
  B.Z.~Kopeliovich, J.~Nemchik, N.N.~Nikolaev and B.G.~Zakharov;
  %``Decisive test of color transparency in exclusive electroproduction of vector mesons,''
  Phys.\ Lett.\ B\textbf{324}, 469 (1994).

\bibitem{Kopeliovich:1981pz} 
  B.Z.~Kopeliovich, L.I.~Lapidus and A.B.~Zamolodchikov;
  %``Dynamics of Color in Hadron Diffraction on Nuclei,''
  JETP Lett.\ \textbf{33}, 595 (1981)
  [Pisma Zh.\ Eksp.\ Teor.\ Fiz.\ \textbf{33}, 612 (1981)].
  %%CITATION = JTPLA,33,595;%%

\bibitem{Nikolaev:1994kk} 
   N.N.~Nikolaev and B.G.~Zakharov;
%  {\it The Pomeron in diffractive deep inelastic scattering},
   J. Exp. Theor. Phys. \textbf{78}, 598 (1994).

\bibitem{Kopeliovich:1999am} 
  B.Z.~Kopeliovich, A.~Schafer and A.V.~Tarasov;
  %``Nonperturbative effects in gluon radiation and photoproduction of quark pairs,''
  Phys.\ Rev.\ D\textbf{62}, 054022 (2000).

\bibitem{Basso:2015pba} 
   E.~Basso, V.P.~Goncalves, J.~Nemchik, R.~Pasechnik and M.~\v{S}umbera;
%  {\it Drell-Yan phenomenology in the color dipole picture revisited}, 
   Phys. Rev. D\textbf{93}, 034023 (2016).

\bibitem{Bertsch:1981py} 
   G.~Bertsch, S.J.~Brodsky, A.S.~Goldhaber and J.F.~Gunion;
%  {\it Diffractive Excitation in QCD},
   Phys. Rev. Lett. \textbf{47}, 297 (1981).

\bibitem{Kopeliovich:1993gk} 
   B.Z.~Kopeliovich, J.~Nemchik, N.N.~Nikolaev and B.G.~Zakharov;
%  {\it Novel color transparency effect: Scanning the wave function of vector mesons},
   Phys. Lett. B\textbf{309}, 179 (1993).

\bibitem{Ryskin:1995hz} 
  M.G.~Ryskin, R.G.~Roberts, A.D.~Martin and E.M.~Levin;
  %``Diffractive J / psi photoproduction as a probe of the gluon density,''
  Z.\ Phys.\ C\textbf{76}, 231 (1997).

\bibitem{Marquet:2007qa} 
  C.~Marquet, R.~B.~Peschanski and G.~Soyez,
  %``Exclusive vector meson production at HERA from QCD with saturation,''
  Phys.\ Rev.\ D {\bf 76}, 034011 (2007).

\bibitem{Dosch:2009fam} 
  H.~G.~Dosch and E.~Ferreira,
  %``Nonperturbative and perturbative aspects of photo- and electroproduction of vector mesons,''
  Eur.\ Phys.\ J.\ C {\bf 51}, 83 (2007).

\bibitem{Jones:2013pga} 
  S.~P.~Jones, A.~D.~Martin, M.~G.~Ryskin and T.~Teubner,
  %``Probes of the small $x$ gluon via exclusive $J/\psi$ and $\Upsilon$ production at HERA and the LHC,''
  JHEP {\bf 1311}, 085 (2013).

\bibitem{Cisek:2014ala} 
  A.~Cisek, W.~Schäfer and A.~Szczurek,
  %``Exclusive photoproduction of charmonia in $\gamma p \to V p$ and $p p \to p V p$ reactions within $k_t$-factorization approach,''
  JHEP {\bf 1504}, 159 (2015).

\bibitem{Armesto:2014sma} 
  N.~Armesto and A.~H.~Rezaeian,
  %``Exclusive vector meson production at high energies and gluon saturation,''
  Phys.\ Rev.\ D {\bf 90}, no. 5, 054003 (2014).

\bibitem{Jones:2015nna} 
  S.~P.~Jones, A.~D.~Martin, M.~G.~Ryskin and T.~Teubner,
  %``Exclusive $J/\psi$ and $\Upsilon$ photoproduction and the low $x$ gluon,''
  J.\ Phys.\ G {\bf 43}, no. 3, 035002 (2016).

\bibitem{GolecBiernat:1998js} 
  K.J.~Golec-Biernat and M.~Wusthoff;
  %``Saturation effects in deep inelastic scattering at low Q**2 and its implications on diffraction,''
  Phys.\ Rev.\ D\textbf{59}, 014017 (1998).

\bibitem{GolecBiernat:1999qd} 
  K.J.~Golec-Biernat and M.~Wusthoff;
  %``Saturation in diffractive deep inelastic scattering,''
  Phys.\ Rev.\ D\textbf{60}, 114023 (1999).

\bibitem{Kogut:1969xa} 
  J.B.~Kogut and D.E.~Soper;
  %``Quantum Electrodynamics in the Infinite Momentum Frame,''
  Phys.\ Rev.\ D\textbf{1}, 2901 (1970).
  
\bibitem{Bjorken:1970ah} 
  J.D.~Bjorken, J.B.~Kogut and D.E.~Soper;
  %``Quantum Electrodynamics at Infinite Momentum: Scattering from an External Field,''
  Phys.\ Rev.\ D\textbf{3}, 1382 (1971).

\bibitem{Kopeliovich:2001hf} 
  B.Z.~Kopeliovich, J.~Raufeisen, A.V.~Tarasov and M.B.~Johnson;
  %``Nuclear effects in the Drell-Yan process at very high-energies,''
  Phys.\ Rev.\ C\textbf{67}, 014903 (2003).

\bibitem{Ryskin:1992ui} 
  M.G.~Ryskin;
  %``Diffractive J / psi electroproduction in LLA QCD,''
  Z.\ Phys.\ C\textbf{57}, 89 (1993).
  
\bibitem{Brodsky:1994kf} 
  S.J.~Brodsky, L.~Frankfurt, J.F.~Gunion, A.H.~Mueller and M.~Strikman;
  %``Diffractive leptoproduction of vector mesons in QCD,''
  Phys.\ Rev.\ D\textbf{50}, 3134 (1994).

\bibitem{Adloff:2000vm} 
  C.~Adloff {\it et al.} [H1 Collaboration];
  %``Elastic photoproduction of J / psi and Upsilon mesons at HERA,''
  Phys.\ Lett.\ B\textbf{483}, 23 (2000).

\bibitem{bronzan-74}
   J.B. Bronzan, G.L. Kane and U.P. Sukhatme;
   Phys. Lett. B\textbf{49}, 272 (1974).

\bibitem{Forshaw:2003ki} 
  J.R.~Forshaw, R.~Sandapen and G.~Shaw;
  %``Color dipoles and rho, phi electroproduction,''
  Phys.\ Rev.\ D\textbf{69}, 094013 (2004).

\bibitem{Terentev:1976jk} 
  M.V.~Terentev;
  %``On the Structure of Wave Functions of Mesons as Bound States of Relativistic Quarks,''
  Sov.\ J.\ Nucl.\ Phys.\ \textbf{24}, 106 (1976)
  [Yad.\ Fiz.\ \textbf{24}, 207 (1976)].

\bibitem{Kopeliovich:2015qna} 
  B.Z.~Kopeliovich, E.~Levin, I.~Schmidt and M.~Siddikov;
  %``Lorentz-boosted description of a heavy quarkonium,''
  Phys.\ Rev.\ D\textbf{92}, 034023 (2015).
  
\bibitem{Goncalves:2006yt} 
  V.P.~Goncalves, M.S.~Kugeratski, M.V.T.~Machado and F.S.~Navarra;
  %``Saturation physics at HERA and RHIC: An Unified description,''
  Phys.\ Lett.\ B\textbf{643}, 273 (2006).

\bibitem{Iancu:2003ge} 
  E.~Iancu, K.~Itakura and S.~Munier;
  %``Saturation and BFKL dynamics in the HERA data at small x,''
  Phys.\ Lett.\ B\textbf{590}, 199 (2004).

\bibitem{Kowalski:2006hc} 
  H.~Kowalski, L.~Motyka and G.~Watt;
  %``Exclusive diffractive processes at HERA within the dipole picture,''
  Phys.\ Rev.\ D\textbf{74}, 074016 (2006).

\bibitem{Soyez:2007kg} 
  G.~Soyez;
  %``Saturation QCD predictions with heavy quarks at HERA,''
  Phys.\ Lett.\ B\textbf{655}, 32 (2007).

\bibitem{Kowalski:2003hm} 
  H.~Kowalski and D.~Teaney;
  %``An Impact parameter dipole saturation model,''
  Phys.\ Rev.\ D\textbf{68}, 114005 (2003).

\bibitem{Rezaeian:2012ji} 
  A.H.~Rezaeian, M.~Siddikov, M.~Van de Klundert and R.~Venugopalan;
  %``Analysis of combined HERA data in the Impact-Parameter dependent Saturation model,''
  Phys.\ Rev.\ D\textbf{87}, 034002 (2013).

\bibitem{Rezaeian:2013tka} 
  A.H.~Rezaeian and I.~Schmidt;
  %``Impact-parameter dependent Color Glass Condensate dipole model and new combined HERA data,''
  Phys.\ Rev.\ D\textbf{88}, 074016 (2013).
  
\bibitem{Bartels:2002cj} 
  J.~Bartels, K.J.~Golec-Biernat and H.~Kowalski;
  %``A modification of the saturation model: DGLAP evolution,''
  Phys.\ Rev.\ D\textbf{66}, 014001 (2002).

\bibitem{deSantanaAmaral:2006fe} 
  J.T.~de Santana Amaral, M.B.~Gay Ducati, M.A.~Betemps and G.~Soyez;
  %``gamma* p cross-section from the dipole model in momentum space,''
  Phys.\ Rev.\ D\textbf{76}, 094018 (2007).

\bibitem{Boer:2007ug} 
  D.~Boer, A.~Utermann and E.~Wessels;
  %``Geometric Scaling at RHIC and LHC,''
  Phys.\ Rev.\ D\textbf{77}, 054014 (2008).
  
\bibitem{Dumitru:2005gt} 
  A.~Dumitru, A.~Hayashigaki and J.~Jalilian-Marian;
  %``The Color glass condensate and hadron production in the forward region,''
  Nucl.\ Phys.\ A\textbf{765}, 464 (2006).
  
\bibitem{Kharzeev:2004yx} 
  D.~Kharzeev, Y.V.~Kovchegov and K.~Tuchin;
  %``Nuclear modification factor in d+Au collisions: Onset of suppression in the color glass condensate,''
  Phys.\ Lett.\ B\textbf{599}, 23 (2004).

\bibitem{JalilianMarian:1997gr} 
  J.~Jalilian-Marian, A.~Kovner, A.~Leonidov and H.~Weigert;
  %``The Wilson renormalization group for low x physics: Towards the high density regime,''
  Phys.\ Rev.\ D\textbf{59}, 014014 (1998).

\bibitem{JalilianMarian:1997dw} 
  J.~Jalilian-Marian, A.~Kovner and H.~Weigert;
  %``The Wilson renormalization group for low x physics: Gluon evolution at finite parton density,''
  Phys.\ Rev.\ D\textbf{59}, 014015 (1998).

\bibitem{Kovner:2000pt} 
  A.~Kovner, J.G.~Milhano and H.~Weigert;
  %``Relating different approaches to nonlinear QCD evolution at finite gluon density,''
  Phys.\ Rev.\ D\textbf{62}, 114005 (2000).
  
\bibitem{Weigert:2000gi} 
  H.~Weigert;
  %``Unitarity at small Bjorken x,''
  Nucl.\ Phys.\ A\textbf{703}, 823 (2002).
  
\bibitem{Balitsky:1995ub} 
  I.~Balitsky;
  %``Operator expansion for high-energy scattering,''
  Nucl.\ Phys.\ B\textbf{463}, 99 (1996).
  
\bibitem{Balitsky:1998kc} 
  I.~Balitsky;
  %``Factorization for high-energy scattering,''
  Phys.\ Rev.\ Lett.\ \textbf{81}, 2024 (1998).

\bibitem{McLerran:1994vd} 
  L.D.~McLerran and R.~Venugopalan;
  %``Green's functions in the color field of a large nucleus,''
  Phys.\ Rev.\ D\textbf{50}, 2225 (1994).

\bibitem{McLerran:1993ka} 
  L.D.~McLerran and R.~Venugopalan;
  %``Gluon distribution functions for very large nuclei at small transverse momentum,''
  Phys.\ Rev.\ D\textbf{49}, 3352 (1994).

\bibitem{Kovchegov:1999yj} 
  Y.~V.~Kovchegov;
  %``Small x F(2) structure function of a nucleus including multiple pomeron exchanges,''
  Phys.\ Rev.\ D\textbf{60}, 034008 (1999).

\bibitem{GolecBiernat:2003ym} 
  K.J.~Golec-Biernat and A.M.~Stasto;
  %``On solutions of the Balitsky-Kovchegov equation with impact parameter,''
  Nucl.\ Phys.\ B\textbf{668}, 345 (2003).

\bibitem{Blaettel:1993rd} 
  B.~Blaettel, G.~Baym, L.L.~Frankfurt and M.~Strikman;
  %``How transparent are hadrons to pions?,''
  Phys.\ Rev.\ Lett.\ \textbf{70}, 896 (1993).

\bibitem{Frankfurt:1993it} 
  L.~Frankfurt, G.A.~Miller and M.~Strikman;
  %``Coherent nuclear diffractive production of mini - jets: Illuminating color transparency,''
  Phys.\ Lett.\ B\textbf{304}, 1 (1993).

\bibitem{Frankfurt:1996ri} 
  L.~Frankfurt, A.~Radyushkin and M.~Strikman;
  %``Interaction of small size wave packet with hadron target,''
  Phys.\ Rev.\ D\textbf{55}, 98 (1997).

\bibitem{Nikolaev:1994cn} 
  N.N.~Nikolaev and B.G.~Zakharov;
  %``BFKL evolution and universal structure function at very small x,''
  Phys.\ Lett.\ B\textbf{327}, 157 (1994).

\bibitem{Gribov:1972ri} 
  V.N.~Gribov and L.N.~Lipatov;
  %``Deep inelastic e p scattering in perturbation theory,''
  Sov.\ J.\ Nucl.\ Phys.\ \textbf{15}, 438 (1972)
  [Yad.\ Fiz.\ \textbf{15}, 781 (1972)].

\bibitem{Altarelli:1977zs} 
  G.~Altarelli and G.~Parisi;
  %``Asymptotic Freedom in Parton Language,''
  Nucl.\ Phys.\ B\textbf{126}, 298 (1977).

\bibitem{Dokshitzer:1977sg} 
  Y.~L.~Dokshitzer;
  %``Calculation of the Structure Functions for Deep Inelastic Scattering and e+ e- Annihilation by Perturbation Theory in Quantum Chromodynamics.,''
  Sov.\ Phys.\ JETP \textbf{46}, 641 (1977)
  [Zh.\ Eksp.\ Teor.\ Fiz.\ \textbf{73}, 1216 (1977)].

\bibitem{Cepila:2015qea} 
  J.~Cepila and J.G.~Contreras;
  %``Rapidity dependence of saturation in inclusive HERA data with the rcBK equation,''
  \textbf{arXiv:1501.06687 [hep-ph]}.

\bibitem{Albacete:2007yr} 
  J.L.~Albacete and Y.V.~Kovchegov;
  %``Solving high energy evolution equation including running coupling corrections,''
  Phys.\ Rev.\ D\textbf{75}, 125021 (2007).

\bibitem{Albacete:2009fh} 
  J.L.~Albacete, N.~Armesto, J.G.~Milhano and C.A.~Salgado;
  %``Non-linear QCD meets data: A Global analysis of lepton-proton scattering with running coupling BK evolution,''
  Phys.\ Rev.\ D\textbf{80}, 034031 (2009).

\bibitem{Albacete:2010sy} 
  J.L.~Albacete, N.~Armesto, J.G.~Milhano, P.~Quiroga-Arias and C.A.~Salgado;
  %``AAMQS: A non-linear QCD analysis of new HERA data at small-x including heavy quarks,''
  Eur.\ Phys.\ J.\ C\textbf{71}, 1705 (2011).

\bibitem{McLerran:1997fk} 
  L.D.~McLerran and R.~Venugopalan;
  %``Boost covariant gluon distributions in large nuclei,''
  Phys.\ Lett.\ B\textbf{424}, 15 (1998).

\bibitem{Iancu:2015joa} 
  E.~Iancu, J.D.~Madrigal, A.H.~Mueller, G.~Soyez and D.N.~Triantafyllopoulos;
  %``Collinearly-improved BK evolution meets the HERA data,''
  Phys.\ Lett.\ B\textbf{750}, 643 (2015).

\bibitem{Barnett:1996yz} 
  R.M.~Barnett {\it et al.};
  %``Particle physics summary,''
  Rev.\ Mod.\ Phys.\ \textbf{68}, 611 (1996).
  
\bibitem{Amendolia:1986wj} 
  S.R.~Amendolia {\it et al.} [NA7 Collaboration];
  %``A Measurement of the Space - Like Pion Electromagnetic Form-Factor,''
  Nucl.\ Phys.\ B\textbf{277}, 168 (1986).

\bibitem{Watt:2007nr} 
  G.~Watt and H.~Kowalski;
  %``Impact parameter dependent colour glass condensate dipole model,''
  Phys.\ Rev.\ D\textbf{78}, 014016 (2008).

% ===== Results =============================================================

\bibitem{Aktas:2005xu} 
  A.~Aktas {\it et al.} [H1 Collaboration];
  %``Elastic J/psi production at HERA,''
  Eur.\ Phys.\ J.\ C\textbf{46}, 585 (2006).

\bibitem{Alexa:2013xxa} 
  C.~Alexa {\it et al.} [H1 Collaboration];
  %``Elastic and Proton-Dissociative Photoproduction of J/psi Mesons at HERA,''
  Eur.\ Phys.\ J.\ C\textbf{73}, 2466 (2013).

\bibitem{Chekanov:2002xi} 
  S.~Chekanov {\it et al.} [ZEUS Collaboration];
  %``Exclusive photoproduction of J / psi mesons at HERA,''
  Eur.\ Phys.\ J.\ C\textbf{24}, 345 (2002).

\bibitem{Chekanov:2004mw} 
  S.~Chekanov {\it et al.} [ZEUS Collaboration];
  %``Exclusive electroproduction of J/psi mesons at HERA,''
  Nucl.\ Phys.\ B\textbf{695}, 3 (2004).

\bibitem{nnn-94}
   N.N.~Nikolaev, B.G.~Zakharov and V.R.~Zoller;
   %\textit{The direct calculation of the slope of the QCD pomeron's trajectory}
   J. Exp. Theor. Phys. Lett. \textbf{60}, 694 (1994).

\bibitem{jan-98}
   J.~Nemchik, N.N.~Nikolaev, E.~Predazzi, B.G.~Zakharov and V.R.~Zoller;
   %\textit{The diffration cone for exclusive vector meson production
   %        in deep inelastic scattering}
   J. Exp. Theor. Phys. \textbf{86}, 1054 (1998).

\bibitem{Buchmuller:1980su} 
  W.~Buchmuller and S.H.H.~Tye;
  %``Quarkonia and Quantum Chromodynamics,''
  Phys.\ Rev.\ D\textbf{24}, 132 (1981).

\bibitem{TheALICE:2014dwa} 
  B.~B.~Abelev {\it et al.} [ALICE Collaboration];
  %``Exclusive $\mathrm{J/}\psi$ photoproduction off protons in ultra-peripheral p-Pb collisions at $\sqrt{s_{\rm NN}}=5.02$ TeV,''
  Phys.\ Rev.\ Lett.\ \textbf{113}, 232504 (2014).
  
\bibitem{Binkley:1981kv} 
  M.~E.~Binkley {\it et al.};
  %``J/psi Photoproduction from 60-GeV/c to 300-GeV/c,''
  Phys.\ Rev.\ Lett.\ \textbf{48}, 73 (1982).

\bibitem{Denby:1983az} 
  B.~H.~Denby {\it et al.};
  %``Inelastic and Elastic Photoproduction of J/$\psi$ (3097),''
  Phys.\ Rev.\ Lett.\ \textbf{52}, 795 (1984).

\bibitem{Breitweg:1998ki} 
  J.~Breitweg {\it et al.} [ZEUS Collaboration];
  %``Measurement of elastic Upsilon photoproduction at HERA,''
  Phys.\ Lett.\ B\textbf{437}, 432 (1998).
  
\bibitem{Chekanov:2009zz} 
  S.~Chekanov {\it et al.} [ZEUS Collaboration];
  %``Exclusive photoproduction of upsilon mesons at HERA,''
  Phys.\ Lett.\ B\textbf{680}, 4 (2009).
  
\bibitem{CMS:2016nct} 
  CMS Collaboration [CMS Collaboration];
  %``Measurement of exclusive Y photoproduction in pPb collisions at $\sqrt{s_{_\mathrm{NN}}} = 5.02~\mathrm{TeV}$,''
  \textbf{CMS-PAS-FSQ-13-009}.
  
\bibitem{Aaij:2015kea} 
  R.~Aaij {\it et al.} [LHCb Collaboration];
  %``Measurement of the exclusive Υ production cross-section in pp collisions at $ \sqrt{s}=7 $ TeV and 8 TeV,''
  JHEP \textbf{1509}, 084 (2015).

\bibitem{Adloff:1997yv} 
  C.~Adloff {\it et al.} [H1 Collaboration];
  %``Photoproduction of psi (2S) mesons at HERA,''
  Phys.\ Lett.\ B\textbf{421}, 385 (1998).
  
\bibitem{Adloff:2002re} 
  C.~Adloff {\it et al.} [H1 Collaboration];
  %``Diffractive photoproduction of psi(2S) mesons at HERA,''
  Phys.\ Lett.\ B\textbf{541}, 251 (2002).

\bibitem{Abramowicz:2016xls} 
  H.~Abramowicz {\it et al.} [ZEUS Collaboration];
  %``Measurement of the cross-section ratio sigma_{psi(2S)}/sigma_{J/psi(1S)} in deep inelastic exclusive ep scattering at HERA,''
  Nucl.\ Phys.\ B\textbf{909}, 934 (2016).

\bibitem{Camerini:1975cy} 
  U.~Camerini {\it et al.};
  %``Photoproduction of the psi Particles,''
  Phys.\ Rev.\ Lett.\ \textbf{35}, 483 (1975).

\bibitem{Barate:1986fq} 
  R.~Barate {\it et al.} [NA14 Collaboration];
  %``Measurement of $J/\psi$ and $\psi^\prime$ Real Photoproduction on $^{6}$Li at a Mean Energy of 90-{GeV},''
  Z.\ Phys.\ C\textbf{33}, 505 (1987).

\bibitem{Binkley:1982yn} 
  M.E.~Binkley {\it et al.};
  %``Psi-prime Photoproduction at a Mean Energy of 150-GeV,''
  Phys.\ Rev.\ Lett.\ \textbf{50}, 302 (1983).

\bibitem{Aubert:1982tt} 
  J.J.~Aubert {\it et al.} [European Muon Collaboration];
  %``Production of charmed particles in 250-GeV $\mu^+$ - iron interactions,''
  Nucl.\ Phys.\ B\textbf{213}, 31 (1983).

\bibitem{Amaudruz:1991sr} 
  P.~Amaudruz {\it et al.} [New Muon Collaboration];
  %``Ratio of J / psi production cross-sections in deep inelastic muon scattering from tin and carbon,''
  Nucl.\ Phys.\ B\textbf{371}, 553 (1992).
  
\bibitem{Shuvaev:1999ce} 
  A.G.~Shuvaev, K.J.~Golec-Biernat, A.D.~Martin and M.G.~Ryskin;
  %``Off diagonal distributions fixed by diagonal partons at small x and xi,''
  Phys.\ Rev.\ D\textbf{60}, 014015 (1999).

\bibitem{Martin:1999wb} 
  A.D.~Martin, M.G.~Ryskin and T.~Teubner;
  %``Q**2 dependence of diffractive vector meson electroproduction,''
  Phys.\ Rev.\ D\textbf{62}, 014022 (2000).
  
\bibitem{spots}
   B.Z.~Kopeliovich and B.~Povh;
   J. Phys. G\textbf{30}, S999 (2004).

\bibitem{drops} 
   B.Z.~Kopeliovich, B.~Povh and I.~Schmidt;
   Nucl. Phys. A\textbf{782}, 24 (2007).

% ================ Appendices ==============================================

\bibitem{Eichten:1979ms} 
  E.~Eichten, K.~Gottfried, T.~Kinoshita, K.D.~Lane and T.M.~Yan;
  %``Charmonium: Comparison with Experiment,''
  Phys.\ Rev.\ D\textbf{21}, 203 (1980).

\bibitem{Eichten:1978tg} 
  E.~Eichten, K.~Gottfried, T.~Kinoshita, K.D.~Lane and T.M.~Yan;
  %``Charmonium: The Model,''
  Phys.\ Rev.\ D\textbf{17}, 3090 (1978);
  Erratum: [Phys.\ Rev.\ D\textbf{21}, 313 (1980)].

\bibitem{Quigg:1977dd} 
  C.~Quigg and J.L.~Rosner;
  %``Quarkonium Level Spacings,''
  Phys.\ Lett.\ B\textbf{71}, 153 (1977).

\bibitem{Martin:1980jx}
  A.~Martin;
  %``A FIT of Upsilon and Charmonium Spectra,''
  Phys.\ Lett.\ B\textbf{93}, 338 (1980).

\bibitem{Martin:1980xh}
  A.~Martin;
  %``Heavy Quark Systems,''
  \textbf{CERN-TH-2876, C80-03-09-25}.

\bibitem{Barik:1980ai}
  N.~Barik and S.N.~Jena;
  %``Fine - Hyperfine Splittings Of Quarkonium Levels In An Effective Power Law Potential,''
  Phys.\ Lett.\ B\textbf{97}, 265 (1980).

\bibitem{thijssen2007}
  J.~Thijssen;
  \textit{Computational Physics}, Cambridge University Press (2007).

\bibitem{Lucha:1998xc} 
  W.~Lucha and F.F.~Schoberl;
  %``Solving the Schrodinger equation for bound states with Mathematica 3.0,''
  Int.\ J.\ Mod.\ Phys.\ C\textbf{10}, 607 (1999).

\end{thebibliography}

\end{document}